\documentclass[prb,aps,superscriptaddress,floatfix,tightenlines,showpacs,twocolumn,longbibliography]{revtex4-1}
\usepackage{graphicx}
\usepackage{dcolumn}   
\usepackage{bm}        
\usepackage{amssymb}   
\usepackage{amsmath}
\usepackage{url}
\usepackage{layout}
\usepackage{epsfig}
\usepackage{graphicx}
\usepackage{epstopdf}
\usepackage{booktabs}
\usepackage{float}
\usepackage[hyperindex,breaklinks]{hyperref}
\usepackage{color}
\usepackage{subfigure}

\def\0{\bm 0}

\begin{document}
\title{Theory of transport property of density wave phases in 
three-dimensional metals and semimetals under high magnetic field}

\author{Xiao-Tian Zhang}
\affiliation{International Center for Quantum Materials, Beijing, 100871, China}
\affiliation{Collaborative Innovation Center of Quantum Matter, Beijing, 100871, China}
\author{Ryuichi Shindou}
\email{rshindou@pku.edu.cn}
\affiliation{International Center for Quantum Materials, Beijing, 100871, China}
\affiliation{Collaborative Innovation Center of Quantum Matter, Beijing, 100871, China}
\begin{abstract}
Three-dimensional (3D) metals/semimetals under magnetic field have an instability 
toward a density wave (DW) ordering which breaks a translational symmetry along the 
field direction. Effective boson models for the DW phases 
take forms of XY models with/without Potts terms. Longitudinal conductivity along 
the field direction is calculated in the DW phases with inclusion of effects of 
low-energy charge fluctuation (phason) and disorder. A single-particle imaginary-time Green 
function is identified with a partition function of 3D XY models in the presence of pairs 
of magnetic monopoles. In terms of the celebrated electromagnetic duality, 
electronic spectral function is calculated near the DW phase transition. 
The calculated result shows that the single-particle spectral function acquires 
an additional low-energy feature due to the strong phason fluctuation. 
Relevance to an in-plane conductance due to surface chiral Fermi arc states 
are also discussed. 
\end{abstract}
\pacs{ }

\maketitle

\section{Introduction} 

An investigation of magnetic-field-induced many-body states in three dimensional (3D)  
metal and semimetal has a distinguished history~\cite{hal2}. 
Unlike the two-dimensional case, where the kinetic energy is completely quenched,
the 3D metal/semimetal under high magnetic field 
has one-dimensional electronic dispersions along the field 
direction. An electron-electron interaction mediates a coupling among them, 
giving rise to a generic $2k_{\rm F}$ instability toward various spontaneous symmetry 
breaking (SSB) phases such as charge density wave, spin density wave and Wigner crystal phases~\cite{hal2,cm,bra,ke,fenton,ms,bc,fuku1}. 
In expectation of these phases, high magnetic field transport experiments have been 
carried out during last a few decades in semimetal materials with 
dilute electron densities, where smaller magnetic fields enable us to 
reach the (quasi) quantum limit.  
These experimental efforts lead to discoveries of unusual consecutive metal-insulator 
transitions in graphite above the quasi quantum limit~\cite{tanuma,iye1,iye2,ys,uji,kope,kumar,fauque,akiba}, 
first-order phase transition into a phase with larger magnetic anisotropy in bismuth~\cite{li,be,ban}, 
and an abrupt field-induced quantum phase transition in Weyl semimetals 
such as tantalumn phosphide~\cite{cz1,cz2}. 
Previous theories \cite{cm,bra,ke,fenton,ms,bc,fuku1,yf,th,mb,aaa,yako,tg,ab} studied   
possible SSB phases in (some of) these systems by means of energy estimations as well as 
functional renormalization group analyses. Meanwhile, few theories \cite{tmg} have been done on  
transport properties of respective SSB phases, which hinder us from deciphering 
identities of the low-temperature phases in the experiments.

In this paper, we present a theory of transport properties of density wave (DW) phases in 
a 3D electron gas under a high magnetic field. The DW phases considered in 
this paper break the translational symmetry along the field direction, and can be 
regarded as a canonical example of SSB phases in 3D metals/semimetals  
under the high field~\cite{bra,fuku1,yf,th,mb,yako,tmg}. 
The quasi-one-dimensional electronic system with an electron interaction 
is transformed to an effective boson model, 
which takes a form of coupled one-dimensional wires of Luttinger liquids. 
Using a renormalization group study, 
we first observe that, in the case of an repulsive electron-electron 
interaction, the model exhibits a finite temperature phase transition as well as 
a $T=0$ quantum phase transition between a normal phase and the DW phase. 
Using the boson model, we next calculate longitudinal optical conductivities in the DW 
phases with inclusions of charge fluctuation (phason excitation) effect and impurity 
(disorder) effect. 
When regularized into a lattice, a partition function of 
the effective boson Hamiltonian can be seen as that of a 
classical 3D XY model, while the imaginary-time single-particle {\it electron} 
Green function can be seen as a partition function of the XY model 
in the presence of two pairs of magnetic dipoles. 
By employing the celebrated duality mapping~\cite{peskin,dh,herbut}, 
we evaluate the single-particle electron Green function and spectral 
function in the DW and normal phases. 
The optical conductivity and spectral function thus calculated 
acquire continuum spectral weights in low-frequency 
regime in the DW phases, which reflect the presence of the phason excitations. 
We show that, in the presence of a bulk-surface coupling,  
the finite spectral weight in the single-particle spectral function in {\it bulk} electronic 
states can be transferred into that of the chiral {\it surface} Fermi arc states. 
We further argue that this results in a temperature-dependent in-plane surface 
conductance, providing a possible explanation for 
a recent transport experiment in graphite~\cite{fauque}.

\subsection{highlight of this paper}
\begin{figure}[t]
\centering
\includegraphics[width=0.47\textwidth]{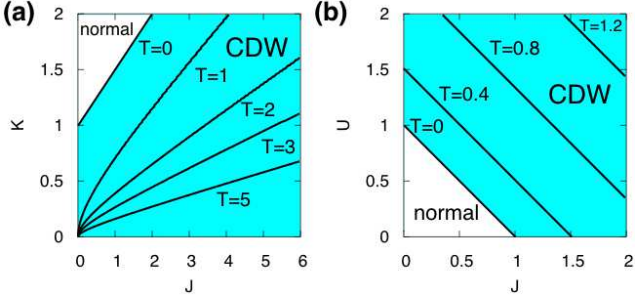}
\caption{(color online) Contour map of three-dimensional phase diagrams subtended by 
(a) $J$, $K$ (Luttinger parameter) and $T$ (temperature) at general electron filling and (b) 
$J$ and $U$ (umklapp term), and $T$ at half electron filling for $K = 1.5 > K_c (=1)$. Phase 
boundaries between normal phase (white color region at $T=0$) and DW phase (blue region) 
are depicted by black curves at different temperatures (read the text for
 units of $J,U,K$ and $T$).} 
\label{4}
\end{figure}

Using the one-dimensional bosonization scheme~\cite{giamarchi}, 
we first bosonize an electron Hamiltonian with a repulsive electron-electron interaction. 
The boson Hamiltonian takes a form of coupled one-dimensional chains. 
Each one-dimensional system is described by a quantum sine-Gordon model with a pair of two 
conjugate phase variables, an electron displacement field $\phi_j(z)$ (along the field direction) 
and electron current field 
$\Pi_{j}(z)$,
\begin{align} 
H &=  \sum_m \int dz \Big\{ \frac{uK \pi }{2} [\Pi_m(z)]^2 +\frac{u}{2\pi K} [\partial_z \phi_m(z)]^2 
\nonumber \\ 
& \ \  -\sum_{j}^{j\ne m} J_{j-m}   \sigma_j^z \sigma_m^z  \cos2[\phi_{j}(z)-\phi_{m}(z)] \Big\}, 
\label{hami-int}
\end{align}
where $j$ denotes the chain index ($j,m=1,2,\cdots,\frac{S}{2\pi l^2}$ with a 
magnetic length $l$ and an area of the system perpendicular to the field $S$). 
$K$ and $u$ stand for Luttinger parameter and renormalized velocity for each one-dimensional 
chain, while  
$\sigma^z_j$ is an Ising variable associated with two Klein factors for left and right movers. 
Due to the repulsive interaction, displacement fields in different chains are 
coupled with each other ferromagnetically;
\begin{align}
J_{m}& \equiv \frac{\sqrt{2\pi} l}{L_x} \!\ J \!\ e^{-\frac{y^2_m}{2l^2}}, \label{Jm-int}   
\end{align}
with $J>0$ and $y_m\equiv 2\pi l^2 m/L_x$ ($m=1,2\cdots$). 
The inter-chain coupling ranges over the magnetic length 
$l$, within which ${\cal O}(L_x/l)$ number of chains are ferromagnetically coupled with one another 
(`long-range' coupling; $L_x$ is a linear dimension of a system size). 
Due to this inter-chain rigidity, the displacement field naturally exhibits a long range order at lower 
temperature or for smaller Luttinger parameter (Fig.~\ref{4}). 
The order is nothing but the DW order which breaks spontaneously the translational symmetry  
along the field direction. At general (incommensurate) electron filling case, where 
$2k_F$ along the field direction is incommensurate to a reciprocal lattice vector along the 
field, the DW order breaks a continuous U(1) translational symmetry. The DW phase is accompanied 
by a gapless Goldstone mode, i.e. phason excitation.

Based on this boson model, we calculate the conductivity in the DW phase 
by taking into account effects of phason excitation and disorder potential. 
The conductivity in the clean limit shows a Drude peak at $\omega=0$ 
at the incommensurate electron filling case, 
${\rm Re}\sigma_{zz}(\omega) = 2A \delta(\omega)$ with $A=e^2 uK/(2\pi l^2)$. 
The Drude peak at $\omega=0$ represents   
the gapless nature of the phason excitation.  

At a commensurate electron filling case, 
where $2k_F$ along the field direction is commensurate to the reciprocal lattice vector, 
the interaction part allows a umklapp process, which adds a phase locking term into 
the effective boson Hamiltonian. For the half filling case, for example, the term takes a form of 
\begin{align}
H_{\rm u} &\equiv -\sum_{j,m} U_{j-m}  \sigma_j^z \sigma_m^z   \int dz 
\cos2[\phi_{j}(z)+\phi_{m}(z)],  \label{hami2-int} \\ 
U_{m}& \equiv \frac{\sqrt{2\pi} l}{L_x} \!\ U \!\ e^{-\frac{y^2_m}{2l^2}}. \label{Um-int} 
\end{align}  
These phase locking terms lock the displacement field on discrete values, 
reducing a symmetry of the Hamiltonian from U(1) to Z$_n$ for the $m/n$ electron filling cases 
respectively ($m,n$ are mutually prime integers). Due to these phase locking terms, the 
phason excitation in the DW phase acquires a finite mass, splitting 
the Drude peak at $\omega=0$ into two resonance peaks at $\omega=\omega_U$,  
${\rm Re}\sigma_{zz}(\omega) = A \delta(|\omega|-\omega_U)$ with $\omega_U= \sqrt{8\pi u K U}$, 
where $U$ stands for the strength of the umklapp term.  
 
\begin{figure}[t]
\centering
\includegraphics[width=0.47\textwidth]{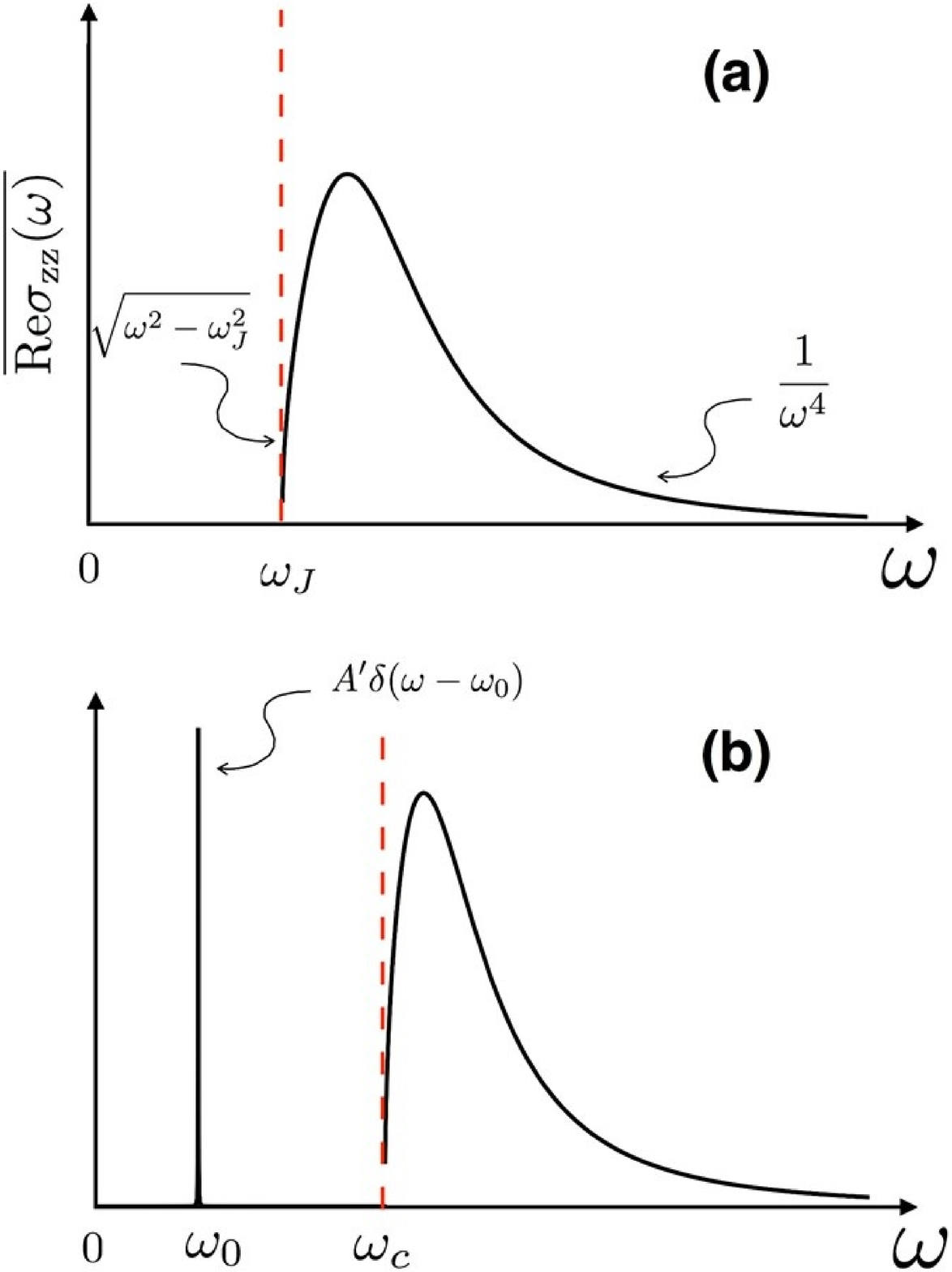}
\caption{(color online) disorder-averaged optical conductivity $\overline{{\rm Re} \sigma_{zz}(\omega)}$ 
as a function of frequency $\omega$. The threshold frequencies are labeled by red dashed lines, above 
which the conductivity shows a continuum spectrum.  (a) DW phase at incommensurate electron filling case. 
(b) DW phase at commensurate electron filling case (weak disorder case, $\bar{g}_y <\bar{g}_{y,c}$; 
see the text). Note that the resonance mode at $\omega=\omega_U$ in the clean limit 
is shifted to $\omega=\omega_0$ due 
to a renormalization by a finite disorder $\overline{g}_y$. $A^{\prime} \equiv 
2A/|F^{\prime}_{-}(\omega_0)|$ where $A\equiv e^2 u K/(2\pi l^2)$ and $F_{-}(\omega)$ and $\omega_0$ 
are defined by eq.~(\ref{Fm-int}). }
\label{1}
\end{figure}

In the presence of finite disorder, the optical conductivity further 
acquires a {\it continuum spectrum} above a threshold frequency 
(Fig.~\ref{1}(a)). It takes a form of 
\begin{align}
\overline{{\rm Re} \sigma_{zz}(\omega)}  
&= \frac{2A}{\pi}\frac{\bar{g}_y} {\sqrt{\omega^2-\omega_J^2}} 
\frac{|\omega| \!\ \Theta(|\omega|-\omega_J)}{\omega^4 
+ \bar{g}_y^2/(\omega^2-\omega_J^2)}, \label{szz-1} 
\end{align}
with $\omega_J  \equiv  \sqrt{4\pi uKJ}$  
in the DW phase at incommensurate electron filling case. $J$ and 
$\bar{g}_y$ stand for the inter-chain coupling strength and 
disorder strength respectively. Remark that the Drude peak in the clean limit 
diminishes immediately once an infinitesimally small $\bar{g}_y$ is introduced. 

In the DW phase at the commensurate filling, 
the conductivity also acquires a continuum spectrum above a threshold 
frequency (Fig.~\ref{1}(b)),   
\begin{align}
& \overline{{\rm Re} \sigma_{zz}(\omega)} =
\frac{2A \omega_0}{|F^{\prime}_{-}(\omega_0)|}  \delta(|\omega|-\omega_{0})  + 
\frac{2A}{\pi} \frac{\bar{g}_y}{\sqrt{\omega^2-\omega_{c}^2}}  \nonumber \\ 
&\hspace{2.5cm}  
\times \frac{|\omega| \!\ \Theta(|\omega|-\omega_{c})}{(\omega^2-\omega_U^2)^2 + 
\bar{g}_y^2/(\omega^2-\omega_{c}^2)},  \label{szz-2} 
\end{align}  
with the threshold frequency $\omega^2_c \equiv \omega^2_J+\omega^2_U/2$. 
Unlike the incommensurate filling case, the resonance mode in the clean 
limit survives even in the presence of finite (but small) disorder; 
the first term in the right hand side of eq.~(\ref{szz-2}). The location of the mode 
(a renormalized mass $\omega_0$) is determined as a root of a 
monotonically-descreasing function,  
\begin{align}
&F_{-}(\omega_0) = 0, \nonumber \\
&F_{-}(\omega) \equiv - \omega^2 + \omega_U^2 - \frac{\bar{g}_y}{\sqrt{\omega_{c}^2 -\omega^2}}. \label{Fm-int} 
\end{align} 
On increasing the disorder strength $\bar{g}_y$, the mass  
becomes smaller. Within the Born approximation, there is a critical disorder 
strength $\bar{g}_{y,c}\equiv \omega^2_U \omega_c$, at which the renormalized mass 
$\omega_0$ becomes zero and above which the resonance mode diminishes. This critical point 
clearly suggests a quantum phase transition from the DW phase to a disorder-driven phase. 

The optical conductivity shown in Fig.~\ref{1} reflects a nature of low-energy collective 
mode (phason excitation) in the DW phases. The 
fluctuation of the displacement and current fields constitute the collective mode, 
which has an energy-momentum dispersion relation $E(k_z,k)$. $k_z$ 
and $k$ are momenta conjugate to $z$ (spatial coordinate along the field) 
and the chain index or coordinate (a spatial coordinate perpendicular to the field) respectively. 
Due to the `long-range' inter-chain coupling, {\it all} the 
collective modes with finite $k$ have a finite mass at $k_z=0$ in 
the thermodynamic limit;
\begin{eqnarray}
E(k_z=0,k\ne 0) = \left\{ \begin{array}{cl} 
 \omega_J & {\rm incommensurate}, \\ 
 \omega_{c} & {\rm commensurate}. \\  
 \end{array} \right.
\end{eqnarray}
With disorder, these finite-$k$ phason modes contribute to the continuum spectrum  
above the threshold frequency. Meanwhile,  
the phason mode at $k=0$ has no mass in the incommensurate filling case 
and has a finite mass in the commensurate 
filling case;
\begin{eqnarray}
E(k_z=0,k = 0) = \left\{ \begin{array}{cl} 
 0 & {\rm incommensurate}, \\ 
 \omega_{U} \!\ ({\rm or} \!\ \omega_0) & {\rm commensurate}. \\  
 \end{array} \right.
\end{eqnarray}
The phason mode at $k=0$ contributes to the Drude peak 
in the incommensurate filling case and the resonance peak 
in the commensurate filling case respectively.

\begin{figure}[t]
\centering
\includegraphics[width=0.47\textwidth]{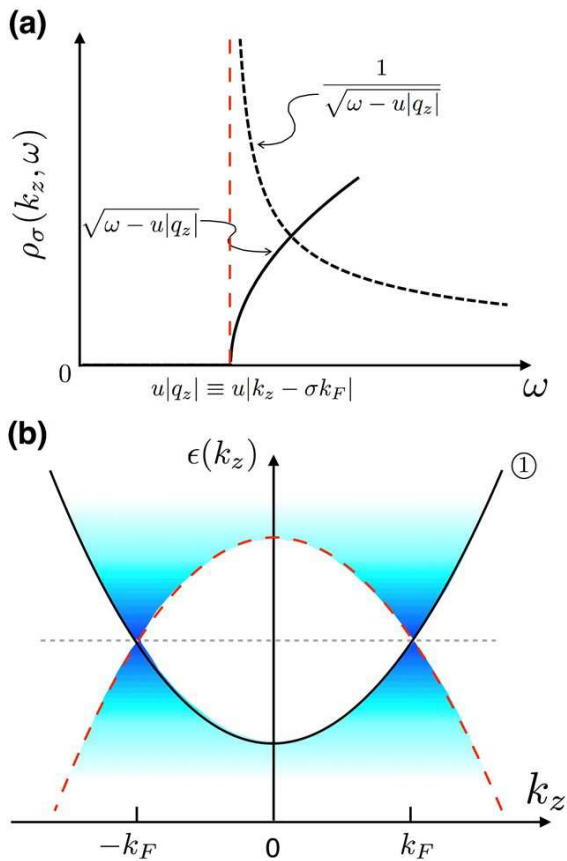}
\caption{(a) Bulk single-particle spectral 
functions $\rho_{\sigma}(k_z,\omega)$ near the two Fermi points ($k_z=\pm k_F$)
in the DW phase (dotted line) and for the normal phase (solid line) 
for the incommensurate electron filling case. $\sigma=\pm$ specifies one 
of the two Fermi points; $k_z = \sigma k_F$. $u$ is a renormalized velocity (see the text). 
(b) Schematic picture of an energy-momentum dispersion for a bulk single-particle 
state $\epsilon_{b}(k_z) = \hbar^2k^2_z/(2m_{*})$ (\textcircled{1}) and dissipative 
region (defined by Eqs.~(\ref{digest1},\ref{digest2}); blue color region).  
$m_{*}$ denotes an effective electron mass 
and straight dotted horizontal line stands for the Fermi energy $\mu$, which defines the 
two Fermi points in the bulk at $k_z=\pm k_F$. $\epsilon_{b}(k_z)$ and $2\mu - \epsilon_{b}(k_z)$ 
(red dotted curved line) bound the dissipative region. Within the dissipative region, the single-particle spectral 
function acquires finite continuum weight as in Fig.~\ref{2}(a).}
\label{2}
\end{figure}

The low-energy phason mode  
induces a strong charge fluctuation, so that it may dramatically modify a naive 
``mean-field''  picture of the single-electron excitation spectrum in the DW phase. 
To explore this possibility, we 
calculate the single-particle spectrum function in terms of the boson Hamiltonian. 
We first note that the $T=0$ partition function of the bosonized 
Hamiltonian can be described by a classical XY model in a cubic lattice, while a single-particle 
imaginary time Green function can be described by a partition function of the 
3D XY model in the presence of two pairs of magnetic monopole and antimonopoles. 
A U(1) phase degree of freedom of the XY model corresponds to the displacement 
field along the field, so that ordered/disordered phases in the XY model correspond  
to the DW/normal phases near the DW phase transition respectively. 

Using the celebrated duality mapping between the XY and frozen 
lattice superconductor (FLS) models~\cite{peskin,dh,herbut}, we calculate an asymptotic behavior 
of the single-particle imaginary-time Green function in the DW/normal 
phases. After an analytic continuation, we obtain the single-particle real-time 
Green function, whose imaginary part is nothing but the single-particle spectral  
function. In the DW phase, the spectral function thus obtained takes a form of 
\begin{eqnarray}
\rho_{\sigma}(q_z,\omega) = \frac{B}{\sqrt{\omega^2-u^2 q^2_z}} \!\ 
\Theta(|\omega|-u |q_z|) + 
\cdots \label{rhoDW}
\end{eqnarray}
with small $q_z=k_z - \sigma k_F$ near the two Fermi points. $\sigma=\pm$ specifies 
one of the two Fermi points. For the normal phase at the incommensurate filling, 
the spectral function is given by   
\begin{eqnarray}
\rho_{\sigma}(q_z,\omega) = B^{\prime} \sqrt{\omega^2-u^2 q^2_z} \!\ \Theta(|\omega|-u |q_z|) + \cdots \label{rhoN}
\end{eqnarray}
where $B$ and $B^{\prime}$ are non-universal parameters. As shown in Fig.~\ref{2}(a), 
the spectral functions in both phases have structures at $\omega = \pm u q_z$ 
with power-law behaviors. The result clearly shows that, due to the strong 
phase fluctuation, the single-particle spectral functions in DW/normal phases 
acquire low-energy continuum spectra at; 
\begin{eqnarray}
|\omega| > u|k_z - \sigma k_F| \label{digest1}
\end{eqnarray}
with $\sigma=\pm$. More generally, the region can be seen as the low-energy 
limit of the following energy-momentum region;  
\begin{eqnarray}
|\omega| > |\epsilon_{b}(k_z) - \mu|. \label{digest2}
\end{eqnarray}
$\epsilon_{b}(k_z)$ denotes the one-dimensional electronic dispersion along the 
field direction, e.g. $\epsilon_{b}(k_z)=\hbar^2k^2_z/(2m_{*})$ with effective electron mass 
$m_{*}$. We dub this region as `dissipative regime' (a blue-colored region in Fig.~\ref{2}(b)). 

\begin{figure}[htbp]
\centering
\includegraphics[width=0.36\textwidth]{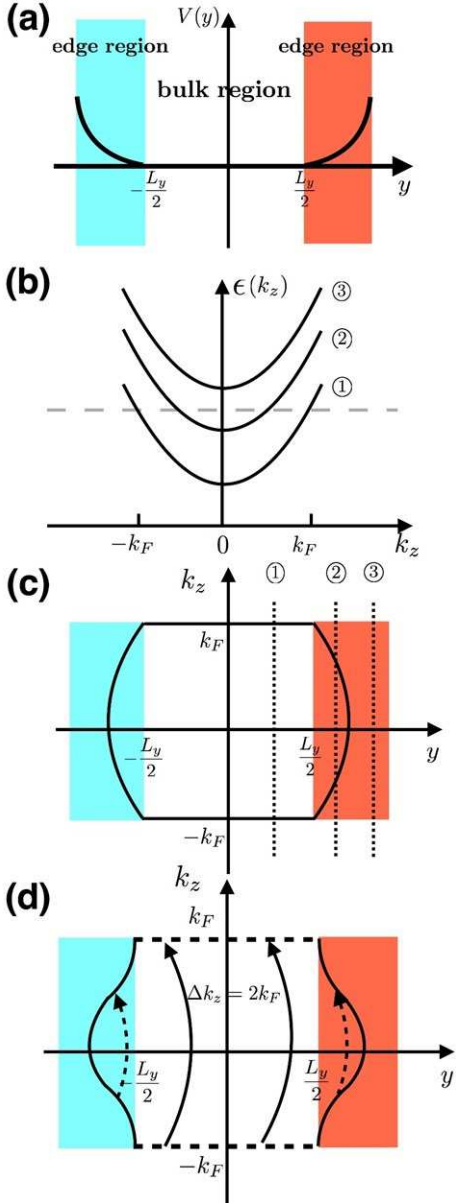}
\caption{(a) Confining potential along the $y$ direction; $V(y)$. $V(y)=0$ in a bulk region 
($|y|<L_y/2$), while $V(y) > 0$ in edge (surface) regions ($|y|>L_y/2$). (b) Schematic picture  
of electronic energy dispersion for a single particle state in the bulk $\epsilon_{b}(k_z)=\hbar^2k^2_z/(2m_{*})$ 
(\textcircled{1}) and that localized at $y=y_j$ with $|y_j|>L_y/2$   
(\textcircled{2},\textcircled{3}); $\epsilon_{s,j}(k_z) = \hbar^2k^2_z/(2m_{*}) + V(y_j)$. 
The subscript `$b$' and `$s$' stand for `bulk' and `surface' respectively. 
The subscript `$j$' stands for the chain index linked with $y_j$ as $y_j\equiv 2\pi j l^2/L_x$ 
($L_x$ is a linear dimension along the $x$-direction). 
The single-particle energy in the edge (surface) region acquires an energy shift due 
to the confining potential $V(y_j)$. The energy shift moves two Fermi points inward, 
forming a Fermi arc state in the edge (surface) regions. (c) Two parallel Fermi lines 
in the bulk and Fermi arc states which connect them. 
\textcircled{1},\textcircled{2},\textcircled{3} in Fig.~(c) correspond to \textcircled{1},\textcircled{2},\textcircled{3} in 
Fig.~(b) respectively. 
(d) Fermi arc states in the presence of the density wave order in the bulk. }
\label{3}
\end{figure}

As an application of our finding above, we consider an in-plane surface 
conductance in the DW/normal phase near the phase transition. The  
in-plane conductance (in $xy$ plane) 
at temperature lower than the cyclotron frequency 
can be dominated by surface transport rather than bulk transport 
(at least in the clean limit). The bulk electronic state 
forms two parallel Fermi lines at $k_z=\pm k_F$ in a 
two-dimensional plane subtended by $k_z$ and the chain coordinate. The two 
are connected with each other by Fermi arc states at the surface regions (Figs.~\ref{3}(a,b,c)). 
The arc states have chiral dispersions, and can be regarded as a bundle 
of chiral edge modes (chiral surface state or chiral surface Fermi arc 
state~\cite{hal1,hal2,bf}). The DW order removes the two Fermi lines, while 
keeps intact the arc states except for their two end points (Fig.~\ref{3}(d)). 
The two end point states repel each other by the $2k_F$ nesting vector, 
such that the Fermi arc state at $k_z=k_F$ is continuously connected 
to the state at $k_z=-k_F$. This leads to a perfect disconnection 
between the arc state at one boundary and that at the 
other side of the boundary. Therefore, one may naively expect 
that the arc states provide a robust in-plane conductance. 

\begin{figure}
\centering
\includegraphics[width=0.36\textwidth]{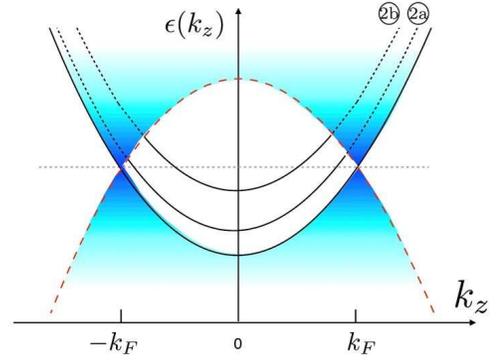}
\caption{Schematic picture of the dissipative region and 
the energy-momentum dispersion for the surface single-particle states 
localized at the edge (surface) regions (\textcircled{2a},\textcircled{2b}). 
The surface states in the dissipative region (dotted curve parts) 
acquire a finite life time due to a surface-bulk coupling. }
\label{2a}
\end{figure}

 Contrary to this naive expectation, however, we show that a finite surface-bulk 
coupling transfers the low-energy continuum spectral weight in the bulk state into the surface Fermi 
arc state, causing a {\it finite life time} to those arc states in the dissipative region. 
Using the Landauer formula, we show that, due to this finite life time, 
the in-plane surface conductance acquire a temperature dependence as; 
\begin{eqnarray}
G_s = \frac{e^2}{h} \sum_{k_z} \bigg(1- \frac{1}{e^{\beta|\epsilon_{b}(k_z)-\mu|}+1}\bigg), \label{sc}
\end{eqnarray}
where the summation over $k_z$ is taken over $[-k_F,k_F]$.  
At the zero temperature, the surface conductance takes a quantized value 
($N_z e^2/h$; $N_z$ is a number of chiral edge modes), as all the surface Fermi 
arc states on the chemical potential are outside the dissipative region (see Fig.~\ref{2a}). 
Nonetheless, the arc states in the dissipative region are 
thermally activated at finite temperature, which leads to a reduction of the surface 
conductance as in eq.~(\ref{sc}). We believe that the derived expression could  
provide a possible explanation of an unusual  
`in-plane metallic behavior', as observed in one of the two SSB phases in graphite. 

\subsection{structure of this paper}
The structure of this paper is as follows. In the next section (Sec.~II), we introduce 
an interacting electron model, its effective boson Hamiltonian, 
and a phase diagram obtained from the renormalization group (RG) study. 
Using the boson Hamiltonian, we calculate in Sec.~III 
the longitudinal (optical) conductivities in the density wave phases 
with disorders. In Sec.~IV, we employ a lattice-regularized version of the effective boson 
model (3D XY model), to calculate the single-particle spectral function by way of 
the electromagnetic duality map between XY and FLS models. In Sec.~V, we argue that 
the chiral surface Fermi arc states can generally have a finite life time   
due to a surface-bulk coupling. We further discuss the temperature-dependence 
of in-plane conductance due to the chiral surface states. 
Sec.~VI is devoted to conclusion and outlook. For completeness, we review how an 
interacting electron Hamiltonian is bosonized into coupled quantum sine-Gordon 
models (appendix A). We give in appendix B a derivation of the finite-temperature RG equation 
for the effective boson Hamiltonian. Detailed procedures of the conductivity 
calculation and of the spectral function calculation are given in appendix 
C and D respectively. 

\section{Hamiltonian, bosonization and a RG phase diagram}
We consider a 3D isotropic metal with parabolic energy band dispersion under 
high magnetic field along the $z$-direction. The kinetic energy part is given by 
\begin{align}
{\cal H}_{\rm kin} &= \int d{\bm r} \Psi^{\dagger}({\bm r}) 
\Big(\frac{{\bm \pi}^2}{2m_{*}} - \mu \Big) \Psi({\bm r}) \nonumber \\   
&= \sum_{n,k_z,j} \bigg(\frac{\hbar^2k^2_z}{2m_{*}} + \Big(n+\frac{1}{2} \Big) \hbar \omega_0 
- \mu \bigg) c^{\dagger}_{n,j,k_z} c_{n,j,k_z}, \label{metal-hami-sup} 
\end{align}
with ${\bm \pi} \equiv -i\hbar (\nabla_{x}-ieBy,\nabla_y,\nabla_z)$, 
an effective mass of electron $m_{*}$, momentum along the 
field direction $k_z$, the cyclotron  frequency $\hbar \omega_0 \equiv e B/m_{*}$, 
Landau level index $n$. $j$ (what we will call the `chain index' later) denotes 
degeneracy within each Landau 
level; $j=1,2,\cdots, L_xL_y/(2\pi l^2)$ and magnetic length $l \equiv \sqrt{\hbar/(eB)}$. 
$L_{x}$ and $L_y$ are linear dimensions of the system size along the $x$ and $y$-direction respectively.
$c^{\dagger}_{n,j,k_z}$ is a creation operator for the $n$-th Landau level with $j$ and $k_z$. 
In the Landau gauge, the electron creation operator can be expanded by the corresponding 
single-particle state as;  
\begin{align}
\Psi^{\dagger}({\bm r}) &\equiv  \sum_{n,j,k_z} \varphi^{*}_{n,j,k_z}({\bm r}) 
\!\ c^{\dagger}_{n,j,k_z} \label{expan1} \\ 
\varphi_{n,j,k_z}({\bm r}) &\equiv  
\frac{e^{ik_x x + ik_z z -(y-y_j)^2/2l^2}}{\sqrt{\sqrt{\pi} l L_x L_z}} 
 \frac{H_n\big((y-y_j)/l\big)}{\sqrt{2^n n !}}  \label{single-particle}
\end{align}
with ${\bm r} \equiv (x,y,z)$, $k_x \equiv 2\pi j/L_x$, $y_j \equiv 2\pi l^2j/L_x$ and 
the Hermite polynomial $H_{n}(y)$. For simplicity, we consider the spinless case. 
We put the chemical potential $\mu$ 
in the lowest Landau level (LLL) and assume that $\hbar \omega_0$ is 
larger than the largest kinetic energy along the field direction. We linearize the 
quadratic energy dispersion around the two Fermi points $k_{z}=\pm k_F$. 
This leads to a Hamiltonian for a slowly-varying field operator $\psi_{\sigma,j}(z)$ 
(defined below),    
\begin{align}
{\cal H}_{\rm kin} = \sum_{j,\sigma=\pm} \int dz \!\ \sigma v_F \psi^{\dagger}_{\sigma,j}(z) 
i\partial_z \psi_{\sigma,j}(z) + \cdots, \label{kin-sup-2}
\end{align}
with a bare Fermi velocity $v_F \equiv \hbar k_F/m_{*}$, Fermi wavelength $k_F$. We take 
$\hbar=1$ henceforth.    
$\sigma=\pm$ specifies one of the two Fermi points,   
\begin{align}
\Psi^{\dagger}({\bm r}) & \equiv  \frac{1}{\sqrt{\sqrt{\pi} l L_x}}  \sum_{j}
e^{-ik_x x -(y-y_j)^2/2l^2} \psi^{\dagger}_{j}(z) + \cdots,  \nonumber \\
\psi^{\dagger}_{j}(z) & \equiv e^{-ik_F z} \psi^{\dagger}_{+,j}(z) + e^{ik_Fz}\psi^{\dagger}_{-,j}(z).  
\label{psi-exp} 
\end{align}

For the interaction part, we consider a short-ranged electron-electron interaction 
among the LLL electrons; 
\begin{align}
\hat{\cal H}^{\prime}
&= \int d^3 {\bm r} \int d^3 {\bm r}^{\prime} \!\ V({\bm r}-{\bm r}^{\prime})  \!\ 
\Psi^\dagger({\bm r}) \Psi^\dagger({\bm r}^{\prime}) 
\Psi({\bm r}^{\prime})  \Psi({\bm r})  \label{int-sup-1}  \\  
&= \frac{g}{L_x} \int dz \int dz^{\prime} \sum_{j,m,n} V_{m-n,j-n}(z,z^{\prime})  \nonumber \\ 
&\hspace{1.0cm} \times \hat{\psi}^\dagger_{n}(z)  \hat{\psi}^\dagger_{j+m-n}(z^{\prime})  
\hat{\psi}_{m}(z^{\prime}) \hat{\psi}_{j}(z)  + \cdots \label{int-sup-2}
\end{align}
with 
\begin{align}
&V_{m-n,j-n}(z,z^{\prime}) = \nonumber \\
&\frac{1}{(2\pi)^2 l_0 l^\prime} 
e^{-\frac{(z-z^{\prime})^2}{2l_0^2}-\frac{(y_j-y_n)^2}{2l^2}-\frac{(y_n-y_m)^2}{2(l^2+l^2_0)}}  
\end{align}
for a short-range interaction potential~\cite{yako,giamarchi}  
\begin{eqnarray}
V({\bm r}-{\bm r}^{\prime}) = \frac{g}{(\sqrt{2\pi}l_0)^3} e^{-\frac{|{\bm r}-{\bm r}^{\prime}|^2}{2l^2_0}} 
\label{short-rep}
\end{eqnarray}  
with an interaction length $l_0$ and ${l^{\prime}}^2 \equiv l^2_0 + l^2$.  

We bosonize the fermion field for each single-particle state localized at $y_j$ as, 
\begin{eqnarray}
\psi_{\sigma,j}(z) = \frac{\eta_{\sigma,j}}{\sqrt{2\pi \alpha}} 
e^{-i(\sigma\phi_{j}(z) - \theta_j(z))}, \nonumber 
\end{eqnarray}  
where an electron displacement field $\phi_j(z)$ and current field $\partial_z \theta_j(z)$ 
are conjugate to each other, 
\begin{eqnarray}
[\phi_{j}(z),\partial_{z^{\prime}} \theta_{m}(z^{\prime})] = i\delta_{j,m} \delta(z-z^{\prime}). \nonumber 
\end{eqnarray}  
$\eta_{\sigma,j}$ is the Klein factor (Majorana fermion) ensuring the anticommutation 
between two fermion fields on different $j$ and $\sigma$; 
$\{\eta_{\sigma,j},\eta_{\sigma^{\prime},m}\}= 
\delta_{\sigma,\sigma^{\prime}} \delta_{j,m}$. $\alpha$ defines 
a short-range cutoff for the spatial coordinate $z$~\cite{giamarchi}. 
Due to the Klein factor, the interaction part cannot be 
fully bosonized without approximation.

To obtain an effective field theory description for the density wave ordering, 
we employ random phase approximation, keeping 
only the Hartree term ($j=n$) and Fock term ($m=n$) in ${\cal H}^{\prime}$. 
This leads to a bosonized Hamiltonian (Appendix A), 
\begin{align} 
H &=  \sum_m \int dz \Big\{ \frac{uK \pi }{2} [\Pi_m(z)]^2 +\frac{u}{2\pi K} [\partial_z \phi_m(z)]^2 
\nonumber \\ 
& \ \  -\sum_{j}^{j\ne m} J_{j-m}   \sigma_j^z \sigma_m^z  \cos2[\phi_{j}(z)-\phi_{m}(z)] \Big\}, 
\label{hami} \\
J_{m}& \equiv \frac{\sqrt{2\pi} l}{L_x} \!\ J \!\ e^{-\frac{y^2_m}{2l^2}}, \label{Jm}   
\end{align}
with $y_m\equiv 2\pi l^2 m/L_x$ and $\pi \Pi_j \equiv \partial_z \theta_j(z)$, 
Luttinger parameter $K$ and renormalized velocity $u$. Ising variable 
$\sigma_m^z=\pm 1$ is associated with the Klein factors of left and right mover 
for each $m$; $\sigma^{z}_m \equiv i \eta_{+,m} \eta_{-,m}$. The Fock term combined 
with the Hartree term give a positive $J_{j-m}$ in total, which gives a rigidity between 
the displacement field on $j$ 
and that on $m$. Since the bosonized Hamiltonian resembles that of coupled one-dimensional 
systems (`chains'), let us refer to $j,m$ as {\it chain} index. The inter-chain rigidity ranges over 
the magnetic length, within which ${\cal O}(L_x/l)$-number of chains are 
ferromagnetically coupled with one another. In the thermodynamic limit ($L_x \rightarrow \infty$),  
the number of chains within $l$ becomes infinite (`infinite-range' coupling). 
Due to the inter-chain rigidity, a long range order sets in for smaller $K$ and 
lower temperature, $\langle \phi_{j}(z) \rangle = \phi$. 
The order is a charge density wave order, which breaks the translational symmetry along $z$.

For a half electron filling, where $2k_F$ is half of a 
reciprocal lattice vector along $z$, the interaction part allows a umklapp process, 
which adds a phase locking term into eq.~(\ref{hami}); 
\begin{align}
H_{\rm u} &\equiv -\sum_{j,m} U_{j-m}  \sigma_j^z \sigma_m^z   \int dz 
\cos2[\phi_{j}(z)+\phi_{m}(z)],  \label{hami2} \\ 
U_{m}& \equiv \frac{\sqrt{2\pi} l}{L_x} \!\ U \!\ e^{-\frac{y^2_m}{2l^2}}. \label{Um} 
\end{align}  
The added term with the rigidity term locks the displacement fields on discrete 
values, $\langle \phi_{j}(z) \rangle =0,\pi/2,\pi,\cdots$ for positive $U$. 
The umklapp term reduces a symmetry of the Hamiltonian from U(1) to Z$_2$; low-energy 
collective excitation in the charge density wave phase acquires a mass.    

For larger Luttinger parameter/temperature $T$, quantum/thermal fluctuation may 
destroy the density wave order. To see this, we derive renormalization group (RG) 
flow equations for coupling constants (appendix B); 
\begin{align}
& \frac{dJ}{dl}= \Big[ 2- 2K\coth\frac{ \Lambda}{2T} \Big]J + KC \big[J^2 +U^2\big], \label{gr-j} \\ 
&\frac{dU}{dl}= \Big[ 2- 2K\coth\frac{ \Lambda}{2T} \Big]U  +2KC J U, \label{rg-u}
\end{align}
with $dT/dl = T$, 
\begin{align}
J\equiv \sum_{m}J_{m}, \ U\equiv \sum_{m}U_m. \label{JU}
\end{align}
$C$ is a numerical constant and $\Lambda$ is an ultraviolet energy cutoff. The Luttinger 
parameter $K$ is not renormalized in the leading order expansion in $1/L_x$. 
For general electron filling case ($U=0$) and 
at $T=0$, the Luttinger parameter has a critical value $K_c(=1)$ above/below which small 
inter-chain rigidity $J$ is irrelevant/relevant (normal/DW phase) respectively (Fig.~\ref{4}).  
For finite $T$, small inter-chain rigidity is renormalized to 
zero for any $K$, while $J$ above a critical strength increases on renormalization. 
The RG equations also suggest that the umklapp term and inter-chain rigidity 
term always help each other to grow into larger values for positive $J$, while small 
umklapp term is renormalized to zero for negative $J$ (Fig.~\ref{4}).    

\section{Longitudinal conductivity in density wave phases}   
\subsection{conductivity in the presence of disorders}
According to the linear response theory, longitudinal conductivity along the field direction 
is given by a retarded correlation function between an electric current operator and electric 
polarization operator,
\begin{eqnarray}
\sigma_{zz}(\omega) = - \frac{i}{ V} \int^{\infty}_{-\infty} dt  
\Theta(t)e^{i\omega t}  {\rm Tr} \Big[\hat{\rho}_G \big[e^{i\hat{K}t} 
\hat{J}_z e^{-i\hat{K}t}, \hat{P}_z  \big]\Big]. \nonumber \\ 
\label{condzz}
\end{eqnarray} 
$V$ is the volume of the system; $V \equiv L_zL_xL_y$. The electric current $\hat{J}_z$ and 
polarization $\hat{P}_z$ are given by the two  phase variables; 
\begin{align}
\hat{J}_z = & \frac{|e| uK}{\pi} \sum_{j} \int dz \!\ \partial_z \hat{\theta}_j(z),   \label{current} \\ 
\hat{P}_z = &  - \frac{|e|}{\pi} \sum_{j} \int dz \!\ \hat{\phi}_{j}(z). \label{displacement}
\end{align} 
$\hat{K}$ and $\hat{\rho}_G$ in Eq.~(\ref{condzz}) are Hamiltonian and a statistical 
density operator respectively (see Eqs.~(\ref{sw},\ref{Ave1},\ref{Ave2}) 
for their definitions of our actual calculations).  

The displacement field exhibits a long range order in the density wave (DW) phases,  
$\langle \hat{\phi}_j(z) \rangle = \phi_0$. The quantum fluctuation around the DW order 
can be described by a linear combination of the current density field 
$\hat{\Pi}_j(z) \equiv \frac{1}{\pi} \partial_z \hat{\theta}_j(z)$ and small fluctuations 
of the displacement field from its ordered value $\hat{\chi}_j(z) \equiv \hat{\phi}_j(z) - \phi_0$. 
An expansion of  the bosonized Hamiltonian Eqs.~(\ref{hami},\ref{hami2}) with respect to these 
fluctuations up to the second order leads to a spin-wave Hamiltonian 
(gaussian approximation),  
\begin{align}
H_{\rm sw} &= \frac{1}{2\pi} \sum_{j} \int dz \!\ \Big\{uK \big(\partial_z \hat{\theta}_j(z)\big)^2  
+ \frac{u}{K} \big(\partial_z \hat{\chi}_j(z)\big)^2 \Big\}  \nonumber \\ 
&\hspace{1cm} + \sum_{j,m} 2J_{j-m} \int dz \!\  (\hat{\chi}_j(z) - \hat{\chi}_m(z))^2 \nonumber \\ 
& \hspace{1.2cm}  + \sum_{j,m} 2 U_{j-m} \int dz  \!\ (\hat{\chi}_j(z)+\hat{\chi}_m(z))^2. 
\label{sw}
\end{align}
Here $U_{j-m}=0$ for the DW phase in the incommensurate electron filling case. 
Correspondingly, the displacement field operator $\hat{\phi}_j(z)$ in 
eq.~(\ref{displacement}) is replaced by $\hat{\chi}_{j}(z)$ henceforth. 
Note that we have omitted the Ising variables $\sigma^z_j$ from  
Eqs.~(\ref{hami},\ref{hami2}). This is because the DW states are 
described by $\sigma^{j}_z=+1$ for all $j$ 
(or by $\sigma^{j}_z=-1$ for all $j$), and because small fluctuations around them 
are not accompanied by flipping these Ising variables. Note also that 
the following argument at the half filling case can be 
generalized to other commensurate electron filling cases.

Electronic disorder potentials are coupled with the displacement field and are generally given by 
\begin{eqnarray}
H_{\rm imp}= \sum_{n=1,2,\cdots} \sum_{j} \int dz A_{j,(n)}(z) 
\cos (2n \hat{\phi}_{j}(z) + \lambda_{j,(n)}(z)). \nonumber \\
\label{imp-a} 
\end{eqnarray} 
Here a cosine term with $n=1$ comes from a single-particle backward scattering process 
\begin{eqnarray}
H^{(1)}_{\rm imp} = \sum_{j} \int dz A_{j,(1)}(z) \big\{ e^{i\lambda_{j,(1)}(z)}
\hat{\psi}^{\dagger}_{+,j}(z) \hat{\psi}_{-,j}(z) + {\rm h.c.}\big\},  \label{single-bcwd}
\end{eqnarray}
while a cosine term with $n=2$ comes from two-particle backward 
scattering process, 
\begin{align}
H^{(2)}_{\rm imp} &= \sum_{j,m} \int dz A_{j,m,(2)}(z) \big\{ 
e^{i\lambda_{j,m,(2)}(z)} \nonumber \\
&\hspace{1cm} 
\times \hat{\psi}^{\dagger}_{+,j}(z) \hat{\psi}^{\dagger}_{+,m}(z) 
\hat{\psi}_{-,m}(z) \hat{\psi}_{-,j}(z) + {\rm h.c.} \big\}.  \nonumber 
\end{align}
It is natural to expect that these backward scattering disorders have significant 
impact on the transport properites. To see their effect on the conductivity, 
$H_{\rm imp}$ is further expanded up to the second order in the fluctuation of 
the displacement field;
\begin{eqnarray}
H_{\rm imp} = \sum_{j} \int dz \!\ \big( X_j(z) \hat{\chi}_j(z) 
+ Y_j(z) \hat{\chi}^2_j(z)\big) + {\cal O}(\chi^3). \nonumber \\ 
\label{imp1}
\end{eqnarray} 
Here $X_j(z)$ and $Y_j(z)$ are given by random amplitudes $A_{j,(n)}(z)$ and random 
phases $\lambda_{j,(n)}(z)$;
\begin{align}
X_j(z) &\equiv 2 \sum_{n=1,2,\cdots} n A_{j,(n)}(z) \sin(2n\phi_0+\lambda_{j,(n)}(z)), \nonumber \\
Y_j(z) & \equiv 2 \sum_{n=1,2,\cdots} n^2 A_{j,(n)}(z) \cos(2n\phi_0+\lambda_{j,(n)}(z)). \nonumber 
\end{align}

We first calculate the conductivity from Eq.~(\ref{condzz}) with 
\begin{align}
\hat{K} &= \hat{H}_{\rm sw} + \sum_{j} \int dz \!\ \big( X_j(z) \hat{\chi}_j(z) + 
Y_j(z) \hat{\chi}^2_j(z)\big), \label{Ave1} \\ 
\hat{\rho_G} &= e^{-\beta \hat{K}}/{\rm Tr}[e^{-\beta \hat{K}}]. \label{Ave2} 
\end{align}
Then we take quenched average over random amplitudes $A_{j,(n)}(z)$ 
and random phases $\lambda_{j,(n)}(z)$ by 
\begin{align} 
&\overline{ \cdots } =  \nonumber \\ 
&\frac{\int {\cal D}X_{j}(z) {\cal D}Y_j(z) 
\cdots e^{-\frac{1}{g_x} \sum_j \int dz \!\ X^2_j(z)  - \frac{1}{g_y} \sum_j \int dz \!\ Y^2_j(z)}}
{\int {\cal D}X_{j}(z) {\cal D}Y_j(z)  e^{-\frac{1}{g_x} \sum_j \int dz \!\ X^2_j(z)  
- \frac{1}{g_y} \sum_j \int dz \!\ Y^2_j(z)}}.  
\label{config}
 \end{align}
$g_x$ and $g_y$ stand for disorder strengths 
associated with disordered fields $X_j(z)$ and $Y_j(z)$ respectively. 
We take the disorder average of the conductivity within a Born approximation, to obtain 
\begin{align}
& \overline{\sigma_{zz}(\omega)} = \overline{Q_{zz} (i\omega_n= \omega + i\eta)}, \label{Mat1-a} \\
& \overline{Q_{zz}(i\omega_n)} = \frac{e^2 uK}{\pi^2 l^2 } \nonumber \\
& \times \frac{\omega_n}{ \omega^2_n + 2u\pi K U(0) -   
\frac{2\pi u K g_y }{L_zN} \sum_{{\bm k}} [M^{-1}_0({\bm k},\omega_n)]_{2,2}}, 
\label{MGFB-a}
\end{align} 
with 
\begin{align}
[M^{-1}_0({\bm k},\omega_n)]_{2,2} &= \frac{\pi u K} {E^2(k_z,k) + \omega^2_n}, \label{M22-a} \\ 
E(k_z,k) &= \sqrt{u^2 k^2_z + 2u\pi K (J(k) + U(k))}. \label{band-a} 
\end{align}  
$Q_{zz}(i\omega_n)$ is a Fourier-transform of imaginary-time correlation function between 
the current density and displacement field (see Appendix C) and $i\omega_n$ is Matsubara frequency 
$i\omega_n=2\pi n/\beta$ with temperature $\beta^{-1}$. 
${\bm k}\equiv (k_z,k)$ and $k_z$ and $k$ are momenta conjugate to the coordinate along the field direction $z$ and 
coordinate associated with the chain index $y_j$ respectively. For example, a Fourier transformation 
of $\chi_{j}(z)$ is given by 
\begin{eqnarray}
\chi_j(z) = \frac{1}{L_z N} \sum_{{\bm k}} e^{ik_z z + ik y_j} \chi({\bm k}), \nonumber  
\end{eqnarray}
with $N\equiv L_x L_y /(2\pi l^2)$. $J(k)$ and $U(k)$ in eq.~(\ref{band-a}) are given by Fourier transforms of 
$J_j$ and $U_j$ with respect to the chain coordinate $y_j$;  
\begin{align}
J(k) &\equiv \sum^{N}_{n=1} J_{n} |1-e^{ik y_n}|^2, \\  \label{nagase0-a} \\
U(k) &\equiv \sum^N_{n=1} U_{n} |1+e^{ik y_n}|^2.   \label{nagase1-a}
\end{align}

\subsection{Low-energy collective excitations in density wave phases}
Low-energy collective excitations in the DW phases consist of fluctuations of 
current and displacement field. They are characterized by an energy-momentum  
dispersion relation $E(k_z,k)$ given in Eq.~(\ref{band-a}). Note that $k_z$ and $k$ are momenta 
conjugate to 
the coordinate $z$ and chain index $y_j$. Since $y_j$ takes discrete values with its 
increment $2\pi l^2/L_x$, the dispersion is periodic in $k$ with respect to 
the first Brillouin zone, $E(k_z,k+\frac{L_x}{l^2}) = E(k_z,k)$.  

Unlike its dispersion along $k_z$, $E(k_z,k)$ for $L_x \gg l$ has a unique dispersion along 
$k$ due to the `infinite-range' nature of the interchain rigidity. Namely, the 
interchain rigidity ranges over the magnetic length $l$, within which {\it all} the 
${\cal O}(L_x/l)$-number of chains are {\it ferromagnetically} coupled with one another. Accordingly, 
the collective modes with $|k|\gg l^{-1}$ always feel the interchain rigidity term 
in an out-of-phase way, giving rise to a finite and constant mass at $k_z=0$ (``optical mode'');
\begin{align}
E(k_z,k) = \left\{\begin{array}{cc}  
\sqrt{u^2 k^2_z + \omega^2_J}  & \ \ \ \ \  {\rm for} \!\ \!\  \frac{1}{l} \ll |k| < \frac{L_x}{2l^2}, U=0, \\
\sqrt{u^2 k^2_z + \omega^2_c}  & \ \ \ \ \  {\rm for} \!\ \!\ \frac{1}{l} \ll |k| < \frac{L_x}{2l^2}, U\ne 0. \\
\end{array}\right.  
\end{align} 
The mass is given by $\omega^2_J=4\pi uK J$ for incommensurate electron 
filling case without the umklapp term ($U=0$) and 
$\omega^2_c=4\pi uK (J+U)$ for the commensurate 
electron filling case with the umklapp term ($U\ne 0$). $J$ and $U$ are defined in 
Eq.~(\ref{JU}). Meanwhile, the collective modes with $|k|\ll l^{-1}$ show usual 
accoustic-phonon behaviours  
(``accoustic mode''); 
\begin{align}
E(k_z,k) = \left\{\begin{array}{cc}  
\sqrt{u^2 k^2_z + \gamma^2 k^2 }  & \ \ \ \ \ {\rm for} \!\ \!\ |k| \ll \frac{1}{l}, U=0, \\
\sqrt{u^2 k^2_z + \gamma^2 k^2 + \omega^2_U}  & \ \ \ \ \ {\rm for} \!\ \!\ |k| \ll \frac{1}{l}, U\ne 0. \\
\end{array}\right.  
\end{align} 
Note that the umklapp term endows the `accoustic mode' with a finite mass at $k_z=k=0$, 
$\omega^2_U=8\pi uK U$.  

\begin{figure}[t]
\centering
\includegraphics[width=0.48\textwidth]{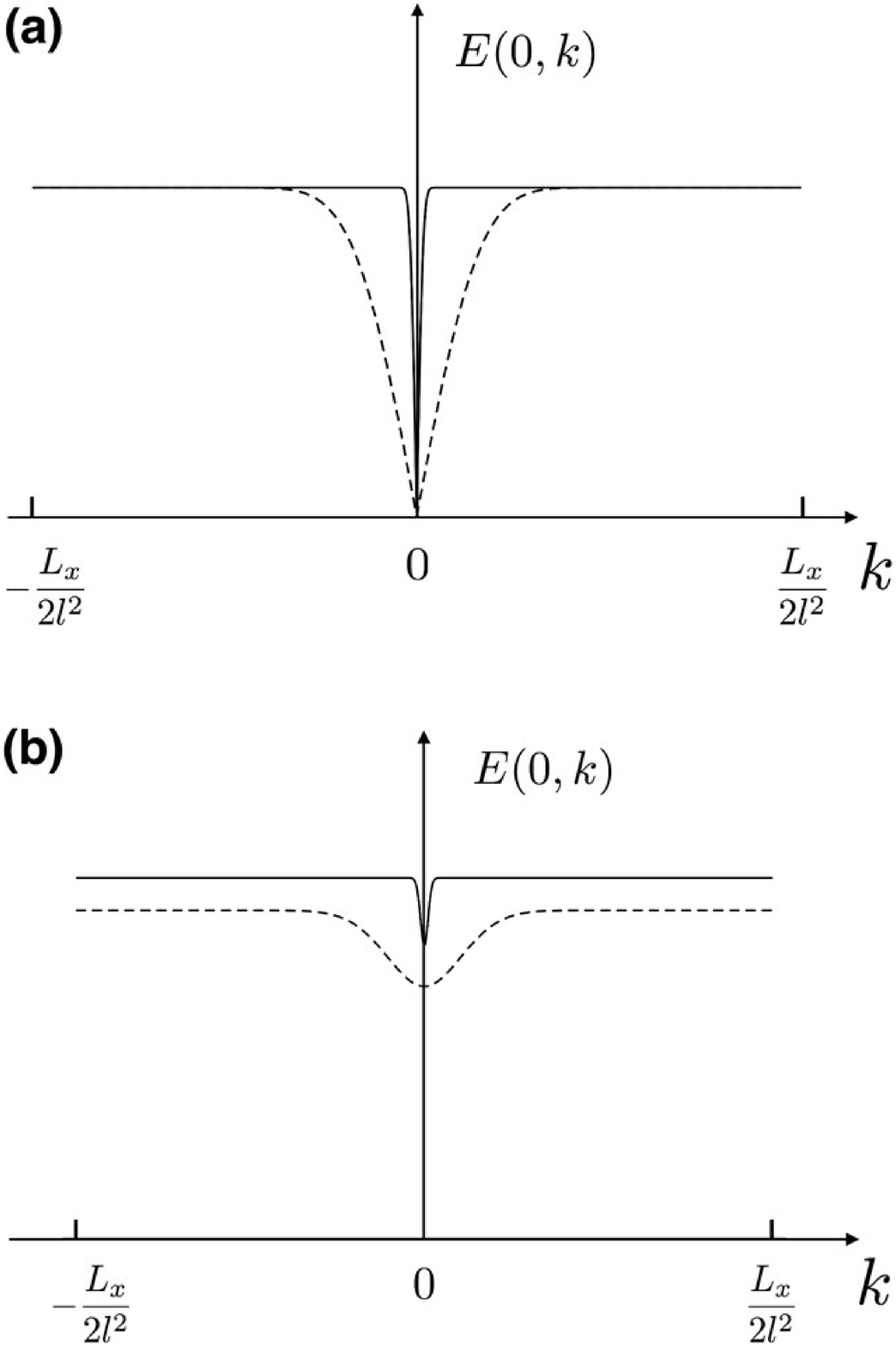}
\caption{(color online) Energy-momentum dispersion relation 
for collective mode with $k_z=0$ for different $L_x$. 
Solid line is for $\lambda=40$ and dashed line is for $\lambda =4$ with 
$\lambda \equiv L_x/(2\pi l)$. We used Eqs.~(\ref{band-a},\ref{nagase0-a},\ref{nagase1-a},\ref{Jm},\ref{Um}) 
for this calculation. (a) incommensurate electron filling case 
(without umklapp term; $U=0$). (b) commensurate electron filling case (with 
umklapp term; $U\ne 0$).}
\label{7}
\end{figure}

In the thermodynamic limit ($L_x\gg l$), the number of the optical modes within 
the first Brillouin zone becomes much larger than the accoustic modes (Fig.~\ref{7}). 
Thereby, the integral for the self-energy part in Eq.~(\ref{MGFB-a}) 
is dominated by the optical-mode contribution in large $L_x$ limit;
\begin{align} 
&\frac{1}{L_z N} \sum_{\bm k} [M^{-1}_0({\bm k},\omega_n)]_{2,2} \nonumber \\
& \hspace{0.5cm} 
= \frac{l^2}{L_x} \int \frac{dk_z}{2\pi} \int^{\frac{L_x}{2l^2}}_{-\frac{L_x}{2l^2}} dk  \!\ 
\frac{\pi u K}{\omega^2_n + E^2(k_z,k)}   \nonumber \\
& \hspace{0.5cm} = \left\{\begin{array}{cc}  
\int \frac{dk_z}{2\pi} \frac{\pi u K}{u^2k^2_z + \omega^2_n + \omega^2_J} 
= \frac{\pi K}{2} \frac{1}{\sqrt{\omega^2_n+ \omega^2_J}} & \!\ \!\ (U=0), \\ 
  \int \frac{dk_z}{2\pi} \frac{\pi u K}{u^2k^2_z + \omega^2_n + \omega^2_c}  
= \frac{\pi K}{2} \frac{1}{\sqrt{\omega^2_n+ \omega^2_c}} & \!\ \!\ (U\ne 0). \\ 
\end{array}\right. \label{S77}
\end{align} 
Substituting Eq.~(\ref{S77}) into eq. Eq.~(\ref{MGFB-a}), we get $\overline{Q_{zz}(i\omega_n)}$; 
\begin{eqnarray}
\overline{Q_{zz}(i\omega_n)} = \frac{e^2uK}{\pi^2 l^2} \frac{\omega_n}{\omega^2_n 
+ 2u\pi K U(0) - \frac{\pi^2 K^2 u g_y}{\sqrt{\omega^2_n+ \omega^2_{*}}}}, \label{S77-s}
\end{eqnarray}
where $\omega_* = \omega_J$ for incommensurate filling case ($U=0$) and 
$\omega_* = \omega_c$ for commensurate filling case ($U\ne 0$). By the analytic 
continuation, $i\omega_n =\omega + i\eta$, we finally obtain the conductivity $\overline{\sigma_{zz}(\omega)}$ 
in these two cases as in eqs.~(\ref{szz-1},\ref{szz-2}) (see the following two subsections). 

\subsection{Conductivity in DW phase (incommensurate electron filling case)}
For the DW phase without the umklapp term (incommensurate electron filling case; 
$U=0$), the Fourier transformed imaginary-time correlation function takes a form of,
\begin{equation}
\overline{Q_{zz}(i\omega_n)} = \frac{e^2 uK}{\pi^2 l^2 } \omega_n \Big\{ \omega^2_n 
- \frac{\pi^2 K^2 u g_y }{ \sqrt{\omega^2_n + \omega_J^2}} \Big\} ^{-1}. 
\end{equation}
In the clean limit ($g_y=0$), the function has the first order pole at $\omega_n=0$,
\begin{equation}
Q_{zz}(i\omega_n) = \frac{e^2 uK}{ \pi^2 l^2 } \frac{1}{\omega_n}.   
\end{equation} 
Thus, the real part of the optical conductivity has a delta function peak 
at $\omega=0$ (`Drude peak'), 
\begin{equation}
{\rm Re} \sigma_{zz}(\omega) = \frac{e^2 uK}{ \pi l^2 } \delta(\omega).   
\end{equation} 
Once finite disorder ($g_y$) is introduced, the Drude peak disappears immediately. 
Instead, it acquires a finite continuum spectrum above a threshold frequency 
$\omega_J$. The threshold frequency is the mass of the optical mode at $k_z=0$. 
This situation can be seen from the conductivity, obtained by 
the analytic continuation of $\overline{Q_{zz}(i\omega_n)}$ in the complex 
$\omega$ plane ($i\omega_n \rightarrow \omega+i0$);  
\begin{align}
\overline{\sigma_{zz}(\omega)} 
&= \frac{e^2 uK}{\pi^2 l^2 } \!\ (-i) \omega \Big\{ -\omega^2 
- \frac{\pi^2 K^2 u g_y }{ \sqrt{|\omega^2-\omega_J^2|}} \nonumber \\
& \times \big[\Theta(\omega_J-|\omega|) 
+ i{\rm sgn}(\omega) \Theta(|\omega|-\omega_J)\big] \Big\}^{-1},  \nonumber 
\end{align}
or its real part, 
\begin{align}
\overline{{\rm Re} \sigma_{zz}(\omega)}&= \frac{e^2 uK}{\pi^2 l^2 } \!\ 
\frac{\pi^2 K^2 u g_y} {\sqrt{\omega^2-\omega_J^2}} 
\frac{|\omega| \Theta(|\omega|-\omega_J)}{\omega^4 
+ \frac{(\pi^2 K^2 u g_y)^2}{\omega^2-\omega_J^2}} .  
\label{gapless}
\end{align}
As is clear from this derivation, the continuum spectrum at $\omega>\omega_J$ 
stems from the optical modes 
with finite $k_z$. The spectrum starts with $\sqrt{\omega-\omega_J}$ near the threshold 
frequency $\omega_J$ and decays with $\omega^{-4}$ in 
high frequency side (Fig.~\ref{8}(b)).  

\subsection{Conductivity in DW phase (commensurate electron filling case)}                
\begin{figure*}[htbp]
\centering
\includegraphics[width=0.9\textwidth]{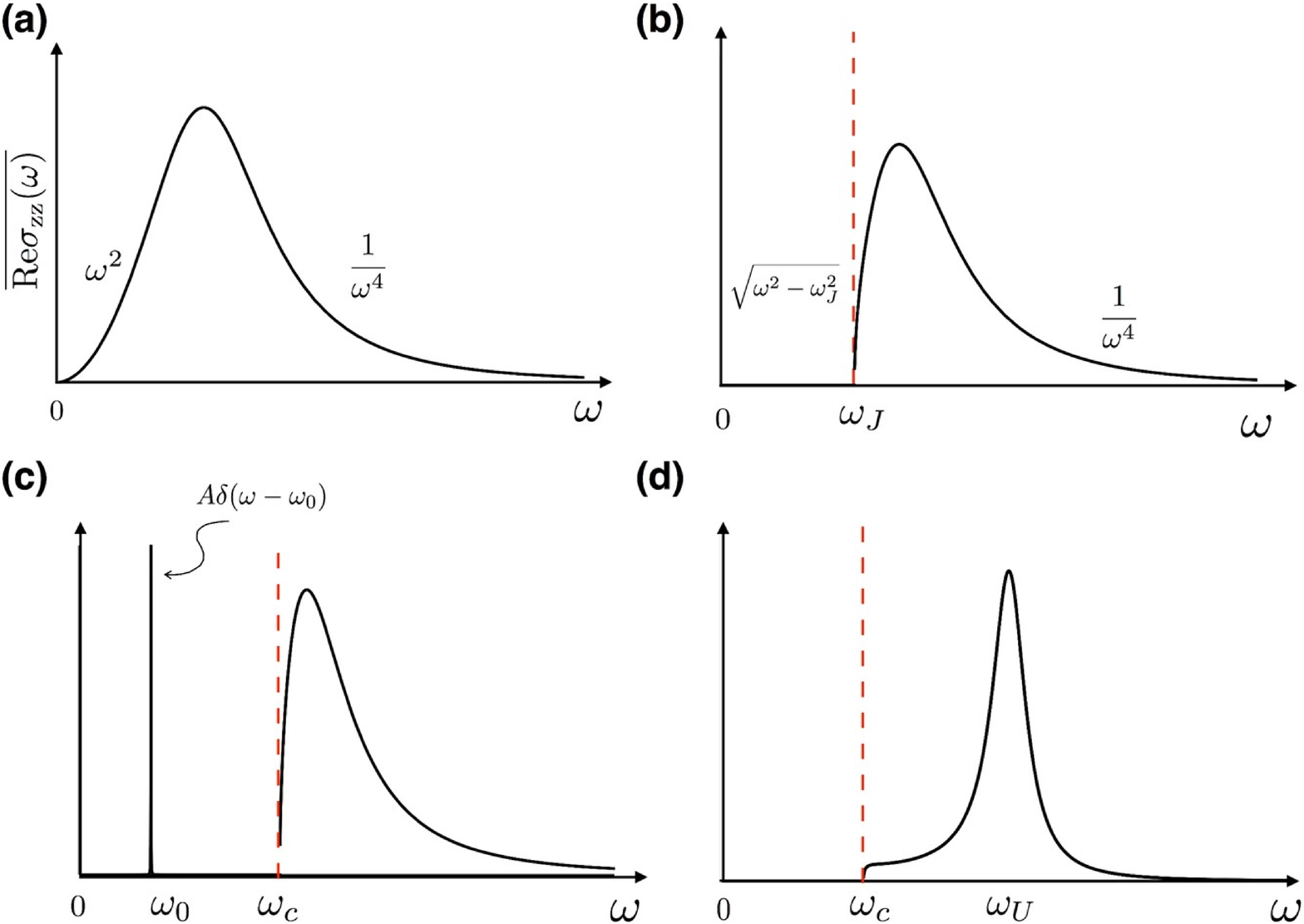}
\caption{(color online) Optical conductivity with disorders 
for different cases. (a) normal phase (decoupled one-dimensional 
chains)~\cite{fuku2}. (b) density wave phase at incommensurate electron filling case ($U=0$).  (c) 
density wave phase at commensurate electron filling case ($U\ne 0$) for weak disorder 
($g_y \le g_{y,c}$). (d) density wave phase at commensurate electron filling case with 
$\omega_U>\omega_c$.}
\label{8}
\end{figure*}

For the density wave phase with the umklapp term (commensurate electron filling case; $U\ne 0$), 
the optical conductivity in the clean limit ($g_y=0$) has a resonance peak at a finite frequency $\omega_U$; 
 \begin{equation}
{\rm Re}\sigma_{zz}(\omega) = \frac{e^2 u K}{ 2\pi l^2 } 
\delta\big(|\omega|-\omega_U \big)    
\end{equation}
where the resonance frequency $\omega_U \equiv \sqrt{8\pi u K U}$ 
stands for the mass of the accoustic mode 
at $k=k_z=0$. Once the disorder is introduced ($g_y \ne 0$), 
the mass is renormalized into a smaller value ($\omega=\omega_0$). 
Besides, the optical conductivity acquires a continuum spectrum above the threshold frequency 
$\omega_{c}$ (Fig.~\ref{8}(c)). The threshold frequency $\omega_c$ is nothing but 
the mass of the optical mode at $k_z=0$ and $k\ne 0$. 
The disorder strength has a critical value $g_{y,c}$ at  
which the renormalized mass $\omega_0$ of the accoustic mode becomes zero and above which 
the resonance peak disappears;
\begin{eqnarray}
g_{y,c} \equiv \frac{\omega_U^2 \omega_{c}}{u \pi^2 K^2}. \nonumber 
\end{eqnarray}  

These results can be obtained in the following way. Firstly, the optical 
conductivity is given by, 
\begin{align}
&\overline{\sigma_{zz}(\omega)} = \nonumber \\
&\ \ \ \frac{\omega} {F_{-}(\omega+i\eta) \Theta(\omega_{c}-|\omega|) + F_{+}(\omega)\Theta(|\omega|-\omega_{c})},  
\end{align}   
with 
\begin{align} 
F_{-}(\omega) &\equiv - \omega^2 + \omega_U^2 - \frac{\pi^2 K^2 u g_y}{\sqrt{\omega_{c}^2 -\omega^2}}, \nonumber \\ 
F_{+}(\omega) &\equiv -  \omega^2 + \omega_U^2 - i \frac{\pi^2 K^2 u g_y}{\sqrt{\omega^2-\omega_{c}^2}} \!\ {\rm sgn} (\omega).  \nonumber 
 \end{align} 
When the disorder strength is weaker than the critical value ($g_y < g_{y,c}$), a real $\omega$ 
solution of $F_{-}(\omega)=0$ exists with $F^{\prime}_{-}(\omega) \ne 0$. This leads to 
\begin{align} 
\overline{{\rm Re} \sigma_{zz}(\omega)} 
&= \frac{e^2 uK}{ \pi l^2 } \!\ 
\frac{\omega_0}{|F^{\prime}_{-}(\omega_0)|} \!\ \delta(|\omega|-\omega_{0}) \Theta(\omega_{c} -|\omega|)  
\nonumber \\ 
& \hspace{-0.5cm}  + \frac{e^2 u^2K^3 g_y}{ l^2\sqrt{\omega^2-\omega_{c}^2}} \frac{|\omega| \!\ \Theta(|\omega|-\omega_{c})}{(\omega^2-\omega_U^2)^2 + \frac{(\pi^2 K^2 u g_y)^2}{(\omega^2-\omega_{c}^2)}}.  \label{gappedweak} 
\end{align}    
The continuum above the threshold frequency is essentially of the same origin as in eq.~(\ref{gapless}). 
When the disorder strength is greater than the critical value ($g_{y}>g_{y,c}$), 
$F_{-}(\omega)=0$ has no real-valued solution. In this case, the optical conductivity has only 
the continuum spectrum, 
\begin{align} 
\overline{ {\rm Re} \sigma_{zz}(\omega)} 
&=  \frac{e^2 u^2K^3 g_y}{ l^2 \sqrt{\omega^2-\omega_{c}^2}} \frac{|\omega \!\ |\Theta(|\omega|-\omega_{c})}{(\omega^2-\omega_U^2)^2 + \frac{(\pi^2 K^2 u g_y)^2}{(\omega^2-\omega_{c}^2)}}.   
\label{gappedstrong}
\end{align}       
 
\subsection{Conductivity in normal phase (decoupled one-dimensional chains)}
In the normal phase ($J=U=0$), the system reduces to  
decoupled 1D Luttinger liquids. In the limit, Fukuyama already calculated essentially 
the same quantity in a context of conductivity in the Peierls-Frohlich state with 
disorders~\cite{fuku2}; 
\begin{equation}
\overline{ {\rm Re} \sigma_{zz}(\omega)}  = 
\frac{e^2 uK}{\pi^2 l^2 } \frac{\pi^2 K^2 u g_y}{\omega^4 + \frac{(\pi^2 K^2 u g_y)^2}{\omega^2}}.  
\label{normal}
\end{equation} 
The optical conductivity reduces to zero at $\omega=0$ with $\omega^2$ and 
decays as $1/\omega^4$ in large $\omega$ region (Fig.~\ref{8}(a)).

\section{Single-particle spectral function in density wave phase} 
A partition function of the effective boson models in eqs.~(\ref{hami},\ref{hami2}) for the 
DW phases can be seen as a (2+1)D XY model with/without Potts term 
(see eqs.~(\ref{before},\ref{after}) for 
the partition functions of the effective boson models), 
\begin{widetext}
\begin{align}
Z = \left\{\begin{array}{lc} 
\int D\Phi_{\bm j} \exp \bigg[\sum_{\bm j} \Big\{ J_{\tau} \cos(\Phi_{\bm j}-\Phi_{{\bm j}+a_{\tau} e_{\tau}}) 
+ J_{z} \cos(\Phi_{\bm j}-\Phi_{{\bm j}+a_{z} e_{z}}) + 
\sum_{n} J_{y,n} \cos(\Phi_{\bm j}-\Phi_{{\bm j}+na_{y} e_{y}}) \Big\} \bigg] & \ \ \ U=0 \\
\int D\Phi_{\bm j} \exp \bigg[\sum_{\bm j} \Big\{ 
J_{\tau} \cdots +  J_{z} \cdots + \sum_{n} J_{y,n} \cdots  
+ \sum_{n} U_{y,n} \cos(\Phi_{\bm j}+\Phi_{{\bm j}+na_{y} e_{y}}) \Big\} \bigg] & \ \ \ U\ne 0 \\ 
\end{array}\right. \label{XY+P}
\end{align}
\end{widetext}
where $\cdots$ parts in the second line 
are same as in the first line. The $U(1)$ phase degree 
of freedom plays role of the displacement field along the field direction 
($z$-direction); $\Phi_{\bm j}= 2\phi_{m}(z,\tau)$. The three-dimensional cubic-lattice coordinate 
${\bm j}=(j_z a_z, j_y a_y, j_{\tau} a_{\tau})$ represents the spatial coordinate along the field direction 
($z$), chain index ($y_j$) and imaginary time ($\tau$) respectively;  
${\bm j}=(z,y_j,\tau)$. The lattice constants of the cubic lattice are 
$a_{z}=ua_{\tau}=\alpha$, $a_y=2\pi l^2/L_x$. Gradient terms along the $z$-direction and 
$\tau$-direction 
are regularized into a cubic lattice as the 
nearest neighbor coupling terms with $J_z=u^2 J_{\tau}=1/(\pi K)$. The coupling   
along the $y$-direction ranges over the magnetic length;
\begin{eqnarray}
J_{y,n} = \frac{\sqrt{2\pi} l}{L_x} J u\alpha^2 e^{-\frac{y^2_n}{2l^2}}, \ \ \ 
U_{y,n} = \frac{\sqrt{2\pi} l}{L_x} U u\alpha^2 e^{-\frac{y^2_n}{2l^2}}. \label{xymodel-r}
\end{eqnarray} 
As is obvious from Eq.~(\ref{XY+P}), the model at incommensurate electron filling case ($U=0$) 
possesses the continuous U(1) symmetry, $\Phi_{\bm j} \rightarrow \Phi_{\bm j}+\varphi$, 
that is the translational symmetry along the field direction. Meanwhile, 
at the commensurate electron filling case ($m/n$ filling case with $n,m$ mutually prime integers), 
the U(1) symmetry reduces to the discrete $Z_{n}$ symmetry due to the umklapp terms 
(Potts terms); $\Phi_{\bm j} \rightarrow \Phi_{\bm j}+\frac{2\pi}{n}$. 
  
Breaking the continuous symmetry spontaneously, the DW phase in the incommensurate 
filling case has a gapless low-energy collective excitation (Fig.~\ref{7}(a)), whose effect 
as well as disorder effect on the conductivity has been discussed in the previous section. 
Since the low-energy excitation is a fluctuation of charge current density and electronic 
displacement, we can naturally expect that such gapless low-energy excitation has also 
significant influence on the single-particle spectral function. 

In expectation of this, we calculate 
in this section the spectral function in the DW phase at incommensurate electron filling case 
and its neighboring normal phase. For a reason which will become clear below, 
we consider a thin torus limit ($L_x \simeq l$) without any disorder. In this limit, 
the interchain rigidity ranges only over a couple of chains, so that we may begin with 
a `nearest-neighbor' 2+1D XY model given by 
\begin{eqnarray} 
Z_{\rm XY} \equiv \int D\Phi_{\bm j} \!\ {\rm exp} \bigg[\sum_{{\bm j},\mu} J 
\cos\big(\Phi_{\bm j}-\Phi_{{\bm j}+a_\mu {\bm e}_{\mu}}\big)\bigg] \label{XY-sup}
\end{eqnarray} 
where the couplings along the three directions are chosen to be same, 
$J_{\tau}=J_{z}=J_{y}=J$ for simplicity. Like in the case with larger 
$L_x$, the above partition function shows a $T=0$ quantum phase transition 
between the DW phase for larger $J$ and the normal phase for smaller $J$~\cite{herbut}. 
The DW/normal phases for larger $L_x$ are expected to be continuously 
connected to the DW/normal phases in the thin torus limit. 
We thus regard that qualitative aspect of these two phases will not change 
on changing $L_x$. 
 
\subsection{Single-particle Matsubara Green Function}
\begin{figure*}[t]
\centering
\includegraphics[width=0.8\textwidth]{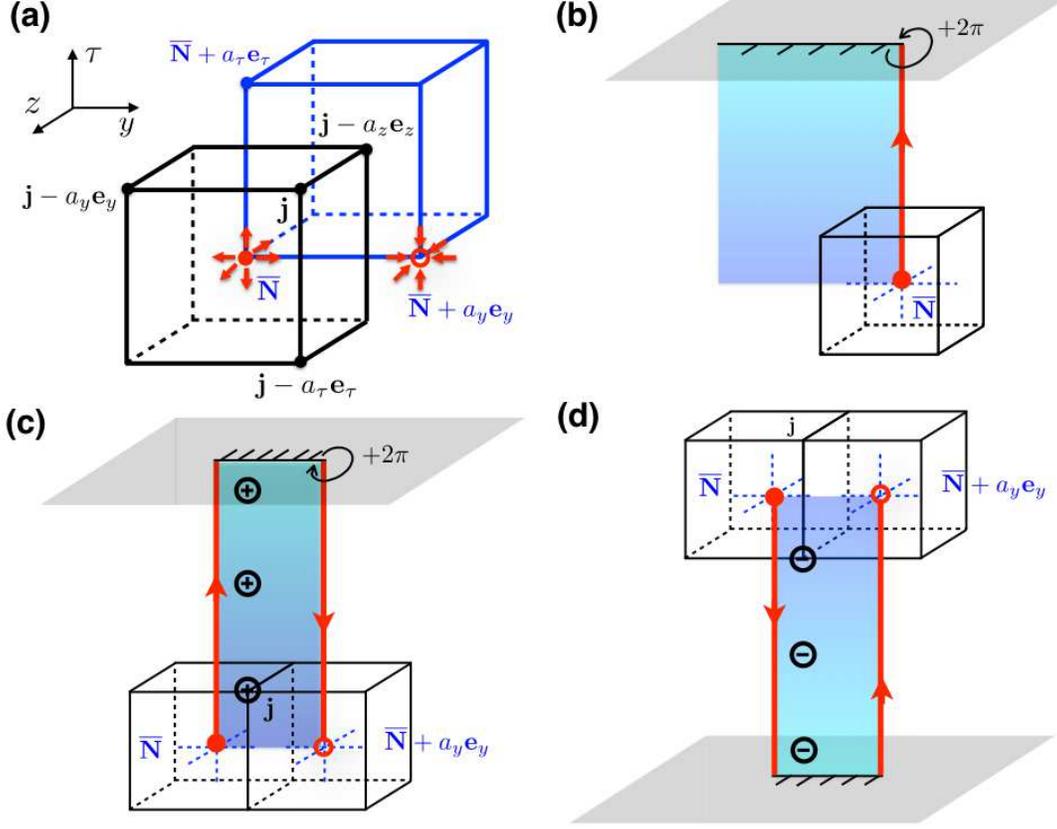}
\caption{(color online) (a) original cubic lattice (black) and dual cubic lattice (blue) (b) 
The magnetic monopole lives on a dual cubic lattice site and emits a quantized Dirac 
string or magnetic flux (red line with arrow). (c,d) A pair of magnetic monopole and 
antimonopole creates a branchcut, on crossing which from the negative $z$ side 
to the positive $z$ side the U(1) phase acquires $\pm 2\pi$ respectively.}
\label{9}
\end{figure*}

A central idea of our calculation of the spectral function in these 
two phases is to relate the single-particle Matsubara  
Green function with a partition function of the 3D XY model in the presence of 
two pairs of magnetic monopole and anti-monopole, 
\begin{widetext}
\begin{align}
& \hspace{-0.2cm} 
{\cal G}_{\sigma,j}(z_1-z_2,\tau_1-\tau_2)  \equiv \frac{1}{{\rm Tr}[e^{-\beta \hat{\cal H}}]} 
{\rm Tr} \big[e^{-{\beta}\hat{\cal H}} {\cal T}_{\tau} 
\big\{ \psi_{H,\sigma,j}(z_1,\tau_1)  \psi^{\dagger}_{H,\sigma,j}(z_2,\tau_2) \big\} \big]    
= \nonumber \\ 
& \hspace{0.1cm} = \!\ - 
\frac{{\rm sgn}(\tau-\tau^{\prime})}{Z_{\rm XY}} \int_{(\Delta\cdot{\bm B})_{\overline{\bm j}}= 
(\delta_{\overline{\bm j},\overline{\bm N}_1} -
\delta_{\overline{\bm j},\overline{\bm N}_1+a_y e_y})-(\delta_{\overline{\bm j},\overline{\bm N}_2} -
\delta_{\overline{\bm j},\overline{\bm N}_2 + a_ye_y})} D\Phi_{\bm j} \!\ \exp \Big[
\sum_{{\bm j},\mu} J_{\mu}
\cos\big(\Phi_{\bm j}-\Phi_{{\bm j}+a_\mu {\bm e}_{\mu}}+2\pi A_{{\bm j},\mu} \big)\Big].  
\label{GreenXY-sup}
\end{align}
\end{widetext}
Here  $\hat{\cal H} \equiv {\cal H}_{\rm kin}+{\cal H}^{\prime}$ denotes 
the interacting electron Hamiltonian (${\cal H}_{\rm kin}$ given by eq.~(\ref{kin-sup-2}) 
and ${\cal H}^{\prime}$ given by eq.~(\ref{int-sup-2}) only with $j=n$ or $m=n$), 
$\psi_{H,\sigma,j}(z,\tau) \equiv e^{\tau \hat{\cal H}} \psi_{\sigma,j}(z) e^{-\tau \hat{\cal H}}$. 
$\sigma=\pm$ specifies left or right mover fermion for each chain $j$. 
The magnetic monopole lives on a dual cubic lattice site denoted by $\overline{\bm N}_{1/2}$, 
while the fermion's creation and annihilation operator live on a original cubic lattice site 
denoted by $(z_{1/2},y_j,\tau_{1/2})$. In Eq.~(\ref{GreenXY-sup}), these two 
coodinates are linked with each other (Fig.~\ref{9}(a)),
\begin{eqnarray}
(z_{\mu},y_j,\tau_{\mu}) = 
\overline{\bm N}_{\mu}+a_y \frac{e_y}{2} + a_z \frac{e_z}{2} + a_{\tau} \frac{e_{\tau}}{2} \label{relation}
\end{eqnarray}
with $\mu=1,2$. 

The magnetic monopole emits a quantized magnetic flux ${\bm B}$ (`Dirac string'). 
The flux lives on a link of the dual cubic lattice,    
penetrating through a center point of a plaquette of the original cubic lattice. 
An associated gauge field  ${\bm A}$ lives on a link 
of the original lattice, penetrating through a center of a plaquette of the dual cubic lattice. 
The gauge field is coupled with the U(1) phase of the XY model as in eq.~(\ref{GreenXY-sup}). 
Accordingly, single monopole at $\overline{\bm N}_1$ 
creates a branchcut for the U(1) phase $\Phi_{\bm j}$ in a region of 
$\tau>\tau_1-a_{\tau}/2$, $z=z_1-a_z/2$ and $y<y_j-a_y/2$ (Fig.~\ref{9}(b)). 
On crossing the branchcut from $z<z_1-a_z/2$ to $z>z_1-a_z/2$, 
$\Phi_{\bm j}$ acquires $-2\pi$ phase winding. Meanwhile, 
single antimonopole at $\overline{\bm N}_1+a_y e_y$  
creates a branchcut in a region of 
$\tau>\tau_1-a_{\tau}/2$, $z=z_1-a_z/2$ and $y<y_j+a_y/2$,  
on crossing which from $z<z_1-a_z/2$ to $z>z_1-a_z/2$, 
$\Phi_{\bm j}$ acquires a $+2\pi$ phase winding. Therefore, a pair of 
the monopole and antimonopole inserted at $\overline{\bm N}_1$ and 
$\overline{\bm N}_1+a_y e_y$ respectively creates a branchcut in 
a region of $\tau>\tau_1-a_{\tau}/2$, $z=z_1-a_z/2$ and $y_j-a_y/2<y<y_j+a_y/2$, 
on crossing which from $z<z_1-a_z/2$ to $z>z_1-a_z/2$ the phase acquires 
$+2\pi$ phase winding (Fig.~\ref{9}(c)). Now that $\Phi_{\bm j} = 2\phi_j(z,\tau)$ 
and $\partial_z \phi_j(z,\tau) = -\pi \rho_{j}(z,\tau)$, the branchcut is nothing but 
an addition of one hole at $z=z_1$, $y=y_j$ and $\tau\ge \tau_1$. Or equivalently, 
an insertion of annihilation operator $\psi_{\sigma,j}(z_1)$ at $\tau = \tau_1$; 
\begin{widetext}
\begin{align}
& \hspace{-0.2cm} 
\frac{1}{{\rm Tr}[e^{-\beta \hat{\cal H}}]} 
{\rm Tr} \big[e^{-{\beta}\hat{\cal H}} {\cal T}_{\tau} 
\big\{ \psi_{H,\sigma,j}(z_1,\tau_1)  \cdots \big\} \big]    
= \nonumber \\ 
& \hspace{0.1cm} = \!\ - 
\frac{{\rm sgn}(\tau-\tau^{\prime})}{Z_{\rm XY}} \int_{(\Delta\cdot{\bm B})_{\overline{\bm j}}= 
(\delta_{\overline{\bm j},\overline{\bm N}_1} -
\delta_{\overline{\bm j},\overline{\bm N}_1+a_y e_y})+\cdots } D\Phi_{\bm j} \!\ \exp \Big[
\sum_{{\bm j},\mu} J_{\mu}
\cos\big(\Phi_{\bm j}-\Phi_{{\bm j}+a_\mu {\bm e}_{\mu}}+2\pi A_{{\bm j},\mu} \big)\Big],   
\label{GreenXY-sup2}
\end{align}
\end{widetext}
with $(z_{1},y_j,\tau_{1}) = \overline{\bm N}_{1}+a_y e_y/2 + a_z e_z/2 + a_{\tau} e_{\tau}/2$. 
Note that the above identification of the magnetic dipole with the annihilation operator 
does not depend on a specific choice of the branchcut. For example, we can also 
regard that the dipole creates a branchcut in a region of 
$\tau<\tau_1-a_{\tau}/2$, $z=z_1-a_z/2$ and $y_j-a_y/2<y<y_j+a_y/2$, 
on crossing which from $z<z_1-a_z/2$ to $z>z_1-a_z/2$ the phase acquires 
$-2\pi$ phase winding, instead of $+2\pi$ (Fig.~\ref{9}(d)). 
Such a branchcut corresponds to an addition of 
particle at $z=z_1$, $y=y_j$ and $\tau\le \tau_1$, which is again equivalent to an insertion of 
an annihilation operator $\psi_{\sigma,j}(z_1)$ at $\tau = \tau_1$. As is obvious from 
the argument, flipping the magnetic dipole exchanges annihilation operator into 
creation operator.   
This leads to eq.~(\ref{GreenXY-sup}). Remark that the extra sign function 
${\rm sgn}(\tau_1-\tau_2)$ in eq.~(\ref{GreenXY-sup}) 
stems from the fermion's time ordering in the left hand side of the same equation 
(${\cal T}_{\tau}$). This maintains a proper boundary condition for 
the Matsubara Green function along the imaginary time 
axis;
\begin{eqnarray}
 {\rm sgn}(\tau) \equiv \left\{ \begin{array}{cc} 
 1 & (0 < \tau < \beta), \\ 
 -1 & (-\beta < \tau < 0). \\  
 \end{array} \right.
\end{eqnarray}

\subsection{Duality mapping to Frozen Lattice Superconductor model}
By the duality transformation~\cite{peskin,herbut,dh}, the 3D XY model is mapped into 
the so-called frozen lattice superconductor model (FLS), where 
a U(1) phase of the superconducting order parameter 
$\theta_{\overline{\bm j}}$ is coupled with an internal magnetic gauge field 
${\bm a}_{{\overline{\bm j}}}$; 
\begin{align}
& Z_{\rm XY} \Longleftrightarrow Z_{\rm FLS}, \nonumber \\ 
&Z_{\rm XY} \equiv \int D\Phi_{\bm j} \exp \bigg[J \sum_{{\bm j},\mu=x,z,\tau}
\cos\Big(\Phi_{\bm j}-\Phi_{{\bm j}+a_\mu {\bm e}_{\mu}}\Big)\bigg],  \nonumber \\ 
& Z_{\rm FLS} \equiv \lim_{t\rightarrow 0}\int D{\bm a}_{\overline{\bm j}} D\theta_{\overline{\bm j} }
\exp \bigg[-S_{\rm FLS}\big[{\bm a}_{\overline{\bm j}},\theta_{\overline{\bm j}}\big]\bigg], \label{zfls} \\ 
&S_{\rm FLS}\big[{\bm a}_{\overline{\bm j}},\theta_{\overline{\bm j}}\big] \equiv 
\frac{1}{2J} \sum_{{\bm j},\mu} (\nabla \times {\bm a})^2_{{\bm j},\mu}  \nonumber \\
& \hspace{2.0cm} - \frac{1}{t} \sum_{\overline{\bm j},\mu} \cos \Big(\theta_{\overline{\bm j}}-
 \theta_{\overline{\bm j}+a_\mu {\bm e}_{\mu}} - 2\pi a_{\overline{\bm j},\mu}). \label{sfls} 
\end{align} 
The U(1) phase lives on the dual cubic lattice site denoted by $\overline{\bm j}$, 
while three components of the gauge 
field ${\bm a}_{\overline{\bm j}} \equiv (a_{\overline{\bm j},z},a_{\overline{\bm j},y},
a_{\overline{\bm j},\tau})$ live on a link of the dual cubic lattice. 
The parameter $t$ plays role of the temperature in the FLS 
model, and is taken to zero or sufficiently small (the model is 
referred to as `frozen' lattice superconductor model).
In the FLS model, smaller $J$ suppresses spatial and temporal fluctuations 
of the magnetic gauge field, so that the superconducting (SC) order sets in 
at low temperature $t$; $\langle e^{i\theta_{\overline{\bm j}}} \rangle \ne 0$. 
The SC phase (`Meissner phase') in the FLS model corresponds to the normal phase 
in the XY model. In the SC phase, the gauge fields are expelled from the SC bulk. 
An effective theory for such phase may be crudely described by a simple 
omission of the magnetic gauge fields from the FLS model; 
\begin{align}
S_{\rm FLS}\big[{\bm a}_{\overline{\bm j}},\theta_{\overline{\bm j}}\big] 
\simeq - \frac{1}{2t} \sum_{\overline{\bm j},\mu} 
 \big(\Delta_{\mu} \theta_{\overline{\bm j}}\big)^2.  
\label{mss-a}
\end{align}
For larger $J$, the fluctuations of the magnetic 
gauge fields become wild enough that the SC order is killed by the gauge fields,
$\langle e^{i\theta_{\overline{\bm j}}} \rangle = 0$. This phase corresponds to 
the DW phase in the XY model. An effective theory for such non-superconducting 
phase in the FLS model is given only by the Maxwell term (`Maxwell phase'):
\begin{align}
S_{\rm FLS}\big[{\bm a}_{\overline{\bm j}},\theta_{\overline{\bm j}}\big] 
\simeq - \frac{1}{2J} \sum_{{\bm j},\mu} 
 (\nabla \times {\bm a})^2_{{\bm j},\mu}. \label{mxw-a}
\end{align}    

By the duality transformation~\cite{herbut}, the partition function of the XY model in the 
presence of a magnetic monopole  and anti-monopole  
is mapped into a correlation function in the frozen lattice superconductor model,
\begin{widetext}
\begin{equation} 
\begin{aligned}
&\frac{1}{Z_{\rm XY}} \int_{(\Delta\cdot{\bm B})_{\overline{\bm j}} 
= \pm \delta_{\overline{\bm j},\overline{\bm N}}
+ \!\ \cdots \cdots \cdots } D\Phi_{\bm j} \!\ \exp \Big[J \sum_{{\bm j},\mu}
\cos\big(\Phi_{\bm j}-\Phi_{{\bm j}+a_{\mu} {\bm e}_{\mu}}+2\pi A_{{\bm j},\mu} \big)\Big]  \\ 
& \ \ \Longleftrightarrow \frac{1}{Z_{\rm FLS}} \int D{\bm a}_{\overline{\bm j}} D\theta_{\bm j} \!\ 
\exp\big[ \pm i\theta_{\overline{\bm N}} + \cdots  \big] \!\  
 \exp \bigg[-\frac{1}{2J} \sum (\nabla \times {\bm a})^2 
 + \frac{1}{t} \sum \cos \Big(\Delta_{\mu} \theta_{\overline{\bm j}} 
 - 2\pi a_{\overline{\bm j},\mu})\bigg] 
 \equiv \big\langle e^{ \pm i\theta_{\overline{\bm N}}} \cdots \big\rangle_{\rm FLS} 
\end{aligned} 
\end{equation}
\end{widetext}
Combined with Eq.~(\ref{GreenXY-sup}), this leads to   
\begin{equation}
\begin{aligned}
&  {\cal G}_{\sigma,j}(z_1-z_2,\tau_1-\tau_2)  = \\
& -{\rm sgn}(\tau-\tau^{\prime}) \!\ 
\big\langle e^{-i\theta_{\overline{\bm N}_1}+i\theta_{\overline{\bm N}_1+a_ye_y} + 
i\theta_{\overline{\bm N}_2} - i\theta_{\overline{\bm N}_2+a_ye_y}} \big\rangle_{\rm FLS}  
\end{aligned} 
\label{GreenFLS} 
\end{equation}
with eq.~(\ref{relation}). 

\subsection{single-partcle spectral function}
Based on these foundations, we calculate the four-point correlation function of the 
FLS model in the SC/non-SC phase, using their respective effective theories 
as in eqs.~(\ref{mss-a},\ref{mxw-a}). We then 
take a Fourier transform of the calculated four-point correlation 
function;
\begin{align}
&{\cal G}_{\sigma,j}(z_1-z_2,\tau_1-\tau_2) 
\equiv \nonumber \\
& \ \frac{1}{\beta L_z} \sum_{q_z,i\omega_n} e^{iq_z(z_1-z_2)-i{\cal E}_n(\tau_1-\tau_2)}
\!\ {\cal G}_{\sigma,j}(q_z,i{\cal E}_n), \label{Fr1-s}
\end{align}
with fermionic Matsubara frequency ${\cal E}_n = (2n+1)\pi/\beta$. 
Here $q_z$ in eq.~(\ref{Fr1-s}) is a momentum $k_z$ measured 
from $\sigma k_F$ with $\sigma=\pm$ and is much 
smaller than $a^{-1}_z$; $q_z=k_z-\sigma k_F$.  
After an analytic continuation of  ${\cal G}_{\sigma,j}(q_z,i{\cal E}_n)$ in the complex $\omega$ plane 
($i{\cal E}_n \rightarrow \omega\pm i\delta$), we obtain the spectral function from, 
\begin{align}
&\rho_{\sigma}(q_z,\omega) \equiv i\Big\{{\cal G}_{\sigma,j}(q_z,i{\cal E}_n=\omega+i\delta) \nonumber \\
& \hspace{2.7cm} 
-{\cal G}_{\sigma,j}(q_z,i{\cal E}_n=\omega-i\delta) \Big\}. \label{rho-G}
\end{align}
The spectral function in the normal phase (which is calculated from the four-point 
correlation function of the FLS model 
in the SC phase) takes a form of 
\begin{align}
\rho_{\sigma}(q_z,\omega) =  2\pi t e^{-C} \!\ \sqrt{\omega^2 - u^2 q^2_z} 
\!\ \Theta(|\omega|-u |q_z|) + \cdots, \label{rho-mss-a}
\end{align}
(see Appendix D1 for its detailed derivation). The spectral function in the DW phase 
(calculated from the correlation function in the non-SC phase) is given by 
\begin{align}
\rho_{\sigma}(q_z,\omega) &= \frac{16 \pi^3 J}{\sqrt{\omega^2 -u^2 q^2_z}} \Theta(|\omega|-u|q_z|) 
+ \cdots, \label{rho-mxw-a}
\end{align}
(see Appendix D2 for its derivation). 
Note that we omit the chain index $j$ in the left hand sides in 
eqs.~(\ref{rho-G},\ref{rho-mss-a},\ref{rho-mxw-a}) 
because the calculated four-point correlation functions are independent from 
the chain coordinate $y_j$. 
Note also that the spectral function thus obtained gives the real and 
imaginary part of the real-time time-ordered Green function; 
\begin{align}
G_{\sigma}(q_z,\omega) & = {\cal P}\int^{\infty}_{-\infty} \frac{d\omega^{\prime}}{2\pi} 
\frac{\rho_{\sigma}(q_z,\omega^{\prime})}{\omega - \omega^{\prime}} \nonumber \\
& \ \ \ - i\pi \tanh \Big(\frac{\beta \omega}{2}\Big) \!\ \rho_{\sigma}(q_z,\omega), \label{realG-fromR} 
\end{align}
where the real-time Green function is defined as 
\begin{align}
iG_{\sigma}(z_1,t_1;z_2,t_2) &\equiv \frac{1}{Z} {\rm Tr}\Big[{\cal T}_{t} 
\big\{ \psi_{H,\sigma,j}(z_1,t_1) \psi^{\dagger}_{H,\sigma,j}(z_2,t_2)\big\} 
\Big], \nonumber \\
G_{\sigma}(q_z,\omega) & \equiv \int dt \int dz e^{-iq_z (z_1-z_2) + i\omega(t_1-t_2)} \nonumber \\
& \hspace{2cm} \times G_{\sigma}(z_1,t_1;z_2,t_2), \nonumber  
\end{align}
and $\psi_{H,\sigma,j}(z,t) \equiv e^{i\hat{\cal H}t}  \!\ \psi_{\sigma,j}(z) \!\ e^{-i\hat{\cal H}t}$. 
$\hat{\cal H} \equiv \hat{\cal H}_{\rm kin}+{\cal H}^{\prime}$ with ${\cal H}_{\rm kin}$ given by 
eq.~(\ref{kin-sup-2}) and ${\cal H}^{\prime}$ given by eq.~(\ref{int-sup-2}) only with $n=m$ or $n=j$.

\section{in-plane conductance} 

Unlike the out-of-plane current operator, in-plane current operators 
($\hat{J}_x$, $\hat{J}_y$) are 
given by a creation/annihilation of the 2nd LLL electron  
and annihilation/creation of the LLL electrons. For the 3D isotropic metal 
under high magnetic field in eq.~(\ref{metal-hami-sup}), they are given by 
\begin{align}
\hat{J}_x + i \hat{J}_y &\equiv \frac{\sqrt{2} e\hbar}{m_{*} l} 
\sum_{j,k_z} c^{\dagger}_{n=0,j,k_z} c_{n=1,j,k_z} + \cdots , \nonumber \\ 
\hat{J}_x - i \hat{J}_y &\equiv \frac{\sqrt{2} e\hbar}{m_{*} l} 
\sum_{j,k_z} c^{\dagger}_{n=1,j,k_z} c_{n=0,j,k_z} + \cdots , \nonumber 
\end{align} 
where $\cdots$ parts stand for higher LL contributions. Accordingly, when the temperature 
is much lower than the cyclotron frequency $\hbar\omega_0$, the in-plane transports 
are dominated by surface transport rather than bulk transport (at least in the 
clean limit). To discuss the surface 
transport concretely, let us choose the Laudau gauge, impose the periodic boundary conditions 
along $x$, $z$-directions, and open boundary condition along $y$-direction, 
introduce a confining potential $V(y)$ which respects the translational symmetries along the 
$z$-direction (Fig.~\ref{3}(a)) ~\cite{hal1}, 
\begin{eqnarray}
\left\{\begin{array}{cc} 
V(y) = 0 &  \ \ \ \ \ {\rm for} \ \ |y|< \frac{L_y}{2}, \\
V(y) > 0 &  \ \ \ \ \ {\rm for} \ \ |y|> \frac{L_y}{2}. \\ 
\end{array}\right. 
\end{eqnarray} 
For $|y|<L_y/2$, the single-particle eigenstate for the LLL 
is given by eq.~(\ref{single-particle}) with $n=0$, whose eigenenergy is $\hbar^2k^2_z/(2m_{*})$. For 
$|y|>L_y/2$, we may also approximately use eq.~(\ref{single-particle}) as an eigenstate, 
provided that the confining potential $V(y)$ is slowly varying compared to 
the magnetic length $l$, $l \times |\partial_y V(y)| \ll |V(y)|$. Such a quasi-eigenstate has an 
energy of $\hbar^2k^2_z/(2m_{*}) + V(y_j)$~\cite{hal1}. When changing 
$y_j$ from the bulk region ($|y_j|<L_y/2$) into the edge regions ($|y_j|>L_y/2$), 
the two Fermi points at $k_z=\pm k_F$ in the bulk region move inward, and merge  
into one point at $|y_j|>L_y/2$ with $V(y_j)=\mu$ (Figs.~\ref{3}(b,c)). 
In other words, the two parallel Fermi lines at $k_z=\pm k_F$ 
are connected with each other by Fermi arc states, which are localized at the two boundaries, 
$|y_j|> L_y/2$. Since $k_xl^2 = y_j$ in eq.~(\ref{single-particle}), the Fermi arc state 
at $y_j > L_y/2$ carries positive $k_x$, and that at  $y_j < -L_y/2$ carries negative $k_x$; 
these arc states are nothing but a bundle of {\it chiral} Fermi edge modes~\cite{hal2,bf}. 

When viewed along the $k_z$-direction, 
the chiral Fermi arc states are connected by $\Delta k_z$, which 
is smaller than $2k_F$ (Fig.~\ref{3}(d)). 
Thus, the arc states except for their two end-point states are robust  
against the density wave formation in the bulk. Only the two end-point states repel each other 
with a help of the density wave order in the bulk, such that the Fermi arc state at  
$k_z=k_F$ is continuously connected with the arc state at $k_z=-k_F$ as a function of 
$k_z$ (Fig.~\ref{3}(d)). This leads to a perfect disconnection between the arc state  
at $y>L_y/2$ and that at $y< -L_y/2$. 

The chiral Fermi arc states often dominate low-$T$ in-plane transports in actual experiments. 
For example, a graphite sample of $50 \!\ \mu$m thickness shows an in-plane resistance of the 
order of $2$ $\Omega \sim 4$ $\Omega$ above the quantum limit~\cite{fauque,akiba}.  
Since an interlayer lattice constant ($a_z$) is on the order of $0.5$nm in graphite, 
the sample with $50\mu$m thickness could have $10^5\!\ (=L_z/a_z)$ number 
of the $k_z$ points within the first Brillouin zone along $k_z$, 
$[-\pi/a_z,\pi/a_z]$. Assuming that 2$k_F$ is several times smaller 
than $2\pi/a_z$, one can expect that the number of chiral Fermi edge modes at 
each surface is on the order of $10^{4}$. Since different edge modes at the same 
surface do not have an electron exchange much, a bundle of chiral Fermi edge modes 
can be regarded as a parallel circuit. Accordingly, an in-plane resistance due to the chiral 
Fermi arc states can be evaluated on the order of $h/e^2 \times 10^{-4} \simeq 2.5\Omega$, 
which is on the same order of the experimental values ($2 \Omega \sim 4 \Omega$). 

The chiral Fermi arc state at $y_j > L_y/2$ and that at $y_j < - L_y/2$ 
are apparently disconnected by the DW order in the bulk. One may therefore expect that 
these chiral surface states provide robust in-plane conductance. Contrary to this 
expectation, we argue in the following that the in-plane conductance due to the surface 
states may have a non-trivial temperature dependence, especially when the bulk 
electronic state has dissipative feature as in Eqs.~(\ref{rho-mss-a},\ref{rho-mxw-a},\ref{realG-fromR}).

\subsection{Coupling between surface and bulk states} 
To see this, let us begin with a simple model which includes a coupling 
between surface and bulk states; 
\begin{equation} \begin{aligned}
{\cal H}_{t} &= {\cal H}_{b} + {\cal H}_{s} + {\cal H}_{c}, \\  
{\cal H}_{s} &= \sum^{|y_j|>\frac{L_y}{2}}_{k_z,j} 
\big(\epsilon_{s,j}(k_z) - \mu \big) \!\ d^{\dagger}_{s,j,k_z} d_{s,j,k_z}, \\ 
{\cal H}_{b}& = \sum^{|y_j|<\frac{L_y}{2}}_{k_z,j} \big(\epsilon_{b}(k_z) - \mu\big) \!\  
d^{\dagger}_{b,j,k_z} d_{b,j,k_z} + {\cal H}^{\prime}, \\ 
{\cal H}_{c} &= \sum^{|y_j|>\frac{L_y}{2}>|y_m|}_{j,m} 
\big\{ T_{j,m} d^{\dagger}_{s,j,k_z} d_{b,m,k_z} + {\rm h.c.} \big\}.   
\end{aligned} 
\label{hami-sup-2}
\end{equation}
Here $\epsilon_{b}(k_z) \equiv \hbar^2 k^2_z/(2m_{*})$ and $\epsilon_{s,j}(k_z)\equiv 
\hbar^2k^2_z/(2m_{*}) + V(y_j)$ are single-particle energies of bulk 
and surface states with $k_z$ and $j$ respectively. The single-particle states for $d^{\dagger}_{s,j,k_z}$ 
and $d^{\dagger}_{b,j,k_z}$ are all in the LLL and 
their wavefunctions are given by eq.~(\ref{single-particle}) 
with $n=0$;
\begin{align}
\left\{\begin{array}{c}
d_{s,j,k_z} = c_{n=0,j,k_z} \ \ \ |y_j|>\frac{L_y}{2}, \\
d_{b,j,k_z} = c_{n=0,j,k_z} \ \ \ |y_j|<\frac{L_y}{2}. \\
\end{array}\right. 
\end{align}
${\cal H}^{\prime}$ denotes the interaction part among electrons in the bulk states.  
We assume that electrons in the surface states are non-interacting. A 
finite spatial gradient of the confining 
potential ($|\partial_y V(y)|$) together with small roughness along the 
$x$-direction induces a 
mixing between $c_{n=0,j,k_z}$ and $c_{n=0,m,k_z}$ with $j\ne m$. This leads to a finite single-particle 
coupling between the surface and bulk states as in ${\cal H}_c$. For simplicity, we assume 
that the coupling preserves the momentum $k_z$.

To see the effect of this surface-bulk coupling, let us introduce real-time time-ordered 
Green functions for surface and bulk states;
\begin{eqnarray}
G_{\mu\nu}(j,t;m,t^{\prime};k_z) \equiv \frac{1}{Z_t} {\rm Tr}\Big[e^{-\beta{\cal H}_t} 
{\cal T}_t \big\{d_{\mu,j,k_z}(t) d^{\dagger}_{\nu,m,k_z}(t^{\prime})\big\}\Big], \nonumber  
\end{eqnarray}
with $\mu,\nu=s,b$ and  $d_{\mu,j,k_z}(t)   
\equiv e^{i {\cal H}_{t} t} d_{\mu,j,k_z} e^{-i {\cal H}_{t}t}$. According to Eq.~(\ref{hami-sup-2}),   
equation of motions (EOMs) for these Green functions are given by,    
\begin{align}
& \big( -i\partial_t +  (\epsilon_{s,j}(k_z)-\mu)\big) 
G_{ss}(j,t;j,t^{\prime};k_z) = - \delta(t-t^{\prime}) \nonumber \\
&\hspace{2.2cm} - \sum^{|y_m|<\frac{L_y}{2}}_{m} T_{j,m} G_{bs}(m,t;j,t^{\prime};k_z),  \nonumber \\  
& \big( -i\partial_t + (\epsilon_{s,j}(k_z)-\mu) \big) G_{bs}(m,t;j,t^{\prime};k_z)  \nonumber \\ 
& \hspace{2.2cm} 
= - \sum^{|y_n|<\frac{L_y}{2}}_{n} T^{*}_{j,n} G_{bb}(m,t;n,t^{\prime};k_z). \label{EOMforG}  
\end{align} 
The coupled EOMs can be further solved in favor for the surface Green function, 
\begin{align}
&iG_{ss}(j,j;k_z,\omega) = \int^{\infty}_{-\infty} ds e^{i\omega s} \!\ iG_{ss}(j,t+s,j,t;k_z)  \nonumber \\ 
& = \frac{i}{\omega- \big(\epsilon_{s,j}(k_z) - \mu\big)} \bigg\{ 
1 - \frac{i}{\omega-\big(\epsilon_{s,j}(k_z) - \mu\big) }  \nonumber \\ 
& \hspace{0.5cm} \times \sum^{|y_n|,|y_m|<\frac{L_y}{2}}_{n,m} 
T_{j,n} T^{*}_{m,j} \!\ iG_{bb}(n,m;k_z,\omega)\bigg\}. \label{EOMforG2}
\end{align} 
When the surface-bulk coupling $T_{j,m}$ is small enough, the bulk Green function 
$G_{bb}(n,m;k_z,\omega)$ in the right hand side could be replaced by the Green function determined 
{\it only} by the bulk Hamiltonian ${\cal H}_b$. Such a Green function is diagonal in $n$ and $m$, 
because ${\cal H}_b$ respects the translational symmetry 
along the $x$-direction;
\begin{equation} \begin{aligned}
G_{bb}(j,t;m,t^{\prime};k_z) = G^{0}_{bb}(j,t;j,t^{\prime};k_z) \delta_{j,m} + {\cal O}(T^2). 
\end{aligned} \end{equation}

When $k_z$ is proximate to the two Fermi points, $k_z \simeq \pm k_F$, 
$G^{0}_{bb}(j,t;j,t^{\prime};k_z)$ in the 
right hand side or its Fourier transform can be further replaced by eqs.~(\ref{realG-fromR},\ref{rho-mss-a}) 
or by eqs.~(\ref{realG-fromR},\ref{rho-mxw-a}); 
\begin{align}
G^{0}_{bb}(j,j;\omega,k_z) &= \int^{\infty}_{-\infty} ds e^{i\omega s} \!\ iG^{0}_{bb}(j,t+s;j,t;k_z), \\ 
& = \left\{\begin{array}{cl} 
G_{+}(k_z-k_F,\omega) & \ \ {\rm for} \!\ \ \ k_{z} \simeq k_F, \\
G_{-}(k_z+k_F,\omega) & \ \ {\rm for} \!\ \ \ k_z \simeq - k_F. \\
\end{array}\right. \label{EOMforG3}
\end{align}
On the one hand, Eq.~(\ref{EOMforG2}) can be rewritten in the following way 
up to the second order in $T_{j,m}$; 
\begin{align}
&G_{ss}(j,j;k_z,\omega) = \nonumber \\ 
&\frac{1}{\omega- \big(\epsilon_{s,j}(k_z)-\mu\big) 
- \sum_{m} |T_{j,m}|^2 \!\ 
G^{0}_{bb}(m,m;k_z,\omega)} + {\cal O}(T^4)  \label{SS2}
\end{align} 

Eq.~(\ref{SS2}) together with Eqs.~(\ref{EOMforG3},\ref{realG-fromR},\ref{rho-mss-a},\ref{rho-mxw-a}) 
dictates that {\it all} the 
chiral surface arc states within the following energy-momentum region (`dissipative region');
\begin{eqnarray}
|\omega| > u|k_z - \sigma k_F| \label{SS3}
\end{eqnarray}
($\sigma=\pm$) acquire a finite life time due to the bulk-surface coupling and the 
dissipative nature of the bulk electronic states in the same region. 
More generally, the region can be seen as the low-energy 
limit of the following momentum-energy region by,  
\begin{eqnarray}
|\omega| > \Big|\frac{\hbar^2k^2_z}{2m_{*}} - \mu \Big| \label{SS4}
\end{eqnarray}
(Fig.~\ref{2}(b) or Fig.~\ref{2a}). 
Those chiral surface states outside the dissipative region have an infinite life 
time at least within our model for the bulk-surface coupling.

\begin{figure}[t]
\centering
\includegraphics[width=0.47\textwidth]{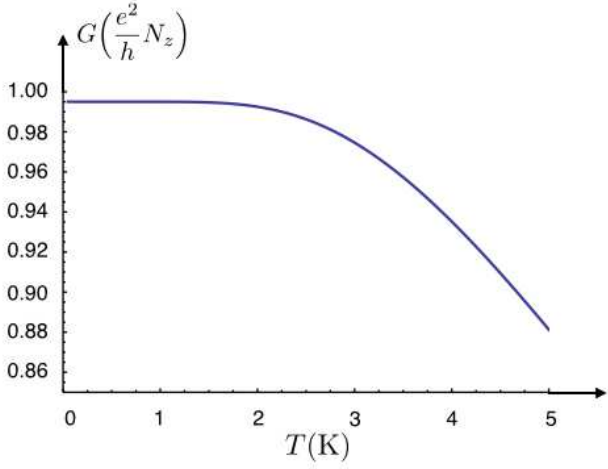}
\caption{(color online) Temperature dependence of the in-plane surface conductance. 
We choose $\epsilon_{b}(k_z)= \hbar^2k^2_z/(2m_{*})$ 
with $m_{*}=m_e/18$ and $\mu=0.355$ meV.}
\label{10}
\end{figure}

\subsection{Temperature dependence of in-plane conductance} 
All the chiral fermi surface states {\it on} 
the Fermi level are outside the dissipative region (Fig.~\ref{2a}). Having an infinite life 
time, they give 100 $\%$ transmission for the in-plane conductance. 
Thus, the surface conductance at $T=0$ is naturally quantized to be 
$G_s = N_z e^2/\hbar$ ($N_z$ is the number of chiral edge modes within 
each surface). When temperature increases, the surface states within the 
dissipative region are thermally activated. Since they have a finite life time, 
the surface conductance may decrease at finite temperature.

To show this crude idea explicitly, we employ the Landauer formula~\cite{datta} for the surface conductance;
\begin{eqnarray}
G_{s} = \frac{e^2}{\hbar L_x} \sum_{j,k_z} T_{j,k_z} 
\Big(-\frac{\partial f}{\partial \epsilon}\Big)_{|\epsilon=\epsilon_{s,j}(k_z)}. \nonumber 
\end{eqnarray}
The summation over chain index $j$ ($y_j \equiv \frac{2\pi l^2}{L_x} j$)  
ranges over a half of the system inlcuding one boundary, 
\begin{eqnarray}
0<y_j < y_{j,{\rm max}} \nonumber
\end{eqnarray}
with $y_{j,{\rm max}} \gg \frac{L_y}{2}$. The summation 
over $k_z$ ranges over the first Brillouin zone of the DW phase; $-k_F < k_z <k_F$.  
For simplicity, we assume that the transmission coefficient of the chiral surface state 
is zero in the dissipative region, while it is in-plane group velocity otherwise;
\begin{eqnarray}
T_{j,k_z} = \left\{\begin{array}{cc} 
0 & {\rm for} \ \ \ \epsilon_{s,j}(k_z)> |\epsilon_b(k_z) - \mu| \\
\frac{\partial \epsilon_{s,j}(k_z)}{\partial k_x} & {\rm for} \ \ \  
\epsilon_{s,j}(k_z) < |\epsilon_b(k_z) - \mu| \\
\end{array}\right. 
\end{eqnarray}
Here $\epsilon_{s,j}(k_z)=\hbar^2k^2_z/(2m_{*}) + V(y_j)$ and $y_j \equiv k_x l^2$. 
With this simplification, the surface conductance can be evaluated as  
\begin{eqnarray}
G_s = \frac{e^2}{h} \sum_{k_z} \bigg(1- \frac{1}{e^{\beta |\epsilon_{b}(k_z)-\mu|}+1}\bigg)
\end{eqnarray}
irrespective of a specific form of the confining potential $V(y)$.  
At the zero temperature, $G_s$ takes the quantized value ($N_ze^2/\hbar$). 
At finite temperature, $G_s$ decreases monotonically on increasing the 
temperature (Fig.~\ref{10}). 

\section{conclusion and outlook}
In this paper, we develop a 
theory of transport properties of DW phases in 3D metal/semimetal 
under high magnetic field, which break the translational symmetries along the field 
direction. Such DW phases have low-energy collective excitations (phason excitations). 
Being comprised of fluctuations of electronic displacement (and current) 
along the field direction, the low-energy collective excitations have 
significant impacts on the longitudinal (optical) conductivity along the field direction. In section~III, 
we calculate the conductivity in the presence of backward scattering type disorders 
and observe that this is indeed the case. With an 
addition of further theoretical studies, the calculated optical conductivities may provide better 
qualitative understandings of non-linear I-V characteristics of the DW phases along the 
field direction. 
  
Note also that both Gaussian and Born approximations used in sec.~III   
become invalid, when the disorders kill the DW phase itself. Especially, the single-particle 
backward scattering disorder introduced in eq.~(\ref{single-bcwd}) 
plays role of ``random magnetic field'' in the XY model, 
while the Imry-Ma's argument~\cite{im,sp,fl} dictates that an infinitesimally small random 
magnetic field kills 
the ordered phase of the XY model completely~\cite{im,sp,fl}. 
Thereby, the DW phase at incommensurate 
electron filling will be likely killed even by an 
infinitesimally small single-particle backward scattering type disorder. This is 
the case especially when the inter-chain rigidity term range only over a couple of chains. 
In the thermodynamic limit, the interchain rigidity ranges over magnetic length, within 
which an extensive number of chains are ferromagnetically coupled with one another 
(see Eqs.~(\ref{Jm}) or Eq.~(\ref{xymodel-r})). An 
Imry-Ma's correlation function analysis suggests that, even in the presence of such a 
longer range interchain coupling, the incommensurate DW phase still {\it cannot} survive under 
the small disorder, unless one assume a rather awkward geometry of the system: 
$l L_x \ge L_z L_y$. This observation suggests us to reconsider the identities of two 
low-temperature resistive phases discovered by transport experiments in graphite in 
the quasi-quantum limit. 

In section IV, we formulate a method of calculating the 
single-particle electron spectral function 
in DW and normal phases and observe that the phason excitation also changes 
low-energy 
feature of the spectral function: giving rise to an additional low-energy continuum spectra in the 
spectral function. In the presence of a surface-bulk coupling, the obtained spectral weight in the bulk 
electronic states 
can be trasferred to the spectral functions of the chiral surface Fermi arc states, giving the latter 
states finite life times. Based on a simple model, we argue that this can result in a temperature-dependent 
in-plane surface conductance (conductance perpendicular to the magnetic field direction, 
carried by the chiral surface states). The obtained result may provide possible explanations 
for a recent in-plane transport measurements in graphite~\cite{fauque}.

\section*{ACKNOWLEDGMENTS}

This work was supported by NBRP of China 
(Grant No. 2014CB920900 and Grant No. 2015CB921104). 

\appendix 

\section{Interacting electron Hamiltonian and its bosonization}


\subsection*{Electronic Hamiltonian for a Weyl semimetal under high magnetic field}
As another possible application of our effective boson model and analyses, we may also consider a 
3D semimetal with two degenerate Weyl nodes at ${\bm k}={\bm K}_{+}$ and 
${\bm k}={\bm K}_{-}$, having linear energy dispersion. The kinetic energy part under the 
magnetic field is given by 
\begin{align}
&{\cal H}_{\rm kin} = \nonumber \\
&\ \sum_{\sigma=\pm} \int d{\bm r} \left(\begin{array}{cc} 
\Psi^{\dagger}_{\sigma,1}({\bm r}) & \Psi^{\dagger}_{\sigma,2}({\bm r}) \\
\end{array}\right) \sigma v_F  \hat{\bm \tau} \cdot {\bm \pi}  
 \left(\begin{array}{c} 
\Psi_{\sigma,1}({\bm r}) \\
 \Psi_{\sigma,2}({\bm r}) \\
\end{array}\right), \label{A}  
\end{align}
where $\Psi_{\sigma,\mu}({\bm r})$ stands for a slowly-varying part of two-component 
field operators ($\mu=1,2$) at a Weyl node of ${\bm k}={\bm K}_{\sigma}$ ($\sigma=\pm$). 
Namely, the electron 
creation operator $\Psi^{\dagger}({\bm r})$ is expanded in terms of Bloch wavefunctions at the Weyl 
nodes $\Phi_{{\bm K}_\sigma,\mu}({\bm r})$ as  
\begin{eqnarray}
\Psi^{\dagger}({\bm r}) = 
\sum_{\sigma=\pm} \sum_{\mu=1,2} \Phi^{*}_{{\bm K}_\sigma,\mu}({\bm r}) 
\Psi^{\dagger}_{\sigma,\mu}({\bm r}) 
+ \cdots. \label{expan3}
\end{eqnarray}
The two-component slowly-varying fields are further expanded in terms of 
eigenstates of  Eq.~(\ref{A}). We keep only an eigenmode for the lowest Landau level;
\begin{eqnarray}
\left(\begin{array}{c}
\Psi_{\sigma,1}({\bm r}) \\
\Psi_{\sigma,2}({\bm r}) \\
\end{array}\right) = \frac{1}{\sqrt{\sqrt{\pi}l L_x}} \sum_{j} e^{ik_x x - \frac{(y-y_j)^2}{2l^2}} 
\left(\begin{array}{c} 
\psi_{\sigma,j}(z) \\
0 \\ 
\end{array}\right) , \label{expan2}
\end{eqnarray}    
with the same definitions for $j=1,2,\cdots, L_xL_y/(2\pi l^2)$, 
the momentum along the $x$-direction $k_x \equiv 2\pi j/L_x$ and $y_j \equiv 2\pi l^2 j/L_x$.  
$l$ denotes the magnetic length. We assume again that 
the cyclotron frequency is much higher than the cutoff energy scale 
associated with the kinetic energy along the $z$-direction, to omit the 
higher Landau level terms. This leads to the kinetic energy part 
given only by the slowly varying 
field operator for the LLL electrons, i.e. $\psi_{\sigma,j}(z)$. The kinetic energy 
part thus obtained takes the exact same form as in eq.~(\ref{kin-sup-2}).  
In principle, the interaction part for the Weyl semimetal 
should be derived by a substitution of Eqs.(\ref{expan3},\ref{expan2}) 
into Eq. (\ref{int-sup-1}). Depending on a form 
of the interaction potential $V({\bm r})$ and Bloch wavefunctions at the nodes, 
this may not necessarily lead to a similar form as in Eq.~(\ref{int-sup-2}). 
For simplicity, we assume the same interaction part as in Eq.~(\ref{int-sup-2}) 
for the Weyl semimetal case.

\subsection*{Bosonization of the interacting electron Hamiltonian}

The electron creation operator for each $j$ is bosonized in terms of the two 
conjugate phase variables~\cite{giamarchi},  
\begin{equation}
\psi_{\sigma,j}(z) =  \frac{\eta_{\sigma,j}}{\sqrt{2\pi \alpha}} e^{-i(\sigma \phi_{j}(z)-\theta_{j}(z))}
\label{FO} 
\end{equation}
where $\eta_{\sigma,j}$ is a Klein factor satisfying the anti-commutation relation 
$\{\eta_{\sigma,j},\eta_{\sigma^{\prime},m}\}= \delta_{\sigma\sigma^{\prime}}
 \delta_{jm}$. $\alpha$ is a short-range cutoff for the spatial coordinate $z$. The two 
phase variables are conjugate to each other,  
satisfying $[\phi_{j}(z^{\prime}),\partial_z \theta_{m}(z)]=i \pi \delta(z-z^\prime) 
\delta_{j,m}$. $\phi_{j}(z)$ is the displacement field along the field direction ($z$); 
spatial derivative of $\phi_j(z)$ with respect to $z$ is the electron density,    
\begin{equation}
\pi \rho_{j}(z)=- \partial_z \phi_{j}(z). 
\label{dd}
\end{equation}  
Besides, the momentum conjugate to the displacement field is the 
current density along the field,   
\begin{equation}
\pi \Pi_{j}(z)=\partial_z \theta_{j}(z).    
\end{equation} 
In terms of these phase variables, the kinetic energy part (eq.(\ref{kin-sup-2})) 
is given by;
\begin{eqnarray}
{H}_{\rm kin} = \sum_{j} \int dz \!\ \frac{v_F}{2\pi} \bigg\{  
\big(\pi \Pi_j(z) \big)^2 + \big(\partial_z \phi_j(z)\big)^2 \bigg\}. \label{kin} 
\end{eqnarray}

Due to the Klein factor (Majorana fermions), the interaction part can not be fully bosonized. 
Those terms in eq.~(\ref{int-sup-2}) with $n\ne m$ and $n\ne j$ are accompanied with 
products of four distinct Majorana fermions. It is generally impossible to bosonize 
simultaneously all such terms. On the one hand, a previous parquet equation study by 
Yakovenko clarified that the short-range repulsive interaction such as in eq.~(\ref{short-rep}) 
leads to a density wave order which breaks the translational symmetry along the 
field direction~\cite{yako,bra,aaa}. 
An order parameter of such DW order is given by a particle-hole pairing 
within the same in-plane momentum $\langle \psi^{\dagger}_{+,j}(z) \psi_{-,j}(z) \rangle$. Such 
DW orders are primarily induced by those interactions terms in eq.~(\ref{int-sup-2}) 
with $n=m$ (Fock term) or $n=j$ (Hartree term), while the others play the secondary 
role. To obtain an effective boson theory for the density wave order phase, we thus 
keep only these Hartree and Fock terms (random phase approximation). This approximation leads to 
a following bosonized effective Hamiltonian;  
\begin{align}
&H_{\rm inc} = \sum_{j} \int dz \bigg\{ \frac{uK\pi}{2} \Pi^2_j(z) + 
\frac{u}{2\pi K} \big(\partial_z \phi_j(z)\big)^2 \nonumber \\ 
&\hspace{0.5cm} 
- \sum^{j\ne m}_{j} J_{j-m} \sigma^z_j \sigma^z_m \cos 2\big[\phi_{j}(z) - \phi_{m}(z) \big]
\bigg\},  \label{incom}
\end{align}
for the incommensurate electron filling case and 
\begin{align} 
&H_{\rm half} = H_{\rm inc} \nonumber \\
& - \sum_{j\ne m} \int dz U_{j-m} \sigma^z_j \sigma^z_m \cos 2\big[\phi_{j}(z) + \phi_{m}(z) 
\big],  \label{half}
\end{align}  
for the half electron filling case.  
Here, the Ising variable $\sigma^z_j$ is defined by two Klein factors at each chain $j$, 
$\sigma^{z}_{j}\equiv i\eta_{+,j}\eta_{-,j}=\pm1$. 
To see how Eqs.~(\ref{incom},\ref{half}) are derived, 
let us investigate the Hartree and Fock terms in the following two subsections. 

\subsubsection*{Hartree term}
The interaction part with $j=n$ is given by,
\begin{equation}
\begin{aligned}
&H_{\rm H} 
=  \frac{g}{L_x} \int dz \int dz^{\prime} \sum_{m,j} V_{j-m,0}(z-z^{\prime}) \nonumber \\  
&\hspace{2.5cm} 
\times \psi^\dagger_{j}(z)  \psi^{\dagger}_{m}(z^{\prime}) \psi_{m}(z^{\prime})  \psi_{j}(z)  \\
&\hspace{0.6cm} 
=  \frac{g}{L_x} \int dz \int dz^{\prime} \sum_{m,j} V_{j-m,0}(z-z^{\prime}) \nonumber \\ 
& \hspace{-0.5cm} 
\times \bigg\{ \big(\rho_{+,j}(z) + \rho_{-,j}(z) \big) \big(\rho_{+,m}(z^{\prime}) + \rho_{-,m}(z^{\prime}) \big) 
\nonumber \\  
& \ \ + e^{-2ik_F (z-z^{\prime})}  \psi^\dagger_{+,j}(z)  \psi^{\dagger}_{-,m}(z^{\prime}) \psi_{+,m}(z^{\prime})  \psi_{-,j}(z)  \nonumber \\ 
&\ \ + e^{2ik_F (z-z^{\prime})}  \psi^\dagger_{-,j}(z)  \psi^{\dagger}_{+,m}(z^{\prime}) \psi_{-,m}(z^{\prime})  \psi_{+,j}(z) \nonumber \\   
& \ \ + e^{-2ik_F (z+z^{\prime})}  \psi^\dagger_{+,j}(z)  \psi^{\dagger}_{+,m}(z^{\prime}) \psi_{-,m}(z^{\prime})  \psi_{-,j}(z)  \nonumber \\ 
&\ \ + e^{2ik_F (z+z^{\prime})}  \psi^\dagger_{-,j}(z)  \psi^{\dagger}_{-,m}(z^{\prime}) \psi_{+,m}(z^{\prime})  \psi_{+,j}(z) 
\bigg\},  
\end{aligned}
\label{H1}
\end{equation}
where density operators are defined as 
\begin{align}
\rho_{\pm,j}(z) \equiv \psi^{\dagger}_{\pm,j}(z) \psi_{\pm,j}(z) 
\end{align}
with $\rho_{j}(z) \equiv \rho_{+,j}(z) + \rho_{-,j}(z) = -(\partial_z \phi_j(z))/\pi$. 
Substituting Eq.~(\ref{FO}) into Eq.~(\ref{H1}), we obtain
\begin{align}
&H_{\rm H} 
=  \frac{\tilde{g}}{L_x} \int dz \int dz^{\prime} \sum_{m,j} V_{j-m,0}(z-z^{\prime}) \nonumber \\  
&\times 
\bigg\{ 4\alpha^2 \big(\partial_z \phi_j(z)\big)\big(\partial_{z^{\prime}} \phi_m(z^{\prime})\big) 
\nonumber \\  
& \ \ \ \  + e^{-2ik_F (z-z^{\prime})}  \sigma^z_j \sigma^z_m e^{2i(\phi_j(z)-\phi_m(z^{\prime}))} 
\nonumber \\ 
& \ \ \ \ + e^{2ik_F (z-z^{\prime})}   \sigma^z_j \sigma^z_m e^{-2i(\phi_j(z)-\phi_m(z^{\prime}))}  
\nonumber \\  
& \ \ \ \  - e^{-2ik_F (z+z^{\prime})}   \sigma^z_j \sigma^z_m e^{2i(\phi_j(z)+\phi_m(z^{\prime}))}   
\nonumber \\
& \ \ \ \ \ 
- e^{2ik_F (z+z^{\prime})} \sigma^z_j \sigma^z_m e^{-2i(\phi_j(z)+\phi_m(z^{\prime}))}   
\bigg\} \label{H2} 
\end{align}
with $\sigma^z_j= i\eta_{+,j}\eta_{-,j}$ and 
\begin{eqnarray}
\tilde{g} \equiv g/(2\pi \alpha)^2. \label{tildeg}
\end{eqnarray}
Note that $V_{j-m,0}(z)$ is short-ranged in $z$ and the associated length $l_0$ is 
typically much shorter than a length scale of the slowly-varying phase 
variables $\phi_j(z)$ and $\theta_j(z)$, i.e. 
$l_0 |\partial_z \phi_j(z)|, l_0 |\partial_z \theta_j(z)|  \ll 1$. As such, we further 
employ a gradient expansion and 
keep only the leading order in $l_0 \partial_z \phi_j(z)$ or $l_0 \partial_z \theta_j(z)$. 
This gives  
\begin{align}
&H_{\rm H} 
=  \frac{\tilde{g}}{L_x} \int dz  \sum_{m,j} \overline{V}_{j-m,0} \!\ 
\bigg\{ 2\alpha^2 \big(\partial_z \phi_j(z)\big)\big(\partial_{z} \phi_m(z)\big) \nonumber \\
& \hspace{0.5cm} 
 +   e^{-2k^2_Fl^2_0} \!\ 
\sigma^z_j \sigma^z_m \cos\Big[2\big(\phi_j(z)-\phi_m(z)\big)\Big]  \nonumber \\
&\hspace{1.0cm} - \sigma^z_j \sigma^z_m 
\cos\Big[2\big(\phi_j(z)+\phi_m(z)\big) - 4k_F z\Big] \bigg\} \label{H3}
\end{align} 
where $\int dz e^{-\frac{z^2}{2l^2_0}\pm 2ik_F z} = \sqrt{2\pi} l_0 e^{-2k^2_Fl^2_0}$ is 
used and  
\begin{eqnarray}
\overline{V}_{j-m,0} = \frac{1}{\pi \sqrt{2 \pi} l^{\prime}} e^{-\frac{(y_j-y_m)^2}{2(l^2+l^2_0)}}. \label{H4}
\end{eqnarray} 
The first term in eq.~(\ref{H3}) gives rise to renormalizations of Luttinger parameter $K$ and 
Fermi velocity $u$. The 2nd term takes the same form as the inter-chain rigidity 
term in Eq.~(\ref{incom}). When the spatial coordinate 
$z$ is lattice regularized, $4k_Fz=2\pi n$ at the half filling, where the third term takes the 
same form as the umklapp term in Eq.~(\ref{half}). At the incommensurate electron filling 
case, the umklapp term can be omitted due to a fast oscillating 
component $e^{-i4k_Fz}$~\cite{giamarchi}.   

\subsubsection*{Fock term}
The interaction part with $m=n$ is given by
\begin{align}
&H_{\rm F} 
= \frac{g}{L_x} \int dz \int dz^{\prime} \sum_{m,j} V_{0,j-m}(z-z^{\prime}) \nonumber \\  
&\hspace{2.5cm} 
\times \psi^\dagger_{m}(z)  \psi^{\dagger}_{j}(z^{\prime}) \psi_{m}(z^{\prime})  \psi_{j}(z)  \nonumber \\ 
& \hspace{0.5cm} =  \frac{g}{L_x} \int dz \int dz^{\prime} \sum_{m,j} V_{0,j-m}(z-z^{\prime}) \nonumber \\  
& \hspace{0.5cm} 
\times \bigg\{ \psi^\dagger_{+,m}(z)  \psi^{\dagger}_{+,j}(z^{\prime}) \psi_{+,m}(z^{\prime})  \psi_{+,j}(z) \nonumber \\ 
& \hspace{1.2cm} 
+  \psi^\dagger_{-,m}(z)  \psi^{\dagger}_{-,j}(z^{\prime}) \psi_{-,m}(z^{\prime}) 
\psi_{-,j}(z) \nonumber \\ 
& \hspace{0.2cm} +  e^{-2ik_F(z-z^{\prime})} 
 \psi^\dagger_{+,m}(z)  \psi^{\dagger}_{-,j}(z^{\prime}) \psi_{+,m}(z^{\prime})  \psi_{-,j}(z) \nonumber \\ 
& \hspace{0.2cm}  
+ e^{2ik_F(z-z^{\prime})} \psi^\dagger_{-,m}(z)  
\psi^{\dagger}_{+,j}(z^{\prime}) \psi_{-,m}(z^{\prime})  \psi_{+,j}(z)  \nonumber \\  
& \hspace{0.8cm} 
+  \psi^\dagger_{+,m}(z)  \psi^{\dagger}_{-,j}(z^{\prime}) \psi_{-,m}(z^{\prime})  \psi_{+,j}(z) 
\nonumber \\   
& \hspace{0.8cm}  
+  \psi^\dagger_{-,m}(z)  \psi^{\dagger}_{+,j}(z^{\prime}) \psi_{+,m}(z^{\prime})  
\psi_{-,j}(z) \nonumber \\  
& \hspace{-0.2cm} 
+ e^{-2ik_F (z+z^{\prime})}  \psi^\dagger_{+,m}(z)  \psi^{\dagger}_{+,j}(z^{\prime}) 
\psi_{-,m}(z^{\prime})  \psi_{-,j}(z)  \nonumber \\ 
& \hspace{-0.2cm} 
+ e^{2ik_F (z+z^{\prime})}  \psi^\dagger_{-,m}(z)  \psi^{\dagger}_{-,j}(z^{\prime}) 
\psi_{+,m}(z^{\prime})  \psi_{+,j}(z) \bigg\}.  \label{F1}
\end{align}
As above, we employ the gradient expansion and keep the leading order in 
small $l_0\partial_z \phi_j(z)$ or $l_0 \partial_z \theta_j(z)$. This gives  
\begin{align}
&H_{\rm F} 
=  \frac{\tilde{g}}{L_x} \int dz  \sum_{m,j} \overline{V}_{0,j-m} \nonumber \\ 
& \times 
\bigg\{ - (1+e^{-2k^2_Fl^2_0}) \!\ \alpha^2 \!\ \big(\partial_z \phi_j(z)\big) 
\big(\partial_z \phi_m(z)\big)  \nonumber \\ 
& \hspace{0.5cm} 
- (1-e^{-2k^2_Fl^2_0}) \!\ \alpha^2 \!\ \big(\partial_z \theta_j(z)\big) 
\big(\partial_z \theta_m(z)\big)  \nonumber \\ 
& \hspace{0.8cm} - \!\  \sigma^z_j \sigma^z_m \cos\Big[2\big(\phi_j(z)-\phi_m(z)\big)\Big] 
\nonumber \\
&\hspace{1.0cm} + \sigma^z_j \sigma^z_m 
\cos\Big[2\big(\phi_j(z)+\phi_m(z)\big) - 4k_F z\Big] \bigg\},  \label{F2}
\end{align} 
with 
\begin{eqnarray}
\overline{V}_{0,m} = \frac{1}{\pi \sqrt{2 \pi} l^{\prime}} e^{-\frac{y^2_m}{2l^2}}. 
\end{eqnarray}
The first two terms in eq.~(\ref{F2}) contribute to renormalizations of the Luttinger 
parameter and the Fermi velocity. The 2nd term and the third term contribute to 
the inter-chain rigidity and the umklapp term respectively.  
When combined together, Eqs.(\ref{kin}), (\ref{H3}), (\ref{F2}) lead to Eq.~(\ref{incom}) for 
the incommensurate electron filling case and Eq.(\ref{half}) for the half electron filling case 
respectively with 
\begin{equation}
\begin{aligned}
J_{m} &= \frac{\tilde{g}}{L_x} \big(\overline{V}_{0,m} - e^{-2k^2_Fl^2_0} \overline{V}_{m,0} \big)  \\ 
&= \frac{\tilde{g}}{L_x}  \frac{1}{\pi \sqrt{2\pi} l^\prime} \Big[ e^{-\frac{y^2_m}{2l^2}}- 
e^{-\frac{y^2_m}{2(l^2+l^2_0)}} e^{-2k_F^2l_0^2} \Big], \\ 
U_{m} &= \frac{\tilde{g}}{L_x}  \big(-\overline{V}_{0,m} + \overline{V}_{m,0} \big)   \\ 
&= \frac{\tilde{g}}{L_x}  \frac{1}{\pi \sqrt{2\pi} l^\prime} \Big[ - e^{-\frac{y^2_m}{2l^2}} +  
e^{-\frac{y^2_m}{2(l^2+l^2_0)}} \Big].\\
\end{aligned}
\label{const1}
\end{equation}

For repulsive interaction case ($\tilde{g} >0$), 
the Fock term favors positive $J_{m}$, while the Hartree 
term favors negative $J_{m}$~\cite{fuku1,yf}. When the interaction length $l_0$ is replaced by 
the Thomas-Fermi length~\cite{fw}, we can typically expect that the Fock term dominates the Hartree 
term due to the small factor; $e^{-2k^2_Fl^2_0} \ll 1$. In the following, we consider this 
case and replace $J_{m}$ by its value of the Fock term, 
\begin{eqnarray}
J_{m} = \frac{\tilde{g}}{L_x} \frac{1}{\pi \sqrt{2\pi} l^{\prime}} e^{-\frac{y^2_m}{2l^2}}. \label{Jjm}
\end{eqnarray}     
$U_{m}$ given in eq.~(\ref{const1}) always takes a positive value, making the phason field to be locked 
on $0,\pi/2,\pi,3\pi/2,\cdots$. For simplicity, we take 
\begin{eqnarray}
U_{m} = \frac{\tilde{g}}{L_x} \frac{1}{\pi \sqrt{2\pi} l^{\prime}} e^{-\frac{y^2_m}{2(l^2+l^2_0)}}. \label{Ujm}
\end{eqnarray}

\section{Derivation of the RG equation and RG phase diagrams}
A partition function for the bosonized Hamiltonian is given by 
\begin{align}
Z &= \sum_{\{\sigma^z_j\}}\int D\phi D \Pi  \exp\Bigg[-\int^{\beta}_{0} d\tau \int dz \sum_{j} 
\nonumber \\
& \Bigg\{-i \!\ \Pi_j(z) \partial_{\tau} \phi_j(\tau) + \frac{uK\pi}{2} \big[\Pi_{j}(z)\big]^2 
+ \frac{u}{2\pi K} \big[\partial_z \phi_j(z)\big]^2  \nonumber \\
& \hspace{0.3cm}  
- \sum_{m\ne j} J_{j-m} \sigma^z_j \sigma^z_m \cos 
\big[2\phi_j(z) - 2\phi_m(z)\big]  \nonumber \\
& \hspace{0.5cm}  
- \sum_{m \ne j} U_{j-m} \sigma^z_j \sigma^z_m \cos \big[2\phi_j(z) + 2\phi_m(z)\big] \Bigg\}
\Bigg]. \label{before}
\end{align}
An integration over the momentum variable $\Pi_{j}(z)$ leads to 
\begin{align}
Z &= \sum_{\{\sigma^z_j\}} \int D\phi  \exp\bigg[- \int^{\beta}_{0} d\tau \int dz \nonumber \\
& \hspace{1.0cm} 
\sum_{j} \frac{1}{2\pi K} \Big\{ \big[\partial_{\tau}\phi_{j}(z)\big]^2 
+ \big[\partial_z \phi_j(z)\big]^2 \Big\} \nonumber \\
& \hspace{0.2cm} + 
\sum^{j\ne m}_{j,m} \Big\{ J_{j-m} \sigma^z_j \sigma^z_m \cos \big[2\phi_j(z) - 2\phi_m(z)\big] 
\nonumber \\ 
&\hspace{0.4cm} 
+ U_{j-m} \sigma^z_j \sigma^z_m \cos \big[2\phi_j(z) + 2\phi_m(z)\big] \Big\} 
\bigg],  \label{after}
\end{align}
where we put $u$ to be 1 for simplicity. The $z$-dependence of the displacement field has a 
short-range cutoff ($\alpha = 2\pi/\Lambda$);
\begin{eqnarray}
\phi_j({\bm r}) = \frac{1}{\beta \!\ L_z} \sum_{i\omega_n} \sum_{|k_z|<\Lambda} 
e^{ik_z z - i\omega_n \tau} \phi_j({\bm q}), 
\end{eqnarray}
with ${\bm r}\equiv (z,\tau)$ and ${\bm q}\equiv (k_z,\omega_n)$. 
The displacement field is decomposed into rapidly-varying mode and slowly-varying mode;
\begin{align}
\phi_j({\bm r}) &= \phi_j^{>}({\bm r})+\phi_j^{<} ({\bm r}) \nonumber \\ 
&= \frac{1}{\beta \!\ L_z} 
\sum_{i\omega_n}\sum_{\Lambda^\prime< |k_z| < \Lambda} e^{i {\bm q}\cdot {\bm r}} \phi_j({\bm q})  
\nonumber \\ 
& \hspace{1.2cm} 
+\frac{1}{\beta \!\ L_z} \sum_{i\omega_n} \sum_{ |k_z| < \Lambda^\prime} 
e^{i{\bm q}\cdot {\bm r}} \phi_j({\bm q}), \label{dec}
\end{align}
with ${\bm q}\cdot{\bm r} \equiv k_z z -\omega_n \tau$, $\Lambda^{\prime} \equiv \Lambda e^{-dl}$ 
and small positive $dl$.   
A further integration over the rapidly-varying mode followed by a rescaling of the $z$ coordinate 
and imaginary time $\tau$, $z\rightarrow z e^{-dl}, \tau\rightarrow \tau e^{-dl}, \beta\rightarrow \beta e^{-dl}$, 
gives a partition function for the slowly-varying mode. The partition function thus obtained 
takes essentially the same form as in eq.~(\ref{after}), while the coupling constant therein such as 
$J_{j-m}$, $U_{j-m}$, $K$ and $\beta$ are renormalized. As shown below, 
the renormalization is described by the following 
renormalization group equations; 
\begin{align}
\frac{d  J_{i-j}}{d l} &= \big[2- 2K\coth\frac{\beta \Lambda}{2} \big]J_{i-j} \nonumber \\
&\hspace{0.5cm} 
+ C(\Lambda)K \sum_n \big[J_{i-n}  J_{n-j}+U_{i-n}U_{n-j} \big], \label{RGforJ} \\ 
\frac{d  U_{i-j}}{d l} &= \big[2- 2K\coth\frac{\beta \Lambda}{2} \big]U_{i-j} \nonumber \\ 
&\hspace{0.5cm} 
+ 2C(\Lambda) K \sum_n J_{i-n} U_{n-j},  \label{RGforU}
\end{align}
with $d\beta/dl = - \beta$. The equations will be derived 
in terms of a perturbative expansion with respect to the 
interchain rigidity term and umklapp term. Perturbative treatments for these two are 
mathematically same. We first derive the equation without the umklapp term (next section). 
We then give a brief derivation of the equations with the umklapp term (next next section).  

\begin{figure*}[t]
\centering
\includegraphics[width=1.0\textwidth]{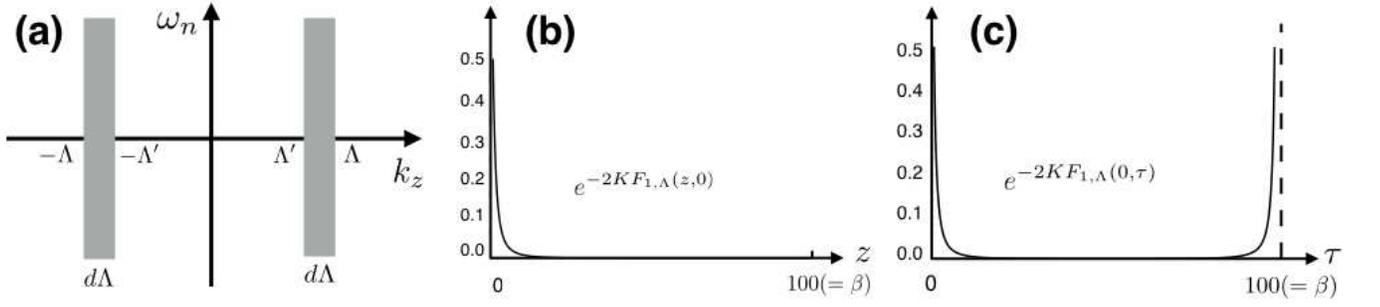}
\caption{(a) Schematic picture for a region of ${\bm q}=(k_z,i\omega_n)$ for  
rapidly-varying mode $\phi^>_j({\bm r})$ (grey color region). 
$d\Lambda = \Lambda dl$ and $\Lambda^{\prime}=\Lambda e^{-dl}$ (see 
the text) (b) $e^{-2KF_{1,\Lambda}(z,0)}$ as function of coordinate variable $z$. (c) 
$e^{-2KF_{1,\Lambda}(0,\tau)}$ as a function of the imaginary time $\tau \in [0,\beta)$. 
$\Lambda=1$, $\beta=100$, and $K=1$.}
     \label{a1}
\end{figure*}

\ 

\subsection*{Incommensurate electron filling case}
The partition function Eq.~(\ref{after}) is expanded in the interchain rigidity term up to the second order,
\begin{widetext}
\begin{align}
\frac{Z}{Z_0}
&=  \frac{1}{Z_0} \sum_{\{\sigma^z_j\}}
\int D\phi e^{-S_0[\phi]}   \Big\{ 1+ \frac{1}{2} \int d^2{\bm r} \sum_{i,j}^{i\ne j} J_{i-j} 
\sigma_i^z \sigma_j^z \sum_{\epsilon=\pm 1}e^{i\epsilon2[\phi_i({\bm r})-\phi_j({\bm r})]}  \nonumber \\ 
&+ \frac{1}{8}\int d^2 {\bm r} \int d^2 {\bm r}^{\prime} 
\sum_{i,j}^{i\ne j}\sum_{m,n}^{m \ne n} J_{i-j}  J_{m-n} \sigma_i^z \sigma_j^z \sigma_m^z \sigma_n^z 
\sum_{\epsilon=\pm 1,\epsilon^\prime=\pm 1} e^{i\epsilon2[\phi_i({\bm r})-\phi_j({\bm r})]} e^{i\epsilon^\prime2[\phi_m({\bm r}^{\prime})-\phi_n({\bm r}^{\prime})]} + {\cal O}(J^3) \Big\} \label{be1} \\
& =  \frac{1}{Z_0^<} \sum_{\{\sigma^z_j\}} \int D\phi^< e^{-S_0[\phi_<]} \  \exp\Bigg[ 
\frac{1}{2}\int d^2 {\bm r}  \sum_{i,j}^{i\ne j} J_{i-j} \sigma_i^z \sigma_j^z \sum_{\epsilon=\pm 1} 
e^{i\epsilon2[\phi_i^<({\bm r})-\phi_j^<({\bm r})]} 
\langle e^{i\epsilon2[\phi_i^>({\bm r})-\phi_j^>({\bm r})]} \rangle_>  \nonumber \\
& + \frac{1}{8}\int d^2{\bm r} \int d^2{\bm r}^{\prime} 
\sum_{i,j}^{i\ne j} \sum_{m,n}^{m \ne n} J_{i-j}  J_{m-n} 
\sigma_i^z \sigma_j^z \sigma_m^z \sigma_n^z 
\sum_{\epsilon=\pm 1,\epsilon^\prime=\pm 1} 
e^{i\epsilon2[\phi_i^<({\bm r})-\phi_j^<({\bm r})]} 
e^{i\epsilon^\prime2[\phi_m^<({\bm r}^{\prime})-\phi_n^<({\bm r}^{\prime})]} \nonumber \\ 
& \Big( \big\langle e^{i\epsilon2[\phi_i^>({\bm r})-\phi_j^>({\bm r})]} 
e^{i\epsilon^\prime2[\phi_m^>({\bm r}^{\prime})-\phi_n^>({\bm r}^{\prime})]} \big\rangle_>- 
\langle e^{i\epsilon2[\phi_i^>({\bm r})-\phi_j^>({\bm r})]} \rangle_> 
\langle e^{i\epsilon^\prime2 [\phi_m^>({\bm r}^{\prime})-\phi_n^>({\bm r}^{\prime})]} \rangle_> 
\Big) \Bigg] 
+ {\cal O}(J^3). \label{af1}
\end{align}
\end{widetext}
$Z_0$ denotes a partition function without $J_{j-m}$, which can be factorized into rapidly-varying 
mode part and slowly-varying mode part;
\begin{eqnarray}
Z_0 = \int d\phi^{<} e^{-S_0[\phi^<]} \cdot 
\int d\phi^{>} e^{-S_0[\phi^>]} \equiv Z^{<}_0 \cdot Z^{>}_0, \label{z0}  
\end{eqnarray}
because  
\begin{align}
S_0[\phi] &= \frac{1}{2\pi K} \int^{\beta}_{0} d\tau \int dz \sum_j \Big\{ 
\big[\partial_{\tau} \phi_j(z)\big]^2 + \big[\partial_z \phi_{j}(z)\big]^2 \Big\} \nonumber \\ 
& = \frac{1}{\beta \!\ L_z} \sum_{j} \sum_{\bm q} \frac{q^2}{K} \phi_j({\bm q}) \phi_j(-{\bm q}). \label{s0}
\end{align}
From Eq.~(\ref{be1}) to Eq.~(\ref{af1}), the integration over the rapid mode has been 
carried out and a functional of the slow mode has been re-exponentiated up to 
the second order in $J_{j-m}$;
\begin{eqnarray}
\langle \cdots \rangle_{>} \equiv \frac{1}{Z^{>}_0} \int d\phi^{>} \cdots e^{S_0[\phi^>]}. \label{def1} 
\end{eqnarray}

The first term in Eq.~(\ref{af1}) gives a one-loop RG correction to the interchain rigidity term, while the 
second term gives a two-loop correction. After the rescaling, the first term takes 
exactly the same form as the interchain rigidity term in eq.~(\ref{after}) with renormalized 
coupling constant;
\begin{eqnarray}
J^{\prime}_{j-m} = J_{j-m} \bigg(\frac{\Lambda}{\Lambda^{\prime}}\bigg)^2  \bigg(1 - 
\frac{2K d\Lambda}{\Lambda} \coth \Big(\frac{\beta \Lambda}{2}\Big)\bigg) + {\cal O}(d\Lambda^2), \nonumber 
\end{eqnarray} 
and $d\Lambda \equiv \Lambda-\Lambda^{\prime}$. This leads to the one-loop RG equation 
\begin{align}
\frac{dJ_{j-m}}{dl} = \bigg[2- 2K \coth\Big(\frac{\beta \Lambda}{2}\Big)\bigg] J_{j-m} \nonumber \\ 
\hspace{1.8cm} + {\cal O}(J^2,d\Lambda^2).  \label{1storder} 
\end{align}

To derive the two-loop RG equation, note first that the second term in eq.~(\ref{af1}) vanishes when 
$i\ne m,n$ and $j\ne m,n$. Finite contributions come from (i) case 
with either one of $i$ and $j$ being same as either one of $m$ and $n$ or (ii) case with both $i$ and 
$j$ same as $m$ and $n$ respectively. The first case generates the two-loop correction to $J_{j-m}$ 
($\epsilon\epsilon^{\prime}=-1$) as well as a new consine term 
$\cos(4\phi_i+2\phi_j-2\phi_m)$ ($\epsilon\epsilon^{\prime}=1$). A scaling 
dimension of the new consine term is $2-6K$ at $T=0$, which is negative for $K\simeq 1$: we 
omit this. The case (i) with $\epsilon\epsilon^{\prime}=-1$ is calculated as 
\begin{widetext}
\begin{align}
\big(i=m, \!\ j\ne n, \!\ \epsilon\epsilon^{\prime}=-1\big) 
&=  \frac{1}{8}\int d^2{\bm r} \int d^2{\bm r}^{\prime} 
\sum^{i\ne j,i\ne n}_{i,j,n} J_{i-j}  J_{i-n} 
\sigma_j^z \sigma_n^z \sum_{\epsilon=\pm1}
e^{i\epsilon2[\phi_i^<({\bm r})-\phi_j^<({\bm r})
-\phi_i^<({\bm r}^{\prime})+\phi_n^<({\bm r}^{\prime})]} \nonumber \\ 
& \hspace{1cm} \big\langle e^{i\epsilon2[\phi_i^>({\bm r})-\phi_j^>({\bm r})- 
\phi_i^>({\bm r}^{\prime})+\phi_n^>({\bm r}^{\prime})]} \big\rangle_> \Big( 1 
-  e^{-4\langle \phi^{>}_i({\bm r}) \phi^{>}_i({\bm r}^{\prime}) \rangle_> }  \Big) \nonumber \\ 
& = \frac{1}{4}\int d^2{\bm r} \int d^2 {\bm r}^{\prime} 
\sum_{i,j,n} J_{i-j}  J_{i-n} 
\sigma_j^z \sigma_n^z 
: \cos\big( 2[\phi_i^<({\bm r})-\phi_j^<({\bm r})
-\phi_i^<({\bm r}^{\prime})+\phi_n^<({\bm r}^{\prime})] \big) : \nonumber \\ 
&\hspace{1cm} e^{-2KF_{1,\Lambda}({\bm r}-{\bm r}^{\prime})} 
4dl KM_{\Lambda}({\bm r}-{\bm r}^{\prime}) + {\cal O}(dl^2), \label{mid2} 
\end{align}
where 
\begin{align}
M_{\Lambda}(z,\tau) &\equiv \cos(\Lambda {z}) \big\{  e^{-|\tau| \Lambda} + 
2 \cosh({\tau} \Lambda) f_B(\Lambda) \big\}, \nonumber \\
e^{-2KF_{1,\Lambda}(z,\tau)} & \equiv \ \ 
\begin{cases}
   \Big(  \frac{B[1+(\Lambda\beta)^{-1}-iz/\beta,1+(\Lambda\beta)^{-1}+iz/\beta ] }{  B[1+(\Lambda\beta)^{-1},1+(\Lambda\beta)^{-1} ] }\Big)^{2K} (1 + \Lambda^2 z^2)^{-K} & 
   \ \ {\rm for} \ \ \tau=0, |1/\Lambda-iz|<\beta, \\
   \Big(  \frac{B[1+(\Lambda\beta)^{-1}-\tau/\beta,1+(\Lambda\beta)^{-1}+\tau/\beta] }{  B[1+(\Lambda\beta)^{-1},1+(\Lambda\beta)^{-1} ] }\Big)^{2K}  (1+\Lambda\tau)^{-2K}  
   &\ \ {\rm for} \ \ z=0, 
\end{cases}
\end{align}
\end{widetext}
with the Bose distribution funciton $f_{B}(x)$ and the Beta function $B(x,y)$.  
From the first line to the second line in eq.~(\ref{mid2}), we 
used $\cos\phi^{<} = :\cos\phi^{<}: e^{-\frac{1}{2}\langle (\phi^<)^2\rangle_<}$ 
and put the cosine term under the normal order denoted by ``$:\cdots :$''~\cite{giamarchi,ng}. 
This enables the Taylor expansion within the normal order. 
Since $e^{-2KF_{1,\Lambda}({\bm r})}$ is localized both in the $z$-direction and in 
the $\tau$-direction (see Figs.~\ref{a1}(b,c)), 
we further employ the gradient expansion within the cosine term, to keep the leading order;
\begin{align}
&\big(i=m, \!\ j\ne n, \!\ \epsilon\epsilon^{\prime}=-1\big) 
= \int d^2{\bm r}   
\sum_{i,j,n} J_{i-j}  J_{i-n} 
\sigma_j^z \sigma_n^z \nonumber \\
& : \cos\big( 2[\phi_j^<({\bm r})-\phi_n^<({\bm r})] \big) : C(\Lambda) K dl + {\cal O}(\Lambda \partial_z \phi_j(z), dl^2)
\end{align}
with
\begin{eqnarray}
C(\Lambda) \equiv \int d^2{\bm r}^{\prime}
e^{-2KF_{1,\Lambda}({\bm r}^{\prime})}  M_{\Lambda}({\bm r}^{\prime}). \label{coflam} 
\end{eqnarray}
This gives the two-loop RG correction to the interchain rigidity term;
\begin{align}
J^{\prime}_{j-m} &= J_{j-m} \bigg( 1 + dl \Big[2 - 2K \coth \Big(\frac{\beta \Lambda}{2}\Big) \Big]   
\bigg) \nonumber \\
&\ \ \ + dl \!\ C(\Lambda) K \sum_i J_{i-j} J_{i-m} + {\cal O}(J^3, dl^2)   
\end{align}
Equivalently, we have 
\begin{eqnarray}
\frac{dJ_{i-j}}{dl} = \Big[2- 2K\coth \Big(\frac{\beta \Lambda}{2}\Big) \Big]J_{i-j} 
+ C(\Lambda)K \sum_n J_{i-n} J_{n-j}  \nonumber \\ 
\label{RGforJ1} 
\end{eqnarray}
   
The second case, (ii) $i=m$ and $j=n$, generates a correction to the Luttinger parameter $K$ 
($\epsilon\epsilon^{\prime}=-1$) as well as a new consine term $\cos(4\phi_i-4\phi_j)$ 
($\epsilon\epsilon^{\prime}=+1$). Having $2-8K$ as its scaling dimension at $T=0$, the 
new consine term is irrelevant around $K\simeq 1$ and we omit this. The second case 
with $\epsilon\epsilon^{\prime}=-1$ can be further calculated as 
\begin{widetext}
\begin{align}
\big(i=m, \!\ j = n, \!\ \epsilon\epsilon^{\prime}=-1\big) 
&=  \frac{1}{8}\int d^2{\bm r} \int d^2{\bm r}^{\prime} 
\sum^{i\ne j}_{i,j} J_{i-j}  J_{i-j} \sum_{\epsilon=\pm1}
e^{i\epsilon2[\phi_i^<({\bm r})-\phi_j^<({\bm r})
-\phi_i^<({\bm r}^{\prime})+\phi_j^<({\bm r}^{\prime})]} \nonumber \\ 
& \hspace{1cm} \big\langle e^{i\epsilon2[\phi_i^>({\bm r})-\phi_j^>({\bm r})- 
\phi_i^>({\bm r}^{\prime})+\phi_j^>({\bm r}^{\prime})]} \big\rangle_> \Big( 1 
-  e^{-8\langle \phi^{>}_i({\bm r}) \phi^{>}_i({\bm r}^{\prime}) \rangle_> }  \Big) \nonumber \\ 
& = \frac{1}{4}\int d^2{\bm r} \int d^2 {\bm r}^{\prime} 
\sum_{i,j} J_{i-j}  J_{i-j}  
: \cos\big( 2[\phi_i^<({\bm r})-\phi_j^<({\bm r})
-\phi_i^<({\bm r}^{\prime})+\phi_j^<({\bm r}^{\prime})] \big) : \nonumber \\ 
&\hspace{1cm} e^{-4KF_{1,\Lambda}({\bm r}-{\bm r}^{\prime})} 
8dl KM_{\Lambda}({\bm r}-{\bm r}^{\prime}) + {\cal O}(dl^2). \label{mid3} 
\end{align}
\end{widetext}
As above, we employ the gradient expansion and keep the leading order. This leads to the two-loop 
correction to the Luttinger parameter;
\begin{align}
&\big(i=m, \!\ j = n, \!\ \epsilon\epsilon^{\prime}=-1\big)  
= 2 \int d^2{\bm r}  
\sum_{i,j} J^2_{i-j} \nonumber \\
& : \sum_{\mu=z,\tau} \big[\partial_{\mu} \phi_i^<({\bm r})- 
\partial_{\mu} \phi_j^<({\bm r})\big]^2 : D(\Lambda) K \!\ dl + \cdots, \nonumber 
\end{align}
with 
\begin{eqnarray}
D(\Lambda) \equiv \int d^2{\bm r} r^2
e^{-4KF_{1,\Lambda}({\bm r})}  M_{\Lambda}({\bm r}). \nonumber 
\end{eqnarray}
Since $J_{i-j}=\frac{\tilde{g}}{L_x} \frac{1}{\pi \sqrt{2\pi} l^{\prime}} e^{-\frac{(y_i-y_j)^2}{2l^2}}$ 
(see Eq.~(\ref{Jjm})), the summation over integer $j$ results in a correction which is  
at most on the order of $1/L_x$. In the leading order in $1/L_x$, 
we thus can ignore this correction. 

\

\subsection*{Half electron filling case}
The partition function Eq.~(\ref{after}) is expanded in the interchain rigidity term 
and umklapp term up to the second order;
\begin{widetext}
\begin{align}
\frac{Z}{Z_0}
= &\frac{1}{Z^{<}_{0}} \sum_{\{\sigma^z_j\}} \int d\phi^{<} \exp\bigg[-S_{0}(\phi^{<}) + \cdots 
+ \frac{1}{2}\int d^2 {\bm r} \sum_{\epsilon=\pm} \sum_{i,j} U_{i-j} \sigma_i^z \sigma_j^z 
e^{i\epsilon2[\phi_i^<({\bm r})+\phi_j^<({\bm r})]} 
\langle e^{i\epsilon2[\phi_i^>({\bm r})+\phi_j^>({\bm r})]}\rangle_> \nonumber \\
& + \frac{1}{8}\int d^2 {\bm r} \int d^2 {\bm r} 
\sum_{\epsilon,\epsilon^\prime=\pm}  
\sum^{i\ne j}_{i,j} \sum^{m\ne n}_{m,n} U_{i-j}  U_{m-n} \!\ \sigma_i^z \sigma_j^z 
\sigma_m^z \sigma_n^z \!\ e^{i\epsilon2[\phi_i^<({\bm r})+\phi_j^<({\bm r})]} e^{i\epsilon^\prime 
2[\phi_m^<({\bm r}^{\prime})+\phi_n^<({\bm r}^{\prime})]} \nonumber \\
& \bigg( \langle e^{i\epsilon2[\phi_i^>({\bm r})+\phi_j^>({\bm r})]} e^{i\epsilon^\prime 
2[\phi_m^>({\bm r}^{\prime})+\phi_n^>({\bm r}^{\prime})]} \rangle_>- 
\langle e^{i\epsilon2[\phi_i^>({\bm r})+\phi_j^>({\bm r})]} \rangle_> \langle e^{i\epsilon^\prime 
2[\phi_m^>({\bm r})+\phi_n^>({\bm r})]} \rangle_>\bigg)  \nonumber \\
& + \frac{1}{4}\int d^2{\bm r}\int d^2 {\bm r}^{\prime} 
 \sum_{\epsilon,\epsilon^\prime=\pm}  \sum^{i\ne j}_{i,j} 
 \sum^{ m\ne n}_{m,n} J_{i-j} U_{m-n} \!\ 
 \sigma_i^z \sigma_j^z \sigma_m^z \sigma_n^z \!\  
 e^{i\epsilon2[\phi_i^<({\bm r})-\phi_j^<({\bm r})]} 
 e^{i\epsilon^\prime 2[\phi_m^<({\bm r}^{\prime})+\phi_n^<({\bm r})]} \nonumber \\
 & \bigg( \langle e^{i\epsilon2[\phi_i^>({\bm r})-\phi_j^>({\bm r})]} e^{i\epsilon^\prime 
 2[\phi_m^>({\bm r}^{\prime})+\phi_n^>({\bm r}^{\prime})]} \rangle_> -\langle e^{i\epsilon2 
 [\phi_i^>({\bm r})-\phi_j^>({\bm r})]} \rangle_> \langle 
 e^{i\epsilon^\prime 2[\phi_m^>({\bm r}^{\prime})+ 
 \phi_n^>({\bm r}^{\prime})]} \rangle_> \bigg) \Bigg] + {\cal O}\big( J^3,U^3,J^2 U, JU^2\big) \label{umklapp1}
\end{align}
\end{widetext}
where $\cdots$ are same as in Eq.~(\ref{af1}). As in the previous section, the first term 
(other than  ``$-S_{0}[\phi^{<}] + \cdots$'') within the exponent gives a one-loop RG 
correction to the umklapp term. The second term with $i=m$, $j\ne n$ and 
$\epsilon\epsilon^{\prime}=-1$ give a two-loop correction to the interchain rigidity 
term. The last term with $i=m$, $j\ne n$ and $\epsilon\epsilon^{\prime}=-1$ gives 
a two-loop correction to the umklapp term. Other cases generate irrelevant consine terms 
or can be omitted up to the leading order in $1/L_x$. Together with Eq.~(\ref{RGforJ1}), 
we finally obtain the RG equations for 
$J_{j-m}$, $U_{j-m}$ and temperature $T$ as in Eqs.~(\ref{RGforJ},\ref{RGforU}). 

\begin{figure*}[t]
\centering
\includegraphics[width=1.0\textwidth]{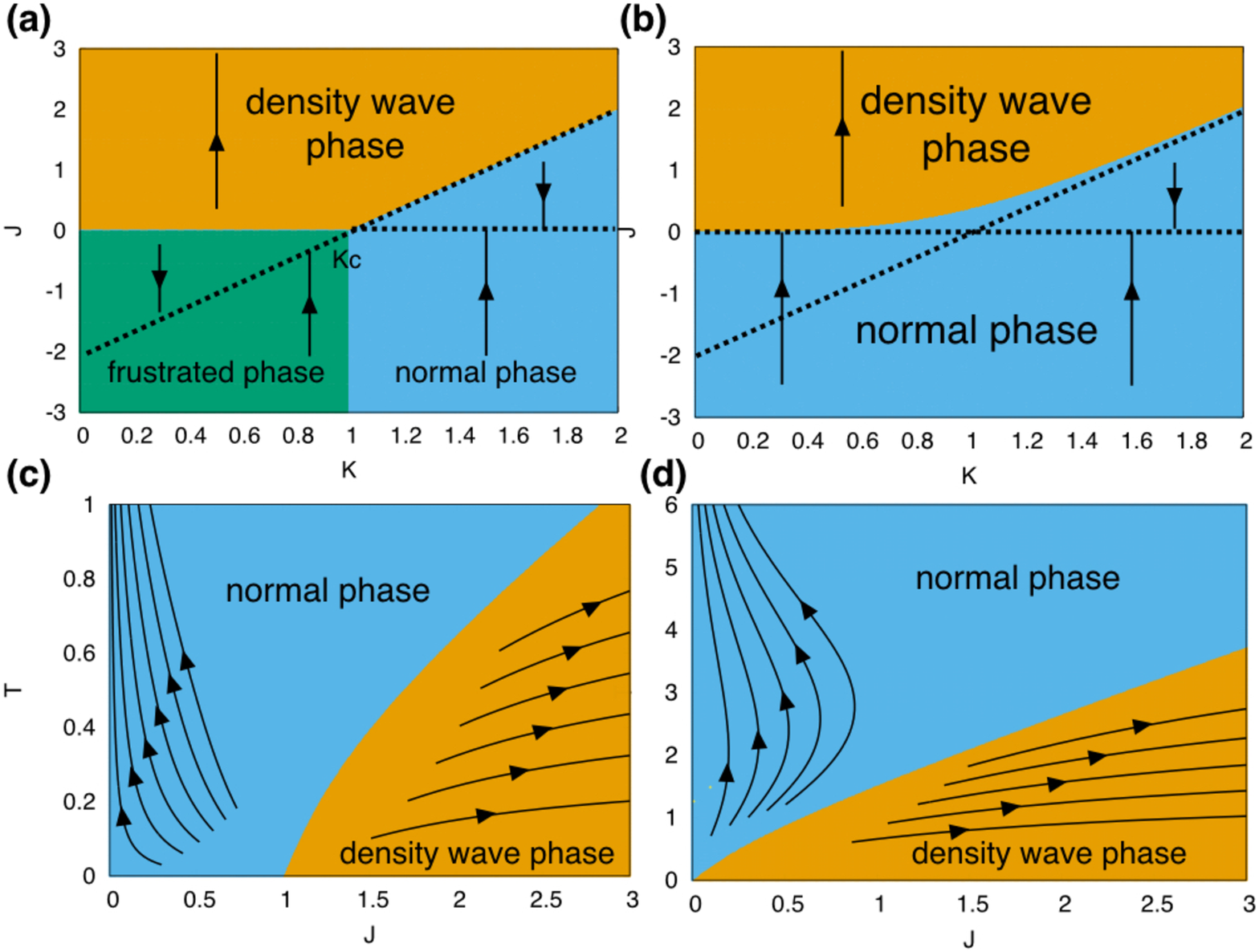}
\caption{RG phase diagrams at incommensurate electron filling case 
(i) normal phase (blue), (ii) DW phase (orange), and (iii) ``frustrated'' phase (green). 
The RG fixed lines (black dashed line), and RG trajectory (black solid line with arrow) are shown. 
(a) $T=0$ $K-J$ phase diagram; (b) Finite temperature (T=0.01) $K-J$ phase diagram. 
(c) $T-J$ phase diagram for $K=1.5$; (d) $T-J$ phase diagram for $K=0.5$.}
\label{a2}
\end{figure*}

\begin{figure}[t]
\centering
\includegraphics[width=0.42\textwidth]{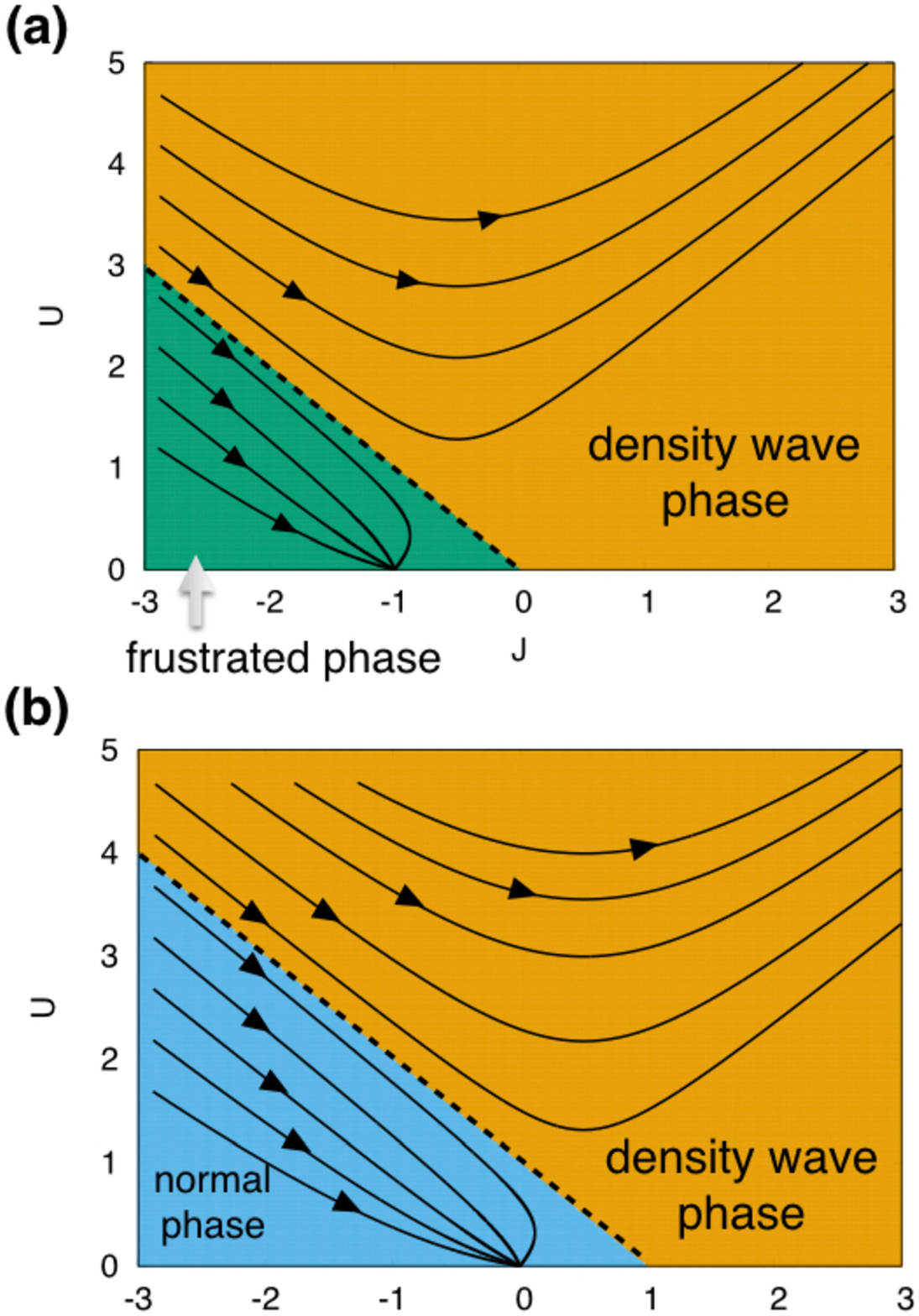}
\caption{$T=0$ $J-U$ phase diagrams at half electron filling case. 
(i) normal phase (blue), (ii) CDW phase (orange), and (iii) ``frustrated'' phase (green). 
(a) $J-U$ phase diagram at $K=0.5$; (b) $J-U$ phase diagram at $K=1.5$.}
\label{a3}
\end{figure}

\subsection*{RG phase diagram} 
Eqs. ~(\ref{RGforJ},\ref{RGforU}) are functional RG equations 
for $J_{m}$ and $U_{m}$ for all $m$. To gain simple ideas of these functional RG equations, 
we reduce them into RG equations for the following two coupling constants,  
\begin{eqnarray}
J \equiv \sum_{m} J_{m}, \ \ U \equiv \sum_{m} U_m. \label{coupling-constant}
\end{eqnarray}
By taking the summation over the integer $j$ in Eqs.~(\ref{RGforJ},\ref{RGforU}), 
we obtain the RG equations for these two coupling constants; 
\begin{align}
& \frac{dJ}{dl}= \Big[ 2- 2K\coth\frac{ \Lambda}{2T} \Big]J + KC \big[J^2 +U^2\big], \label{gr-j-s} \\ 
&\frac{dU}{dl}= \Big[ 2- 2K\coth\frac{ \Lambda}{2T} \Big]U  +2KC J U, \label{rg-u-s}
\end{align}
with $dT/dl = T$.

 Without the umklapp term ($U=0$), three distinct fixed points are identified at $T=0$ phase diagram; 
(i) $J=0$ (decoupled one-dimensional chains: normal-phase fixed point) 
(ii) $J=+\infty$ (the displacement field shows a long-range order, 
$\langle\phi_j(z)\rangle =\phi_0$; DW-phase fixed point) 
(iii) $J=-J_0<0$ (a `frustrated' fixed point). We call the fixed point with negative $J$ to be `frustrated' 
fixed point, because there are no obvious classical solutions of $\phi_j(z)$ which minimize 
the interchain interaction term with negative $J$; for negative $J$, the extensive number 
of chains are antiferromagnetically coupled with one another within the magnetic length $l$. 
At $T=0$, the Luttinger parameter has a critical value $K=K_c$ above which 
infinitesimal small $J$ is always irrelevant, and below which small positive/negative $J$ goes 
to the DW fixed point/`frustrated' fixed point respectively. At finite temperature, the critical 
value $K_c$ goes to zero (Fig.~\ref{a2}(b)). There is a finite-temperature 
phase transition between the DW phase with a fixed point of $J=+\infty$ and $T\ne 0$ 
and the normal phase with a fixed point of $J=0$ and $T= +\infty$ (Figs.~\ref{a2}(c,d)).  

RG phase diagrams with the umklapp term are determined by three fixed points, (i) 
$(J,U)=(0,0)$ normal phase fixed point, (ii) $(J,U)=(+\infty,+\infty)$ DW-phase fixed point 
and (iii) $(J,U)=(-|J_0|,0)$ frustrated fixed point. As shown in Figs.~\ref{a3}(a,b), the interchain rigidity 
and the umklapp term help with each other for positive $J$, so that they grow up into 
the DW-phase fixed point. For negative $J$, the umklapp term is always 
renormalized to zero, which leads to either the normal phase fixed point ($K>K_c$) 
or the frustrated fixed point $(K<K_c)$.   
  
\section{conductivity calculation}
 
\subsection{Imaginary-time time-ordered correlation function and its path integral}
In terms of respective Lehmann representation, the real-time correlation 
function in eq.~(\ref{condzz}) is obtained from an imaginary-time time-ordered function 
by an analytic continuation in the complex $\omega$ plane,
\begin{align}
\sigma_{zz}(\omega) &= Q_{zz} (i\omega_n= \omega + i\eta), \label{Mat1} \\
Q_{zz}(i\omega_n) &\equiv \int^{\beta }_{0} d\tau \!\ Q_{zz}(\tau) \!\ e^{i\omega_n \tau}, \label{Mat2} \\ 
Q_{zz}(\tau) & \equiv - \frac{1}{ V} 
{\rm Tr} \big[\hat{\rho}_G \!\ T_{\tau} \{\hat{J}_z(\tau)  \!\  
\hat{P}_{z}\}\big] , \label{Mat3}
\end{align}
with $\hat{J}_z(\tau) \equiv e^{\hat{K}\tau} \hat{J}_z 
e^{-\hat{K}\tau}$. $\hat{J}_z$ and $\hat{P}_z$ are given by 
\begin{align}
\hat{J}_z = & \frac{|e| uK}{\pi} \sum_{j} \int dz \!\ \partial_z \hat{\theta}_j(z),   \nonumber \\ 
\hat{P}_z = &  - \frac{|e|}{\pi} \sum_{j} \int dz \!\ \hat{\chi}_{j}(z). \nonumber 
\end{align}
$\hat{\rho}_G$ and $\hat{K}$ are given by eqs.~(\ref{Ave1},\ref{Ave2}).

The imaginary-time correlation function $Q_{zz}(\tau)$ can 
be further calculated by a path integral,
where the quantum statistical ensemble average with respect to 
Eqs.~(\ref{Ave1},\ref{Ave2}) can be carried 
out by a Gaussian integral over complex variables $\chi_{j}(z,\tau)$ and 
$\theta_{j}(z,\tau)$,
\begin{equation}
\begin{aligned}
Q_{zz}(\tau) = \frac{e^2 uK}{\pi^2  V} \sum_{j,m} \int dz \int dz^{\prime} \!\ 
R_{jm}(\tau;z,z^{\prime}), \ \\ 
R_{jm}(\tau;z,z^{\prime}) \equiv \frac{\int {\cal D} \chi {\cal D}\theta \!\ 
e^{-S} \!\ \partial_z \theta_{j}(z,\tau) \!\ \chi_{m}(z^\prime,0)}{\int {\cal D} \chi {\cal D}\theta \!\ 
e^{-S}}.    
\end{aligned}
\label{MGF}
\end{equation}
Here the action $S$ is given by 
\begin{align}
- S 
&= - \frac{1}{2\beta  L_z N} 
\sum_{{\bm K}} \left(\begin{array}{cc} \theta^{*}({\bm K}) & \chi^{*}({\bm K}) 
\end{array}\right) 
\big[{\bm M}_0({\bm K})\big] 
\left(\begin{array}{c} \theta({\bm K}) \\ 
\chi({\bm K}) \end{array}\right)  \nonumber \\ 
& \hspace{0.2cm} - \frac{1}{L_z N} \sum_{{\bm k}} X(-{\bm k}) \chi({\bm k},i\omega_n=0) 
\nonumber \\
&\hspace{0.5cm} 
- \frac{1}{\beta  (L_z N)^2} \sum_{i\omega_n,{\bm k},{\bm k}^{\prime}} Y(-{\bm k}^{\prime}) 
\chi^{*}({\bm k}-{\bm k}^{\prime},i\omega_n)\chi({\bm k},i\omega_n), \label{action}   
\end{align}
with a $2\times 2$ matrix   
\begin{equation}
\begin{aligned}
& [{\bm M}_0({\bm K})] \equiv \left(\begin{array}{cc} 
\frac{uK k^2_z}{\pi} & \frac{ik_z \omega_n}{\pi} \\
\frac{ik_z \omega_n}{\pi} & \frac{uk^2_z}{\pi K} + 2J(k) + 2U(k) \\   
\end{array}\right),  
\end{aligned}
\label{M0-exp}
\end{equation}
and 
\begin{equation}
\begin{aligned}
J(k) &\equiv \sum^{N}_{n=1} J_{n} |1-e^{ik y_n}|^2, \\  
U(k) &\equiv \sum^N_{n=1} U_{n} |1+e^{ik y_n}|^2.   
\end{aligned}
\label{nagase}
\end{equation} 

In eq.~(\ref{action}), we used  
\begin{align}
\Big( \chi_{j}(z,\tau), \theta_j(z,\tau)\Big) 
& \equiv \frac{1}{\beta  L_z N} \sum_{\bm K} e^{-i\omega_n \tau + ik_z z + ik y_j} \nonumber \\ 
&\hspace{0.5cm} \times \Big( \chi({\bm k},i\omega_n), \theta({\bm k},i\omega)\Big), \nonumber \\ 
\Big(X_j(z),Y_j(z)\Big)  &\equiv \frac{1}{L_z N} \sum_{\bm k} e^{ ik_z z + ik y_j} 
\Big(X({\bm k}),Y({\bm k})\Big), \nonumber  
\end{align}
with $N \equiv L_x L_y/(2\pi l^2)$, the magnetic length $l$, 
${\bm K}\equiv ({\bm k},i\omega_n) \equiv (k_z,k,i\omega_n)$,  ${\bm K}^{\prime} 
\equiv ({\bm k}^{\prime},i\omega^{\prime}_n) \equiv (k^{\prime}_{z},k^{\prime},i\omega^{\prime}_n)$. 
$k_z$ and $k$ are wave vectors conjugate to $z$ and $y_j$ (chain index) respectively. 
$i\omega_n$ is bosonic Matsubara frequency $i\omega_n=2n\pi/\beta$. 
In the ${\bm K}$ space, the correlation function is given by 
\begin{align}
R_{j,m}(\tau;z,z') &\equiv \frac{1}{(\beta L_z N)^2} \sum_{{\bm K},{\bm K}^{\prime}}
e^{i\omega_n \tau - ik_z z - ik y_j} e^{ik^{\prime}_z z^{\prime} 
+ ik^{\prime} y_m} \nonumber \\
& \hspace{-1.5cm} 
\times (-ik_z) \frac{\int {\cal D}\chi {\cal D} \theta \!\ \!\ \theta^{*}({\bm k},i\omega_n) 
\chi({\bm k}^{\prime},i\omega^{\prime}_n) \!\ 
e^{-S}}{\int {\cal D}\chi {\cal D} \theta \!\ 
e^{-S}}. \label{rjm}
\end{align}

\subsection{Generating function formulation} 
The integrand in Eq.~(\ref{rjm}) can be generated from a partition function in the presence 
of external fields,
\begin{align}
&\frac{\int {\cal D}\chi {\cal D} \theta \!\ \!\ \theta^{*}({\bm k},i\omega_n) 
\chi({\bm k}^{\prime},i\omega^{\prime}_n) \!\ 
e^{-S}}{\int {\cal D}\chi {\cal D} \theta \!\ 
e^{-S}} \nonumber \\
&\hspace{0.1cm} 
= \bigg(\frac{\partial^2 \ln Z[A,B]}{\partial B({\bm K}) \partial A^{*}({\bm K}^{\prime})}
\bigg)_{|A,B\equiv 0} \nonumber \\
& \hspace{0.3cm} +  \bigg(\frac{\partial \ln Z[A,B]}{\partial B({\bm K})}\bigg)_{|A,B\equiv 0}  
\bigg(\frac{\partial \ln Z[A,B]}{\partial A^{*}({\bm K}^{\prime})}\bigg)_{|A,B\equiv 0},  \label{gff}
\end{align}
where the partition function is given by  
\begin{align} 
&\ \ \ \ Z[A,B] \equiv \int {\cal D}\chi {\cal D}\theta e^{-S[A,B]},   \nonumber \\
&- S[A,B] \equiv - S + \sum_{{\bm K}} 
\Big[ \left(\begin{array}{cc} 
B^{*}({\bm K}) & A^{*}({\bm K})\end{array}\right) \left(\begin{array}{c} 
\theta({\bm K}) \\ 
\chi({\bm K}) \\
\end{array}\right) \nonumber \\
&\hspace{1cm} +  \left(\begin{array}{cc} 
\theta^{*}({\bm K}) & \chi^{*}({\bm K}) \end{array}\right) 
\left(\begin{array}{c} 
B({\bm K}) \\ 
A({\bm K}) \\
\end{array}\right)\Big], \nonumber  
\end{align} 
with $S$ in eq.~(\ref{action}). A Gaussian integration over 
$\theta({\bm K})$ and $\chi({\bm K})$ leads to a free energy,  
\begin{align}
&\ln Z[A,B] = \ln Z[A=0,B=0]   \ \ +  
\nonumber \\ 
&\sum_{\alpha,\beta=1,2}\sum_{{\bm K},{\bm K}^{\prime}} 
\left(\begin{array}{cc} B^{*}({\bm K}) , & A^{*}({\bm K}) - {\bm X}^{*}({\bm K}) \end{array}\right)_{\alpha} 
\nonumber \\ 
& \hspace{0.8cm} 
 \times [{\bm M}^{-1}]_{(\alpha,{\bm K}|\beta,{\bm K}^{\prime})} \left(\begin{array}{c} 
B({\bm K}^{\prime}) \\ 
A({\bm K}^{\prime}) - {\bm X}({\bm K}^{\prime}) \\
\end{array}\right)_{\beta}, 
\end{align}
with 
\begin{equation}
{\bm X}({\bm K}) \equiv  \frac{1}{2L_z N} X({\bm k}) \delta_{\omega_n,0}, 
\label{Xfield}
\end{equation}
and 
\begin{align}
& [{\bm M}]_{(\alpha,{\bm K}|\beta,{\bm K}^{\prime})} 
\equiv \frac{1}{2 \beta L_z N} [{\bm M}_0({\bm K})]_{\alpha,\beta} 
\delta_{{\bm K},{\bm K}^{\prime}}  \nonumber \\ 
& \hspace{1.5cm} + \frac{1}{\beta  (L_z N)^2} 
Y({\bm k}-{\bm k}^{\prime}) \delta_{\omega_n,\omega^{\prime}_n} 
\delta_{\alpha,2}\delta_{2,\beta},  \nonumber \\ 
& [{\bm M}^{-1}]_{(\alpha,{\bm K}|\beta,{\bm K}^{\prime})} 
\equiv 2 \beta L_z N  \delta_{\omega_n,\omega^{\prime}_n} \Bigg\{ 
[{\bm M}^{-1}_0({\bm k},\omega_n)]_{\alpha,\beta} \delta_{{\bm k},{\bm k}^{\prime}} \nonumber \\
& \hspace{0.1cm} - \frac{2}{L_z N}  [{\bm M}^{-1}_0({\bm k},\omega_n)]_{\alpha,2}
[{\bm M}^{-1}_0({\bm k}^{\prime},\omega_n)]_{2,\beta} 
Y({\bm k}-{\bm k}^{\prime})  \ + \nonumber \\ 
& \hspace{0.1cm}   \sum^{\infty}_{m=2}\frac{(-1)^m 2^m}{(L_z N)^m} 
\sum_{{\bm k}_3,\cdots,{\bm k}_{m+1}}[{\bm M}^{-1}_0({\bm k},\omega_n)]_{\alpha,2} 
[{\bm M}^{-1}_0({\bm k}_1,\omega_n)]_{2,2}\nonumber \\ 
& \hspace{0.2cm}  \cdots \cdots  
[{\bm M}^{-1}_0({\bm k}_{m-1},\omega_n)]_{2,2}  [{\bm M}^{-1}_0({\bm k}^{\prime},\omega_n)]_{2,\beta}  
\times \nonumber \\ 
&  \hspace{0.3cm} \times Y({\bm k}-{\bm k}_1)  
\cdots Y({\bm k}_{m-2}-{\bm k}_{m-1}) Y({\bm k}_{m-1}-{\bm k}^{\prime}) \Bigg\}.     
\label{M}
\end{align}
The $Y$-field does not carry frequency and couples only between two 
$\chi$-fields. Thus, $[{\bm M}^{-1}]_{(2,{\bm K}|1,{\bm K}^{\prime})}$ is 
always proportional to $[{\bm M}^{-1}_{0}({\bm k}^{\prime},i\omega_n)]_{2,1}$. Since 
$[{\bm M}^{-1}_{0}({\bm k}^{\prime},i\omega_n)]_{2,1}$ vanishes at $i\omega_n=0$
and ${\bm X}({\bm K})$ vanishes at $i\omega_n\ne 0$ 
(see eq.~(\ref{M0-exp},\ref{Xfield}) respectively), the 2nd term in the right hand side 
of eq.~(\ref{gff}) does not contribute; 
\begin{align}
\frac{\int {\cal D}\chi {\cal D} \theta \!\ \!\ \theta^{*}({\bm k},i\omega_n) 
\chi({\bm k}^{\prime},i\omega^{\prime}_n) \!\ 
e^{-S}}{\int {\cal D}\chi {\cal D} \theta \!\ 
e^{-S}} & 
= \big[{\bm M}^{-1}\big]_{(2,{\bm K}^{\prime}|1,{\bm K})}.  \label{integrand}
\end{align} 
\begin{figure*}[t]
\centering
\includegraphics[width=0.9\textwidth]{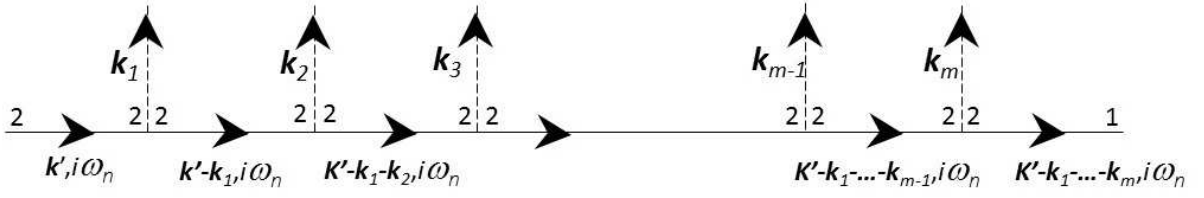}
\caption{(color online) Feynmann diagram for the conductivity along the 
field direction. Solid line with arrow represents the phason fields; 
$\chi$ field denoted by ``2'' and $\theta$ field denoted by ``1''. 
Dotted line with arrow represents the impurity fields ($Y$-field).}
\label{5}
\end{figure*}
\begin{figure*}[t]
\centering
\includegraphics[width=0.9\textwidth]{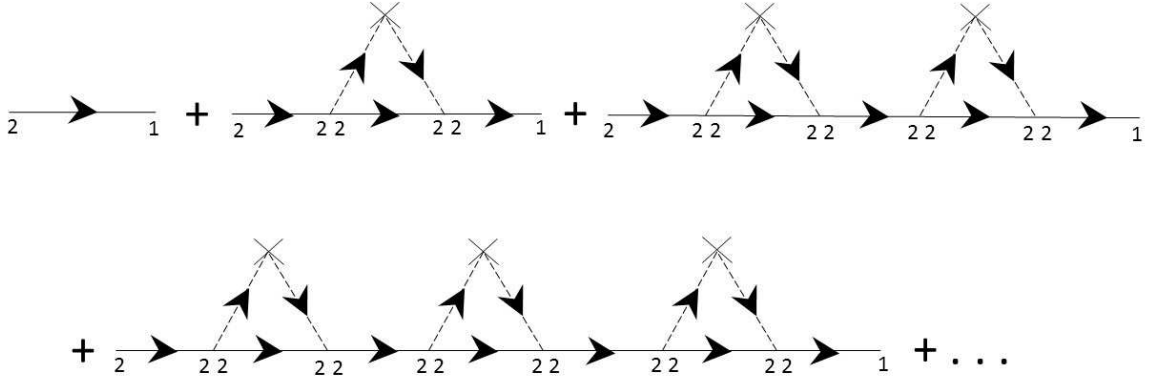}
\caption{(color online) Feynman diagrams for Born approximation, in which the 
self-energy is evaluated in the lowest order approximation.}
\label{6}
\end{figure*}
\subsection{Quenched disorder average and Born approximation}
The conductivity in a given disorder realization was calculated so far.  
In the following, the conductivity will be further averaged 
over different disorder realizations. For simplicity, we take this average by   
\begin{align} 
&\overline{ \cdots } =  \nonumber \\ 
&\frac{\int {\cal D}X_{j}(z) {\cal D}Y_j(z) 
\cdots e^{-\frac{1}{g_x} \sum_j \int dz \!\ X^2_j(z)  - \frac{1}{g_y} \sum_j \int dz \!\ Y^2_j(z)}}
{\int {\cal D}X_{j}(z) {\cal D}Y_j(z)  e^{-\frac{1}{g_x} \sum_j \int dz \!\ X^2_j(z)  
- \frac{1}{g_y} \sum_j \int dz \!\ Y^2_j(z)}}.  
\label{config}
 \end{align}
To this end, we have only to 
take the disorder average of $\big[{\bm M}^{-1}\big]_{(2,{\bm K}^{\prime}|1,{\bm K})}$ 
given in eq.~(\ref{M}). The average leads to a pair contraction among 
even integer number of the $Y$-fields in Eq.~(\ref{M}), such that a pair 
of two impurity lines ($Y$-field lines) contracted share a same momentum, 
\begin{align} 
&\overline{ Y({\bm k}_1) Y({\bm k}_2) \cdots Y({\bm k}_{2M})} = 
\Big(\frac{g_y L_z N}{2}\Big)^M \times \nonumber \\ 
& \sum_{\sigma} \delta_{{\bm k}_{\sigma(1)}+{\bm k}_{\sigma(2)},0} 
 \delta_{{\bm k}_{\sigma(3)}+{\bm k}_{\sigma(4)},0} \cdots 
 \delta_{{\bm k}_{\sigma(2M-1)}+{\bm k}_{\sigma(2M)},0}. \label{contraction}
\end{align} 
Here the summation over $\sigma$ denotes a sum of all possible permutation among 
$2M$ impurity lines under the following 
conditions: $\sigma(2j-1) < \sigma(2j)$ $(j=1,2,\cdots,M)$ 
and $\sigma(1) < \sigma(3) < \cdots \sigma(2M-1)$. 
Those terms with odd integer number  impurity lines vanish 
under the average. The momentum conservation at every impurity 
vertex makes the two external momenta in Eq.(\ref{M}) to be same, 
${\bm k}={\bm k}^{\prime}$. 

Eqs. (\ref{MGF},\ref{rjm},\ref{integrand},\ref{M},\ref{contraction}) lead to 
\begin{align} 
&\overline{Q_{zz}(\tau)} = \frac{2 e^2u K}{\pi^2  V} 
\sum_{j,m} \int dz \int dz' \frac{1}{\beta L_z N}
\sum_{{\omega_n},{\bm k}} e^{i\omega_n \tau} \nonumber \\
& \ \ e^{-ik_z(z-z')-ik_x(y_j-y_m)} (-ik_z)\bigg\{1 + \sum^{\infty}_{M=1} \frac{(2g_y)^M}{(L_zN)^{M}} 
\ \times  \nonumber \\ 
& \sum_{\sigma} \sum_{{\bm k}_1,{\bm k}_2,\cdots,{\bm k}_{2M}} \delta_{{\bm k}_{\sigma(1)}+{\bm k}_{\sigma(2)},0} 
\cdots \delta_{{\bm k}_{\sigma(2M-1)}+{\bm k}_{\sigma(2M)},0} \ \times \nonumber \\
&\hspace{0.1cm} 
[M^{-1}_0({\bm k},\omega_n)]_{2,2} [M^{-1}_0({\bm k}-{\bm k}_1,\omega_n)]_{2,2} \cdots \nonumber \\ 
& \hspace{0.2cm} 
[M^{-1}_0({\bm k}-{\bm k}_1\cdots-{\bm k}_{2M-1},\omega_n)]_{2,2} \bigg\} 
[M^{-1}_0({\bm k},\omega_n)]_{2,1}.  \label{Qzzomega-1}       
 \end{align}  
With Eq.~(\ref{Mat2}), we may rewrite this as, 
\begin{align} 
&\overline{Q_{zz}(i\omega_n)} = \frac{2e^2 uK}{\pi^2 V}  
\sum_{k_z,k,l} \int dz^{\prime\prime} e^{-ik_z z^{\prime\prime}-ik y_l} \nonumber \\  
&\frac{\pi \omega_n}{u^2 k^2_z + 2u\pi K\big[J(k) + U(k)\big] + \omega^2_n} 
\bigg\{1 + \sum^{\infty}_{M=1} \frac{(2g_y)^M}{(L_zN)^{M}}   \nonumber  \\ 
& \sum_{\sigma} \sum_{{\bm k}_1,{\bm k}_2,\cdots,{\bm k}_{2M}} \!\ \delta_{{\bm k}_{\sigma(1)}+{\bm k}_{\sigma(2)},0} 
 \cdots \delta_{{\bm k}_{\sigma(2M-1)}+{\bm k}_{\sigma(2M)},0} \!\ \times \nonumber \\
& \hspace{0.3cm} 
[M^{-1}_0({\bm k},\omega_n)]_{2,2} [M^{-1}_0({\bm k}-{\bm k}_1,\omega_n)]_{2,2} 
\cdots \nonumber \\ 
& \hspace{1.2cm} 
\cdots [M^{-1}_0({\bm k}-{\bm k}_1\cdots-{\bm k}_{2M-1},\omega_n)]_{2,2} \bigg\}  \nonumber \\  
&= \frac{e^2 uK}{\pi^2 l^2 } \!\   
\frac{\omega_n}{2u\pi K U(0) + \omega^2_n}  \bigg\{1 + \sum^{\infty}_{M=1} \frac{(2g_y)^M}{(L_zN)^{M}}   
\nonumber \\  
& \sum_{\sigma} \sum_{{\bm k}_1,{\bm k}_2,\cdots,{\bm k}_{2M}} \!\ \delta_{{\bm k}_{\sigma(1)}+{\bm k}_{\sigma(2)},0} 
\cdots \delta_{{\bm k}_{\sigma(2M-1)}+{\bm k}_{\sigma(2M)},0} \times \nonumber \\ 
& \hspace{0.2cm} 
[M^{-1}_0({\bm k}=0,\omega_n)]_{2,2} [M^{-1}_0(-{\bm k}_1,\omega_n)]_{2,2} \cdots \nonumber \\ 
& \hspace{1.5cm} 
\cdots [M^{-1}_0(-{\bm k}_1\cdots-{\bm k}_{2M-1},\omega_n)]_{2,2} \bigg\},  \label{Qzzomega-2}
\end{align}  
with $J(0)=0$. From eq.~(\ref{Qzzomega-1}) to eq.~(\ref{Qzzomega-2}), we took 
the integral over the center-of-mass coordinates, $Z\equiv \frac{z+z^{\prime}}{2}$ 
and $n \equiv \frac{j+m}{2}$;
\begin{align}
\int dz \int dz^{\prime} = \int dZ \int dz^{\prime\prime} = L_z \int dz^{\prime\prime}, \!\ 
\sum_{j,m} = N \sum_l,  \nonumber   
\end{align} 
with $z^{\prime\prime} \equiv z-z^{\prime}$ and $l \equiv j-m$. From the first line 
to the second line in eq.~(\ref{Qzzomega-2}), 
we took the integral over the relative coordinates $z^{\prime\prime}$ and $l$, making the 
external momenta to be zero, $k_z=k=0$.  

Generally, it is hard to carry out analytically 
the summation over all possible permutations 
in Eq.(\ref{Qzzomega-2}). To gain a simple idea, 
we employ the lowest order approximation for a 
self-energy (Born approximation; compare Fig.~\ref{5} with Fig.~\ref{6}). 
This leads to  
\begin{align} 
&\overline{Q_{zz}(i\omega_n)} 
= \frac{e^2 uK}{\pi^2 l^2 } \!\   
\frac{\omega_n}{2u\pi K U(0) + \omega^2_n}  \nonumber \\
&\hspace{0.4cm} \times \sum^{\infty}_{M=0}  \Big( \frac{2g_y}{L_zN}  [M^{-1}_0(0,\omega_n)]_{2,2} 
\sum_{\bm k} [M^{-1}_0({\bm k},\omega_n)]_{2,2} \Big)^M \nonumber \\ 
&= \frac{e^2 uK}{\pi^2 l^2 } \frac{\omega_n}{ \omega^2_n + 2u\pi K U(0) -   
\frac{2\pi u K g_y }{L_zN} \sum_{{\bm k}} [M^{-1}_0({\bm k},\omega_n)]_{2,2}}, 
\label{MGFB}
\end{align} 
with 
\begin{align}
[M^{-1}_0({\bm k},\omega_n)]_{2,2} &= \frac{\pi u K} {E^2(k_z,k) + \omega^2_n}, \label{M22} \\ 
E(k_z,k) &= \sqrt{u^2 k^2_z + 2u\pi K (J(k) + U(k))}. \label{band} 
\end{align}
Eqs.~(\ref{MGFB},\ref{M22},\ref{band}) are nothing but eqs.~(\ref{MGFB-a},\ref{M22-a},\ref{band-a}) 
respectively.

\subsection{chemical potential type disorder case} 
We can also carry out the same calculation of the conductivity with a 
chemical potential type disorder, 
\begin{equation}
H^{\rm chem}_{\rm imp} = \sum_{j} \int dz \rho_{j}(z) \partial_z \chi_{j}(z).  
\end{equation} 
Following the same process up to Eq.~({\ref{integrand}}), the Fourier transformed 
Matsubara Green function is given by,
\begin{align}
&\frac{\int {\cal D}\chi {\cal D} \theta \!\ \!\ \theta^{*}({\bm k},i\omega_n) 
\chi({\bm k}^{\prime},i\omega^{\prime}_n) \!\ 
e^{-S}}{\int {\cal D}\chi {\cal D} \theta \!\ 
e^{-S}}  
= \big[{\bm M}^{-1}_0\big]_{(2,{\bm K}|1,{\bm K})} \delta_{{\bm K},{\bm K}^{\prime}} \nonumber \\
&+ \sum_{{\bm K}_1,{\bm K}_2} 
\big[{\bm M}^{-1}_0\big]_{(2,{\bm K}^{\prime}|2,{\bm K}_2)} 
{\bm W}({\bm K}_2) {\bm W}^{*}({\bm K}_1) 
\big[{\bm M}^{-1}_0\big]_{(2,{\bm K}_1|1,{\bm K})}   \label{s89}
\end{align} 
with the impurity field ${\bm W}({\bm K}) \equiv -\frac{1}{2L_z N} (ik_z) \rho({\bm k})\delta_{\omega_n,0}$. 
Due to the same reason given above Eq.~(\ref{integrand}),
the second term vanishes in eq.~(\ref{s89}).  
Thus, we see that the chemical potential type disorder 
has no influence on the (optical) conductivity. 

\section{single-particle spectral function calculation}
The four-point correlation function in the frozen lattice 
superconductor (FLS) model is calculated in the 
two limiting cases; one is in the non-superconducting 
(SC) phase (`Maxwell phase') and 
the other is in the superconducting (SC) phase (`Meissner phase'). 
We evaluate the correlation function by respective free 
theories. By use of eq.~(\ref{GreenFLS}), 
we obtain a Fourier transform of the single-particle 
Matsubara Green function, and, by an analytic continuation, we obtain a 
single-particle spectral function. The next subsection 
begins with the four-point correlation function in the 
SC phase of the FLS model, which gives the spectral function in the 
normal phase in the XY model. The next next subsection begins with 
the correlation function in the non-SC phase of FLS model, which 
gives the spectral function in the DW phase in the XY model.  
 
\subsection{Meissner phase in the FLS model (normal phase in the XY model)}
In the Meissner phase, the U(1) phase of the SC order parameter exhibits a  
long range order and the gauge field is expelled from the SC bulk. 
Thus, we omit the coupling between gauge field and U(1) phase, 
and expand the cosine term with respect to a gradient of the U(1) phase;
\begin{align}
& S_{\rm FLS}\big[{\bm a}_{\overline{\bm j}},\theta_{\overline{\bm j}}\big] 
\simeq - \frac{1}{2t} \sum_{\overline{\bm j},\mu} 
 \big(\Delta_{\mu} \theta_{\overline{\bm j}}\big)^2.  
 \label{mss}
\end{align}

With this Gaussian theory, the four-points correlation function is evaluated,   
\begin{align}
&\langle e^{i\theta_{\overline{\bm N}_1}-i\theta_{\overline{\bm N}_1+a_y e_y} 
- i\theta_{\overline{\bm N}_2} + i\theta_{\overline{\bm N}_2+a_y e_y}} \rangle \nonumber \\
&\equiv \exp \bigg[-C+t f(z_1-z_2,\tau_1-\tau_2)\bigg],  \label{whatis4}
\end{align}
with Eq.~(\ref{relation}). The constant $C$ in eq.~(\ref{whatis4}) is given by 
\begin{align}
&\ \ C = \nonumber \\
&\ \ \frac{t}{\beta L_z L_y} \sum_{k_z,k_y,\omega_n} \frac{|1-e^{ik_y a_y}|^2}{6-2\cos k_y a_y 
- 2\cos k_z a_z - 2\cos \omega_n a_\tau}. \nonumber 
\end{align} 
$f(z,\tau)$ and its Fourier component is calculated as
\begin{align}
&f(z,\tau) 
\equiv \frac{1}{\beta L_z} \sum_{\omega_n,q_z} e^{iq_z z - i\omega_n \tau}  
{\cal F}(q_z,i\omega_n)  \nonumber \\
&{\cal F}(q_z,i\omega_n)  \nonumber \\
&=\frac{1}{L_y}\sum_{q_y}
\frac{2-2\cos q_y a_y}{6-2\cos q_y a_y - 2\cos q_z a_z - 2\cos \omega_n a_\tau}  \nonumber \\
&=  1 - \sqrt{\frac{4-2\cos q_z a_z - 2\cos \omega_n a_\tau}{8-2\cos q_z a_z - 2\cos 
\omega_n a_\tau}}  \nonumber \\ 
&= 1- \frac{\sqrt{q^2_z + \omega^2_n} }{2} + \cdots, \label{whatisf}
\end{align}
with bosonic Matsubara frequency, $i\omega_n=2n\pi/\beta$. 
In the final line of eq.~(\ref{whatisf}), 
we took $u=1$ and the lattice constants $a_z,a_{\tau}$ to 
be unit for simplicity. 

Regarding $t$ to be small, we further expand Eq.~(\ref{whatis4}) in $t$. 
The expansion together with Eq.~(\ref{GreenFLS}) leads to 
the following expression of the Matsubara Green function, 
\begin{eqnarray} 
{\cal G}_{\sigma,j}(z,\tau) = -e^{-C} {\rm sgn}(\tau) \!\ \Big\{1 +t f(z,\tau)  + {\cal O}(t^2) \Big\},  
\end{eqnarray} 
whose Fourier-transform is given by a convolution of the Fourier transform 
of the sign function and that of $f(z,\tau)$, 
\begin{align}
{\cal G}_{\sigma,j}(q_z,i{\cal E}_n) 
&\equiv \int^{\infty}_{-\infty} d z \!\ \!\  \frac{1}{2} \int^{\beta}_{-\beta} d \tau \!\ \!\  
e^{-iq_z z + i{\cal E}_n \tau} {\cal G}_{\sigma,j}(z,\tau),  \nonumber \\
&\hspace{-1.9cm} = 2 e^{-C} \!\ \delta(q_z) \frac{1}{i{\cal E}_n} \nonumber \\
& \hspace{-1.7cm} 
- \frac{t \!\ e^{-C}}{\beta} \sum_{{\cal E}^{\prime}_n} \frac{1}{i{\cal E}^{\prime}_n}
\sqrt{q^2_z + ({\cal E}_n-{\cal E}^{\prime}_n)^2} + \cdots  + {\cal O}(t^2).  \label{mid33}
\end{align}
Here we used   
\begin{align} 
\frac{1}{2}
\int^{\beta}_{-\beta}{\rm sgn}(\tau) \!\ e^{i{\cal E}_n \tau} d\tau 
= -\frac{1}{i{\cal E}_n} \big(1 -\cos {\cal E}_n \beta\big) 
= -\frac{2}{i{\cal E}_n},  
\end{align}    
with fermionic Matsubara frequency 
$i{\cal E}_{n}\equiv (2n+1)\pi/\beta$ and 
$i{\cal E}^{\prime}_{n}\equiv (2n^{\prime}+1)\pi/\beta$. The summation over the 
frequency in the right hand side of eq.~(\ref{mid33}) can be carried out 
by an integral in the low-temperature limit ($\beta \rightarrow 0$). 
After the analytic continuation of ${\cal G}_{\sigma,j}(q_z,i{\cal E}_n)$ 
($i{\cal E}_n \rightarrow \omega \pm i\eta$), we obtain the spectral 
function,
\begin{align}
\rho_{\sigma}(q_z,\omega) &\equiv \frac{1}{i} \Big\{ {\cal G}_{\sigma,j}(q_z,i{\cal E}_n = \omega-i\eta) 
\nonumber \\ 
& \hspace{0.7cm} - {\cal G}_{\sigma,j}(q_z,i{\cal E}_n = \omega+i\eta) \Big\},  \nonumber \\ 
&=4\pi e^{-C} \delta(q_z) \delta (\omega) \nonumber \\
&\hspace{0.7cm} + 2\pi t e^{-C} \!\ \sqrt{\omega^2 - q^2_z} 
\!\ \Theta(|\omega|-|q_z|). \label{rho-mss}
\end{align}
With the Fermi velocity $u$ recovered, we obtain eqs.~(\ref{rhoN},\ref{rho-mss-a}).
The imaginary part of the real-time time-ordered Green function is 
given by the spectral function,
\begin{align}
&{\rm Im} \!\ G_{\sigma}(q_z,\omega) \equiv  
- \pi \tanh \Big(\frac{\beta \omega}{2}\Big)  \rho_{\sigma}(q_z,\omega) \nonumber \\
& =  - \pi t e^{-C} \!\ \tanh \Big(\frac{\beta \omega}{2}\Big)  
\sqrt{\omega^2-u^2 q^2_z} \!\ \Theta(|\omega|-u |q_z|).  \label{IM1}  
\end{align}
By the above derivation, the usage of these expressions  
may be limited to low-energy and long wavelength region; $q_z\ll a_z^{-1}, \omega \ll u a_z^{-1}$.

\subsection{Maxwell phase in the FLS model (DW phase in the XY model)} 
In the Maxwell phase, the superconducting order is destroyed by the gauge fluctuation,
while the gradient of the U(1) phase is locked into the associated gauge field under the 
`frozen' limit ($t\rightarrow 0$);
\begin{align}
& {\cal G}_{\sigma,j}(z_1-z_2,\tau_1-\tau_2) = \nonumber \\ 
&- {\rm sgn}(\tau_1-\tau_2) 
  \langle e^{i\theta_{\overline{\bm N}_1}-i\theta_{\overline{\bm N}_1+a_y e_y} 
- i\theta_{\overline{\bm N}_2} + i\theta_{\overline{\bm N}_2+a_y e_y}} \rangle \nonumber \\ 
& = - {\rm sgn}(\tau_1-\tau_2) 
\big\langle e^{i2\pi a_{\overline{\bm N}_1,y} - i2\pi a_{\overline{\bm N}_2,y}}
  \big\rangle_{\rm FLS}. \label{whatisG} 
\end{align} 
Eq.~(\ref{whatisG}) is evaluated with respect to the free theory for the non-superconducting 
phase of the FLS model; 
\begin{align}
& S_{\rm FLS}\big[{\bm a}_{\overline{\bm j}},\theta_{\overline{\bm j}}\big]  \simeq - \frac{1}{2J} \sum_{{\bm j},\mu} 
 (\nabla \times {\bm a})^2_{{\bm j},\mu}. \label{mxw}
\end{align}    
According to this free Maxwell term, two vortex segments separated by $R$ 
interact via the Coulombic force $1/R$~\cite{peskin}; 
the Matsubara Green function is given by, 
\begin{align}
&{\cal G}_{\sigma,j}(z_1-z_2,\tau_1-\tau_2) = \nonumber \\ 
& \ \  - {\rm sgn}(\tau_1-\tau_2)
\frac{4\pi^2 J}{\sqrt{(z_1-z_2)^2 + (\tau_1-\tau_2)^2}}. \label{WhatisG2}  
\end{align} 
The Fourier transform of ${\cal G}_{\sigma,j}(z,\tau)$ is given by the following 
convolution,
\begin{eqnarray}
{\cal G}_{\sigma,j}(q_z,i{\cal E}_n) = \frac{1}{\beta}\sum_{{\cal E}^{\prime}_n}  
\frac{1}{i{\cal E}^{\prime}_n} 
\frac{8\pi^2 J}{\sqrt{q^2_z + ({\cal E}^{\prime}_n-{\cal E}_{n})^2}}  \label{s115}
\end{eqnarray} 
with the Matsubara frequency, ${\cal E}_n=(2n+1)\pi/\beta$. 
The summation over the fermionic Matsubara frequency in the right hand side of Eq.~(\ref{s115}) 
is carried out by an integral in the low temperature limit ($\beta \rightarrow 0$). After the 
analytic continuation of ${\cal G}_{\sigma,j}(q_z,i{\cal E}_n)$ in the complex plane, 
we obtain the spectral function and the imaginary part of the real-time time-ordered 
Green function as follows; 
\begin{equation}
\begin{aligned}
\rho_{\sigma}(q_z,\omega) &= \frac{16 \pi^3 J}{\sqrt{\omega^2 -u^2 q^2_z}} \Theta(|\omega|-u|q_z|), \\
{\rm Im} \!\ G_{\sigma}(q_z,\omega) &= 
-\tanh \Big(\frac{\beta \omega}{2} \Big)
\frac{16\pi^3 J}{\sqrt{\omega^2-u^2 q_z^2}} \!\ \Theta(|\omega|-u|q_z|),     
\end{aligned}
\label{IM2}
\end{equation}
(see eq.~(\ref{rhoDW},\ref{rho-mxw-a})).

\end{document}